%% file: bh_sro.tex
\documentclass[twocolumn,tighten]{aastex63}

\usepackage{libertineotf}

\usepackage{amsmath}
\usepackage{booktabs}

\usepackage{bbding}
\newcommand{\hollowstar}{\raisebox{-.2em}{\text{\FiveStarOpen}}}
\newcommand{\filledstar}{\raisebox{-.2em}{\text{\FiveStar}}}

\newcommand{\unit}[1]{\ensuremath{\,\mathrm{#1}}}
\newcommand{\ie}{\textit{i.e.}, }

\newcommand{\rsat}{{\rm sat}}
\newcommand{\rsym}{{\rm sym}}
\newcommand{\entropy}{\,k_B\,{\rm baryon}^{-1}}

\newcommand{\rbounce}{{\rm bounce}}
\newcommand{\rbaryon}{{\rm baryon}}
\newcommand{\rgrav}{{\rm grav}}
\newcommand{\rPNS}{{\rm PNS}}
\newcommand{\rshock}{{\rm shock}}
\newcommand{\rmax}{{\rm max}}
\newcommand{\rBH}{{\rm BH}}
\newcommand{\tr}{{\rm tr}}

\newcommand{\rsnm}{{\rm SNM}}
\newcommand{\rpnm}{{\rm PNM}}
\newcommand{\rTOV}{{\rm TOV}}

\newcommand{\rPNM}{{\rm PNM}}
\newcommand{\rSNM}{{\rm SNM}}

\newcommand{\reff}{{\rm eff}}

\newcommand{\LS}[1]{${\rm LS}_{#1}$}

\shorttitle{EOS effects on BH formation}
\shortauthors{Schneider et al.}

\begin{document}


\title{Equation of State and Progenitor Dependence of Stellar-Mass Black-Hole Formation}

\input{sections/affiliations}

\input{sections/abstract}



\input{sections/intro}

\input{sections/model}


\input{sections/gr_vs_grep}


\input{sections/results}

\input{sections/conclusions}

\input{sections/acknowledgements}
%
%

\bibliography{bh_sro}

\end{document}

%% file: sections/affiliations.tex
\correspondingauthor{Andr\'{e} da Silva Schneider}
\email{andre.schneider@astro.su.se}

\author[0000-0003-0849-7691]{Andr\'{e} da Silva Schneider}
\affil{The Oskar Klein Centre, Department of Astronomy,
Stockholm University, AlbaNova, SE-106 91 
Stockholm, Sweden}

\author[0000-0002-8228-796X]{Evan O'Connor}
\affil{The Oskar Klein Centre, Department of Astronomy,
Stockholm University, AlbaNova, SE-106 91 
Stockholm, Sweden}

\author{Elvira Granqvist}
\affil{The Oskar Klein Centre, Department of Astronomy,
Stockholm University, AlbaNova, SE-106 91 
Stockholm, Sweden}

\author{Aurore Betranhandy}
\affil{The Oskar Klein Centre, Department of Astronomy,
Stockholm University, AlbaNova, SE-106 91 
Stockholm, Sweden}

\author[0000-0002-5080-5996]{Sean M. Couch}
\affil{Department of Physics and Astronomy, Michigan State University, East Lansing, MI 48824, USA}
\affil{Department of Computational Mathematics, Science, and Engineering, Michigan State University, East Lansing, MI 48824, USA}
\affil{Joint Institute for Nuclear Astrophysics-Center for the Evolution of the Elements, Michigan State University, East Lansing, MI 48824, USA}
\affil{National Superconducting Cyclotron Laboratory, Michigan State University, East Lansing, MI 48824, USA}


\date{\today}

%% file: sections/abstract.tex
\begin{abstract}

The core collapse of a massive star results in the formation of a proto-neutron star (PNS). 
If enough material is accreted onto a PNS it will become gravitationally unstable and further collapse into a black-hole (BH). 
We perform a systematic study of failing core-collapse supernovae in spherical symmetry for a wide range of presupernova progenitor stars and equations of state (EOSs) of nuclear matter. 
We analyze how variations in progenitor structure and the EOS of dense matter above nuclear saturation density affect the PNS evolution and subsequent BH formation. 
Comparisons of core-collapse for a given progenitor star and different EOSs show that the path traced by the PNS in mass-entropy phase space $M_\rgrav^\rPNS-\tilde{s}$ is well correlated with the progenitor compactness and almost EOS independent, apart from the final endpoint. 
Furthermore, BH formation occurs, to a very good approximation, soon after the PNS overcomes the maximum \textit{gravitational} mass supported by a hot NS with constant entropy equal to $\tilde{s}$. 
These results show a path to constraining the temperature dependence of the EOS through the detection of neutrinos from a failed galactic supernova.

\end{abstract}

\keywords{black hole physics -- equation of state -- hydrodynamics -- stars: evolution -- stars: neutron -- supernovae: general}

%% file: sections/intro.tex
\section{Introduction}

The existence of stellar mass black-holes (BHs) has long been inferred from observations of X-ray transients \citep{kubota:98, esin:98, belczynski:12, wiktorowicz:13} and has recently been boosted by multiple gravitational wave (GW) detections of binary BH mergers \citep{abbott:19}. 
The most likely channels in which these BHs form involve the core collapse of massive stars with zero-age main sequence (ZAMS) masses in the range $10~M_\odot \lesssim M_{\rm ZAMS} \lesssim 100~M_\odot$ \citep{burrows:88, oconnor:11}. 
These massive stars may form binary BH systems when found in dense stellar systems \citep{portegieszwart:00, rodriguez:15} or when they have low metallicities and share a common envelope evolution \citep{belczynski:16, mandel:16, stevenson:17, vitale:17}.

The overall picture of how these stellar mass BHs come to be is well understood, even if some of the details are still not. 
At the end of their nuclear fusion cycle, the electron-degenerate iron core of massive stars grows in mass from the ash of the silicon shell burning happening just outside the core.
When it surpasses the effective Chandrasekhar mass, which not only depends on the electron fraction of the core but also its thermal structure \citep{woosley:02}, the gravitational force becomes too strong to resist and the core collapses.
During collapse, stellar core density increases by several orders of magnitude until nuclear densities are reached. 
At this stage, the equation of state (EOS) stiffens, core collapse momentarily halts forming a proto-neutron star (PNS) and sending a shock wave through the still infalling outer core. 
As the shock propagates it loses momentum and stalls due to photodissociation of heavy nuclei and neutrino losses \citep{colgate:66, bethe:85, bethe:90}. 
Neutrino heating and other processes may revive the shock and unbind the stellar mantle resulting in a core-collapse supernova (CCSN) \citep{janka:01, janka:12, ott:18}. 
If the shock is not revived, or even if it is but not enough material is unbound, fallback accretion onto the PNS occurs until the repulsive nuclear forces cannot prevent another gravitational instability and the PNS collapses into a stellar-mass BH \citep{burrows:86, burrows:88, oconnor:11}.

The mechanisms by which CCSNe occur, their explosion energy and ejected mass, if any, all depends on yet poorly constrained details of the equation of state (EOS) of dense matter \citep{lattimer:81, swesty:94, lattimer:00, hempel:12}. 
For a given progenitor even small changes in the EOS \citep{schneider:19} or neutrino transport \citep{melson:15,burrows:18} may dictate whether a successful or failed supernova occurs. 
Thus, accurate predictions of the outcome of CCSNe can be quite difficult. 
Multidimensional simulations predict a variety of CCSN outcomes that include successful and failed explosions which, in the former, can lead to the formation of a NS or a BH \citep{janka:12, burrows:18, ott:18, oconnor:18, pan:18, obergaulinger:19, burrows:20}. 
However, because successful explosions depend on multidimensional physics, explosions in spherically symmetric simulations only occur for the lightest core-collapse progenitors, $M\lesssim10\,M_\odot$ \citep{kitaura:06, fischer:10}; or by artificially enhancing neutrino cross reaction rates \citep{fischer:10, fischer:12, martinez-pinedo:12}, artificially driving PNS contraction \citep{ugliano:12,ertl:16}, or including approximate schemes to emulate multi-dimensional effects \citep{couch:19}. 
Because our main interest is on PNS collapse into a BH, we limit our simulations to spherical symmetry without facilitating explosions.

In the current study we build upon the work of \citet{oconnor:11}, which studied EOS effects in the failed collapse of 106 pre-SNe progenitor stars and 4 different EOSs. 
Here, we perform simulations of failed CCSNe using 49 EOSs from \citet{schneider:19} and 51 supernova progenitors selected from the works of \citet{woosley:02}, \citet{woosley:07}, and \citet{sukhbold:18}. 
By performing a systematic study of failed CCSNe we are able to significantly improve upon our understanding of how different features of the EOS affect PNS evolution until the moment of BH formation. 
Works by \citet{hempel:12, yasin:18, schneider:19} hint that the temperature dependence of the EOS may be the main source of uncertainty in the outcome of CCSNe. 
Particularly, \citet{hempel:12, steiner:13} found, studying the collapse of a single pre-SN progenitor and 7 different EOSs, that the time for BH formation during a failed core-collapse and its initial gravitational mass is proportional to the maximum gravitational mass supported by a NS with constant entropy $s=4\entropy$. 
From our set of simulations we are able to explain the results seen in many previous works of the relations between the EOS of dense matter and progenitor structure and BH formation, including \cite{sumiyoshi:09, oconnor:11, nakazato:12, hempel:12, steiner:13, char:15, pan:18}. 
Furthermore, we show that some characteristics of failed core-collapse are mostly progenitor dependent, \ie almost EOS independent, while others depend as much on the zero-temperature EOS as on its finite-temperature components.

This work is structured as follows. 
In Sec.~\ref{sec:model} we discuss the pre-supernova progenitor stars, the EOSs, and the setup of our simulations. 
We perform the bulk of our simulations using a general relativistic effective potential (GREP). 
Thus, to test the accuracy of this approach we compare BH formation scenarios using a GREP to full general relativistic calculations in Sec.~\ref{sec:gr_vs_grep}.
Results of our simulations and insight gained from them are discussed in Sec.~\ref{sec:results}. 
We conclude in Sec.~\ref{sec:conclusions}. 

%

%% file: sections/model.tex
\section{Model}
\label{sec:model}

\subsection{Progenitors}
\label{ssec:progenitors}

In our study, we perform core collapse simulations for 51 different progenitor stars and analyze PNS evolution up to the point of BH formation. 
For our main results, we analyze six progenitors stars compiled from \citet{woosley:02} and \citet{woosley:07} and listed in Table~\ref{tab:progenitors}. 
These progenitors cover a wide range of compactnesses $0.2\lesssim\xi_{2.5}\lesssim0.9$ as defined by \citet{oconnor:11},
\begin{equation}\label{eq:compactness}
 \xi_M = \left.\frac{M/M_\odot}{R(M_{\rbaryon}=M)/1\,000\unit{km}}\right|_{t=t_0}.
\end{equation}
The compactness $\xi_M$ of a pre-SN progenitor star is, to first order, connected to the time it takes for the coordinate mass $M$ to be accreted by a newly formed PNS \citep{oconnor:11}. 
Thus, $\xi_M$ is also a good indicator of the time for the PNS to collapse into a BH provided we set $M\simeq M_{\rmax}^{\rPNS}$, where $M_{\rmax}^{\rPNS}$ is the maximum mass that can be supported by the PNS.
Following \citet{oconnor:11}, we take $M=2.5M_\odot$. 
The compactness, $\xi_{2.5}$, may be computed at different times during the stellar evolution. 
While we use the values from pre-SN progenitors at the start of collapse, $t_0=0$, so that the values are independent of the EOS used to simulate the core collapse, \citet{oconnor:11} compute $\xi_{2.5}$ at the moment of core bounce, $t=t_{\rm{bounce}}$ and, thus, their values are sensitive to the EOS. 
Therefore, the values quoted here for $\xi_{2.5}$ may differ from the ones of \citet{oconnor:11} for the same progenitor. 
These differences become larger the more compact the progenitor is.

\begin{figure*}[htb]
\plottwo{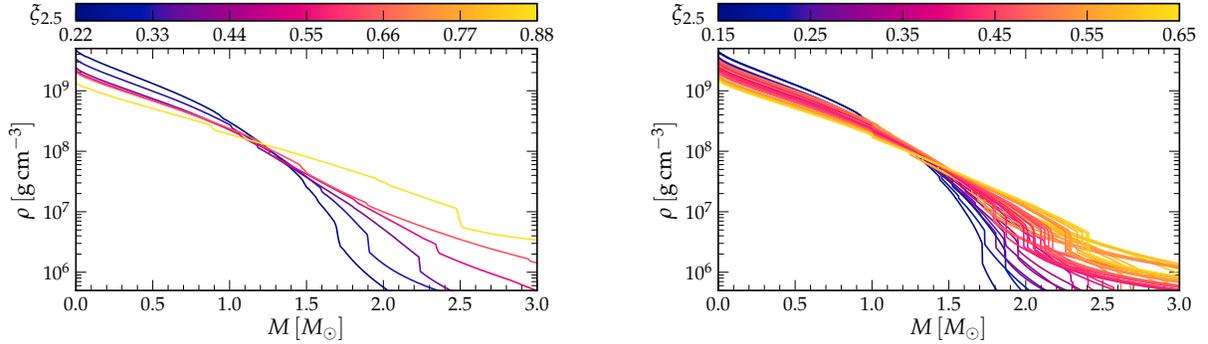}{{figures/swh18_preSN}.pdf}
\caption{\label{fig:preSN} Density as a function of baryonic mass coordinate and compactness $\xi_{2.5}$, Equation~\eqref{eq:compactness}. Plots are for six models from \citet{woosley:02} and \citet{woosley:07} (left, see also Table \ref{tab:progenitors}) and 45 models of \citet{sukhbold:18} with masses between 16-$M_\odot$ and 60-$M_\odot$ (right). 
}
\end{figure*}

We simulate the collapse of the six progenitor stars of  \citet{woosley:02} and \citet{woosley:07} using 49 unique EOSs of \citet{schneider:19}, discussed below. 
Furthermore, we add to our analysis data from the core-collapse of 45 progenitors from \citet{sukhbold:18} that were produced with a reduced (to 1/10 of the default value) mass loss rate and ZAMS masses in the range of $16\,M_\odot\leq M \leq 60\,M_\odot$ in increments of $1\,M_\odot$. 
These pre-SN progenitors cover compactness in the range $0.15\lesssim\xi_{2.5}\lesssim 0.65$. 
For the simulations performed with these 45 progenitors we use a single EOS. 
These simulations are run to test the insights gained from the simulations using the six progenitors in Table~\ref{tab:progenitors} with many different EOSs and to test the impact of the effective gravitational potential on BH formation properties.
The density profiles of the pre-SN progenitor stars are shown in Figure~\ref{fig:preSN}.

\begin{table}[hbt!]
\centering
\caption{ \label{tab:progenitors} Pre-SN progenitor name, zero age main sequence mass $\left(M_{\mathrm{ZAMS}}\right)$, total pre-supernova mass $\left(M_{\mathrm{pre-SN}}\right)$, iron core mass $\left(M_{\mathrm{Fe}}\right)$, and compactness parameters $\xi_{2.5}$ \citep{oconnor:11}. 
Models that start with the letter \texttt{s} have solar metalicity, \texttt{u} have $10^{-4}$ solar metalicity and \texttt{z} have zero metalicity.
Models with WH07 and WHW02 in their name are from \citet{woosley:07} and \citet{woosley:02}, respectively. 
Compactness, $\xi_{2.5}$, values are for pre-SN progenitors at the start of collapse and, thus, differ from the ones of \citet{oconnor:11}, which were computed for a single EOS at the moment of core bounce. 
}
\begin{tabular}{ccDDDD}
\hline
\hline
{Name} & {$M_{\mathrm{ZAMS}}$} &
\multicolumn2c{$M_{\mathrm{pre-SN}}$} &
\multicolumn2c{$M_{\mathrm{Fe}}$} &
\multicolumn2c{$\xi_{2.5}$} \\
{ } & {$[M_\odot]$} & 
\multicolumn2c{$[M_\odot]$} &
\multicolumn2c{$[M_\odot]$} &
\multicolumn2c{ } \\
\hline
\decimals
\texttt{s50WH07}    & 50 &  9.76 & 1.50  & 0.221 \\
\texttt{s25WH07}    & 25 & 15.8  & 1.60  & 0.330 \\
\texttt{z25WHW02}   & 25 & 25.0  & 1.81  & 0.389 \\
\texttt{s40WH07}    & 40 & 15.3  & 1.83  & 0.544 \\
\texttt{u40WHW02}   & 40 & 40.0  & 1.90  & 0.638 \\
\texttt{u75WHW02}   & 75 & 74.1  & 2.03  & 0.879 \\
\hline
\end{tabular}
\end{table}

We focus on PNS evolution as a function of compactness due to our main interest lying in failed CCSNe and BH formation. 
\citet{ertl:16,ebinger:18} have shown that one characteristic parameter of the pre-SN progenitor star, such as the compactness, is not enough to address which progenitors lead to successful or failed CCSNe explosions. 
In fact, as we show in Section~\ref{ssec:swh18}, stars with very similar compactnesses may have quite different accretion histories and, thus, particular likelihoods of leading to a successful explosion that depend on their structures.


\subsection{Equations of State}
\label{ssec:EOS}

For each of the progenitors from \citet{woosley:02} and \citet{woosley:07} we run simulations using 49 unique EOSs from \citet{schneider:19} computed using the open-source code SROEOS of \citet{schneider:17}. 
These EOSs consider only nucleons (protons and neutrons) in the nuclear sector. 
Nuclear matter is assumed to be in thermal equilibrium with a background gas of photons, electrons, and positrons, and to be locally charge neutral. 
Starting from a baseline Skyrme-type EOS which agrees with current nuclear physics constraints \citep{margueron:18, danielewicz:02}, see Table~\ref{tab:constraints}, \citet{schneider:19} constructed four sets of EOSs that vary in at most two experimentally accessible physical parameters: 
(1) set $s_M$ varied the effective mass of nucleons;
(2) set $s_S$ varied the symmetry energy of nuclear matter and its slope;
(3) set $s_K$ varied the incompressibility of symmetric nuclear matter (SNM) and pure neutron matter (PNM); and (4) set $s_P$ varied the pressure of SNM and PNM according to constraints of \citet{danielewicz:02}.

In this work we focus on two EOS sets, $s_M$ and $s_P$. 
Set $s_M$ allows variations in the effective masses of SNM at nuclear saturation density, $m^\star = m_n^\star(n=n_\rsat,y=1/2) \simeq m_p^\star(n_\rsat,1/2)$, and the effective-mass splitting for PNM, $\Delta m^\star = m^\star_n(n_\rsat,0) - m^\star_p(n_\rsat,0)$. 
Here $n$ is the baryon density, $n_\rsat=0.155\unit{fm}^{-3}$ the saturation density, and $y$ the proton fraction, \ie the ratio of protons to nucleons. 
In Skyrme-type models, the effective mass is computed from
\begin{equation}\label{eq:meff}
 \frac{\hbar^2}{2m_t^\star} = \frac{\hbar^2}{2m_t} + \alpha_1 n_t + \alpha_{2} n_{-t}\,,
\end{equation}
where $\alpha_i$ are parameters of the model, $m_t$ and $n_t$ are, respectively, the vacuum mass and the number density of nucleon of type $t$. 
If $t=n$ then $-t=p$ and vice versa. 
These EOSs were constructed such that all EOSs in set $s_M$ have very similar zero-temperature components regardless of choice for $m^\star$ and $\Delta m^\star$.
The finite-temperature component of the EOS, on the other hand, is sensitive to the effective masses of nucleons \citep{prakash:97, steiner:13, constantinou:14} and, thus, differs amongst EOSs in set $s_M$.

The second set analyzed, $s_P$, allows variations in the pressure of zero-temperature SNM and PNM at $4n_\rsat$ based on constraints by \citet{danielewicz:02}. 
The range of pressures allowed at $n=4n_\rsat$ is such that all EOSs can produce cold non-rotating beta-equilibrated neutron stars (NSs) with gravitational mass $M_\rgrav\geq2\,M_\odot$. 
We stress that all EOSs in set $s_P$ have the same effective masses and, therefore, have the same contributions from the thermal component of the EOS. 
In sets $s_M$ and $s_P$ the effective masses and pressures at $n=4n_\rsat$ can, respectively, vary from the baseline EOS values by $0$, $\pm1$ and $\pm2$-$\sigma$ from their average values, see Table~\ref{tab:constraints}. 
Thus, each set contains 25 EOSs with the baseline EOS being present in both sets.

\begin{figure}[htb]
\plotone{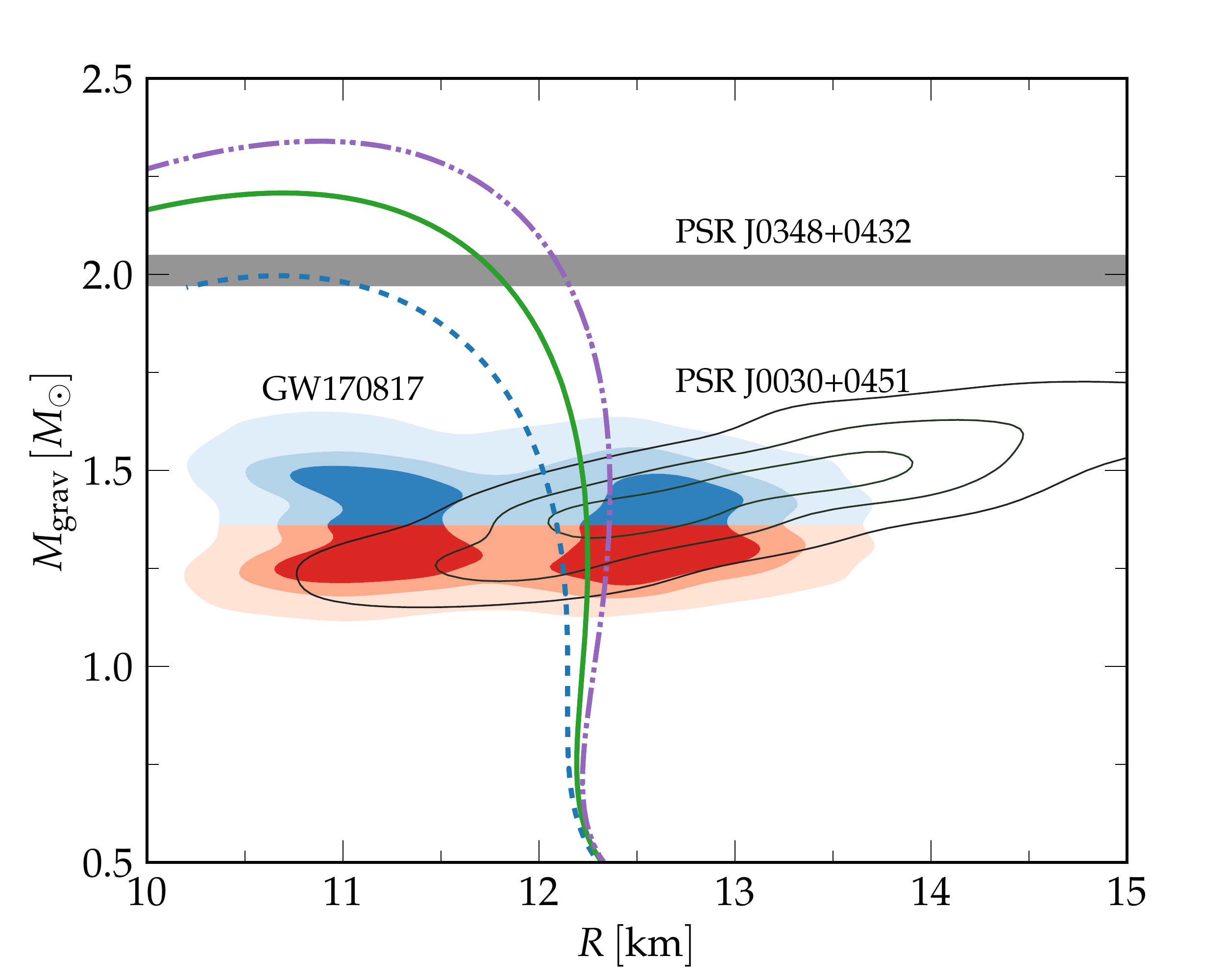}
\caption{\label{fig:mass_rad} Mass-radius relationship for our baseline EOS and the softest and stiffest EOSs in set $s_P$. 
Also plotted are observational constraints from the LIGO-Virgo detection of GW170817 \citep{abbott:19a}, 
the $2.01\pm0.04\,M_\odot$ mass of PSR J0348+0432 as measured by \citet{antoniadis:13}, 
and the mass-radius of PSR J0030+0451 from the NICER collaboration \citep{miller:19}. 
Different shades in the GW170817 data and the contour lines of the PSR J0030+0451 data 
denote the 50\%, 68\%, and 95\% confidence intervals of the observations. 
}
\end{figure}

We focused on these two sets as variations in the effective mass (pressure) keeping pressure  (effective mass) constant probe how core collapse and BH formation are affected by thermal (non-thermal) components of the EOS \citep{yasin:18, schneider:19, schneider:19a}. 
Often, EOSs are discussed in terms of their softness/stiffness. 
Here we make a distinction between what it means for an EOS to be soft or stiff depending on the set we are referring to. 
Since for set $s_M$ the zero temperature component of the EOS is very similar for all EOSs in the set, we refer to the stiffest EOS to be the one where the pressure rises the fastest with increasing temperature. 
Meanwhile, for set $s_P$ the temperature component of all EOSs is the same and, therefore, we refer to the stiffest EOS to be the one where the pressure rises the fastest with increasing density. 
The NS mass-radius relationship for our baseline EOS as well as the softest and the stiffest zero-temperature EOSs in set $s_P$ are shown in Figure~\ref{fig:mass_rad}. 
By construction all EOSs in set $s_M$ show very similar mass-radius curves \citep{schneider:19}. 
Finally, in order for all EOSs to have the same low density behavior, they are joined to an EOS of 3\,335 nuclei in nuclear statistical equilibrium (NSE) at a transition density of $n_\tr = 10^{-3} \unit{fm}^{-3}$ with width $n_\delta = 0.33$ \citep{schneider:17}.

\begin{table}[h!]
\renewcommand{\thetable}{\arabic{table}}
\centering
\caption{\label{tab:constraints} 
Nuclear matter properties of our baseline EOS from constraints of \citet{margueron:18}, \citet{nattila:16}, and \citet{danielewicz:02} and references therein, see \citet{schneider:19} for details. 
1-$\sigma$ variations used to compute EOS sets $s_M$ and $s_P$ are also shown.}
\begin{tabular}{ccDDc}
\hline
\hline
{Set}  &
{Quantity}  &
\multicolumn2c{Baseline}  &
\multicolumn2c{Variation $\sigma$} &
{Unit} \\
\hline
\decimals
$-$ &$n_\rsat$             & 0.155  & .  & $\unit{fm}^{-3}$\\
    &$\epsilon_\rsat$      & -15.8  & .  & \footnotesize{$\unit{MeV\,baryon}^{-1}$}\\
$-$ &$\epsilon_\rsym$      & 32.    & .  & \footnotesize{$\unit{MeV\,baryon}^{-1}$}\\
    & $L_\rsym$            & 45.    & .  & \footnotesize{$\unit{MeV\,baryon}^{-1}$}\\
$-$ & $K_\rsat$            & 230.   & .  & \footnotesize{$\unit{MeV\,baryon}^{-1}$}\\
    & $K_\rsym$            & -100.  & .  & \footnotesize{$\unit{MeV\,baryon}^{-1}$}\\
\hline   
$s_M$&$m^\star$           & 0.75   &   0.10 & $m_n$ \\
     &$\Delta m^\star$    & 0.10   &   0.10 & $m_n$ \\
\hline   
$s_P$&$P^{(4)}_{\rsnm}$   & 125.    &   12.5 & {$\unit{MeV\,fm}^{-3}$} \\
     &$P^{(4)}_{\rpnm}$   & 200.    &   20.0 & {$\unit{MeV\,fm}^{-3}$} \\
\hline
\end{tabular}
\end{table}

\subsection{\textsc{Flash} Simulation Setup}
\label{ssec:FLASH}

For our set of progenitors and EOSs we perform spherically symmetric core-collapse simulations using the \textsc{Flash} simulation framework \citep{fryxell:00, dubey:09}. 
We use a similar \textsc{Flash} setup to that of \citet{oconnor:18} and \citet{couch:19}. 
We review some general features of \textsc{Flash} here and point to references that contain further details about the code and systems of equations solved. 
In the simulations, gravity is treated using an approximate general relativistic effective potential \citep{marek:06, oconnor:18a, oconnor:18b}. 
The equations of motion of the system are solved using a newly-implemented hydrodynamics solver \citep{couch:19,couch:19a} while neutrino transport is computed using the so-called ``M1'' transport \citep{shibata:11, cardall:13, oconnor:15, oconnor:18a, oconnor:18b}. 
Both the hydrodynamics and the radiation transport share a common time step that is set by the light-crossing time of the smallest grid zone and a Courant factor. 
We use a spherical grid with adaptive mesh refinement (AMR) for our simulations that extends out to $2\times10^9\unit{cm}$. 
On the coarsest level there are 320 grid zones with a length of $62.5\unit{km}$. 
We allow up to 10 total levels of refinement resulting in a smallest grid zone with a length of $\simeq122\unit{m}$. 
This small grid zone length improves modeling of the high density region near BH formation time.
We use a Courant factor of 0.8 that results in a hydrodynamic time step of $\simeq0.3\times10^{-6}\unit{s}$. 
We do two radiation substeps per hydrodynamic step and, therefore, the effective CFL factor for the radiation substeps is 0.4. 
We refine based on a combination of the second spatial derivatives of the density and pressure. 

Neutrino-matter interactions that couple neutrino transport to the core-collapse hydrodynamics are computed using the \textsc{NuLib} library \citep{oconnor:15}. 
We consider the same neutrino-matter interactions as \citet{oconnor:15, schneider:19} and reproduced in Table \ref{tab:nu} for completeness.
For each EOS table, a consistent set of neutrino opacities is generated using \textsc{NuLib}. 
We use 18 logarithmically spaced energy groups, with a highest-energy bin of $\simeq250\unit{MeV}$.

\begin{table}[h!]
\renewcommand{\thetable}{\arabic{table}}
\centering
\caption{\label{tab:nu} 
List of neutrino reactions included in simulations.  Interactions taken from NuLib \citep{oconnor:15} and modeled off of \cite{bruenn:85,burrows:06} with corrections from \cite{horowitz:02}.
Interactions with $\nu$ are flavor insensitive, while interactions with $\nu_i$ are flavor sensitive. 
$\nu_x$ denotes heavy-lepton neutrinos while $N$ denotes a nucleon (either neutron or proton) and ${}^A_ZX$ a specific element with $Z$ protons and mass number $A$.
Thermal processes are included via an effective emissivity \citep{oconnor:15}.
}
\small
\begin{ruledtabular}
\begin{tabular}{c  c}
\multicolumn{2}{c}{Production} \\
\hline
Charged-current interactions & 
Thermal Processes \\
\hline
$\nu_e + n \longleftrightarrow p + e^-$ & 
$e^-+e^+\longleftrightarrow \nu_x + \bar\nu_x$ \\ 
$\bar\nu_e + p \longleftrightarrow n + e^+$ & 
$N+N\longleftrightarrow N+N$ \\ 
$\nu_e+{}^A_ZX\longleftrightarrow e^- + {}^A_{Z+1}X$ & 
\hspace{2cm}$+\nu_x + \bar\nu_x$\\
\hline
\multicolumn{2}{c}{Scattering} \\
\hline
Iso-energetic scattering & 
Inelastic Scattering \\
\hline
$\nu + \alpha\longleftrightarrow\nu + \alpha$ & 
$\nu_i + e^- \longleftrightarrow {\nu_i}' + {e^-}'$ \\ 
$\nu_i + p     \longleftrightarrow\nu_i + p$ & \\
$\nu_i + n     \longleftrightarrow\nu_i + n$ & \\
$\nu +{}^A_ZX\longleftrightarrow\nu +{}^A_ZX$ & \\
\end{tabular}
\end{ruledtabular}
\end{table}

%% file: sections/gr_vs_grep.tex
\section{GR vs. GREP}
\label{sec:gr_vs_grep}

In the \textsc{FLASH} code, gravity is included in the hydrodynamic equations using an effective gravitational potential, $\Phi_\reff$, that approximates general relativistic (GR) effects via the ``Case A'' formalism of \cite{marek:06}.
This formalism has never been rigorously tested in the regime of BH formation and, therefore, we do that here.

After the original proposal by \cite{rampp:02}, the GREP was improved by \citet{marek:06} who tested several empirical modifications to the standard TOV potential in order to find one that reduced discrepancies in the early evolution of a $15\,M_\odot$ progenitor model of \citet{woosley:95} when compared to full GR calculations with the \textsc{CoCoNuT} \citep{dimmelmeier:02, dimmelmeier:02a, dimmelmeier:05} and the \textsc{Agile-BoltzTran} codes \citep{liebendorfer:01, liebendorfer:02, liebendorfer:04, liebendorfer:05}. 
For the simulated collapse, \citet{marek:06} found that the best suited modification was their ``Case A'', where a factor of $\Gamma$ was introduced in the integrand of the TOV mass equation.
The ``Case A'' GREP equations are given here for reference,
\begin{align}\label{eq:PhiTOV}
 \Phi_\reff(r)=-4\pi G \int_0^r &
 \left(\frac{m_\rTOV}{4\pi}+r'^3(P+p_\nu)/c^2\right)\nonumber\\
 &\times\frac{1}{\Gamma^2}\left(\frac{\rho+(e+P)/c^2}{\rho}\right)\frac{dr'}{{r'}^2}\,.
\end{align}
Here, $P$ is the matter pressure, $p_\nu$ is the neutrino pressure, $\rho$ is the rest-mass density, $e$ is the internal energy density, and $c$ is the speed of light.  The effective $\rTOV$ mass is given by,
\begin{equation}\label{eq:mTOV} 
 m_\rTOV(r) = 4\pi\int_0^r r'^2\left(\rho+(e+E+\frac{vF}{c^2\Gamma})/c^2\right) \times {\boldsymbol{\Gamma}} dr^\prime\,,
\end{equation} 
where $E$ is the neutrino energy density, $v$ is the fluid velocity, $F$ is the neutrino flux, and the metric function $\Gamma$ is given by
\begin{equation}\label{eq:Gamma} 
 \Gamma=\sqrt{1+\left(\frac{v}{c}\right)^2-\frac{2Gm_\rTOV}{c^2r^\prime}}\,.
\end{equation}
In Equation \eqref{eq:mTOV}, we highlight in bold the additional factor of $\Gamma$ which is not present in the standard TOV equations. In what follows, we ignore the term proportional to ${vF}$ as its contribution to core-collapse dynamics is negligible, ${vF}\ll E$. 

This change, as motivated by \cite{marek:06}, reduces the TOV mass, $m_\rTOV$, in the potential, since $\Gamma<1$.
This compensates for the lack of distinction between the proper volume and the coordinate volume that acts to overestimate $m_\rTOV$.
The ``Case A'' effective potential is used throughout the literature for including effective general relativistic gravity in Newtonian hydrodynamic simulations.
This includes several BH formation studies \citep{hudepohlthesis:14,pan:18,walk:19}.
While there has been comparisons to published literature simulations in select cases \citep{hudepohlthesis:14,oconnor:18}, the ``Case A'' effective potential has never been systematically tested against fully general relativistic simulations for BH formation.
Since we focus on BH formation, in this section we systematically compare core-collapse evolutions for many progenitors from \citet{sukhbold:18}\footnote{We choose the model set with reduced mass loss (by a factor of 10), which results in particularly compact supernova progenitors.} with different compactnesses using the general-relativistic radiation-hydrodynamics code \textsc{GR1D} \citep{oconnor:10, oconnor:15} and the \textsc{FLASH} code using the modified TOV effective potential. 
In the \textsc{GR1D} runs we use the same baseline EOS (Section~\ref{ssec:EOS}) and the same \textsc{NuLib} table generated with paramters described in Section~\ref{ssec:FLASH}.
The computational grid in \textsc{GR1D} is set to have 700 grid cells, constant cell size of $200\unit{m}$ out to a radius of $20\unit{km}$, and then geometrically increasing cell size to an outer radius where the baryonic density is $2\times10^3\unit{g\,cm}^ {-3}$. 
In order to match the initial setup used in \textsc{FLASH} we map stellar mass rest-mass density $\rho$, proton fraction $y_p$, and temperature $T$ from the progenitor star to \textsc{GR1D}.
To evolve stably until late times in \textsc{GR1D}, we neglect velocity terms and inelastic scattering in this set of comparison runs in both \textsc{GR1D} and \textsc{FLASH}.  
For the purposes of testing the effective potential this approximation is justified.

\begin{figure}[!htb]
\includegraphics[width=0.45\textwidth]{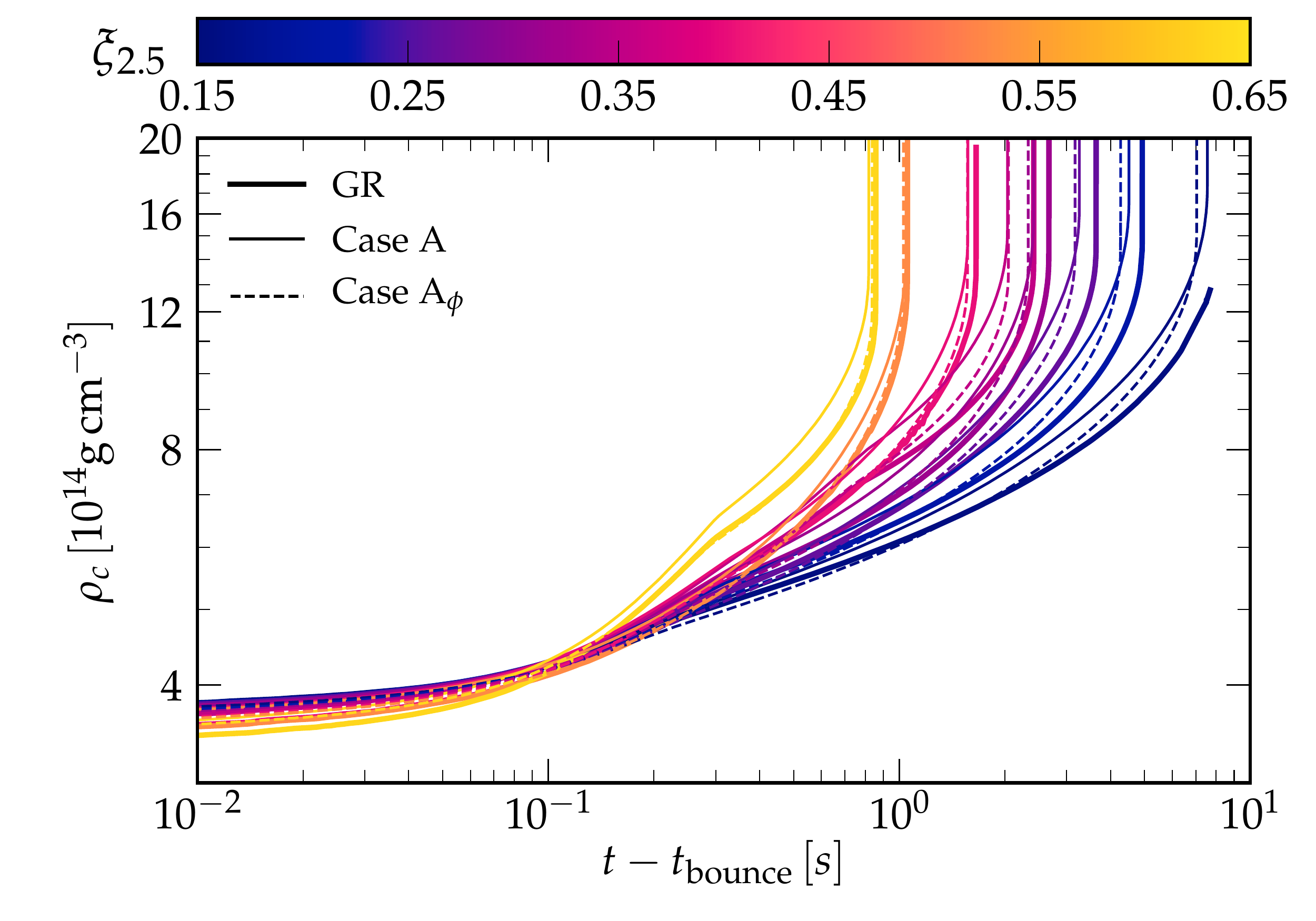}
\vspace{-0.25cm}
\caption{\label{fig:GREP} Comparison of central density evolution for full GR simulations using \textsc{GR1D} and two different GREP approximations using \textsc{FLASH}, see text. 
We limit the plot to 8 pre-SN progenitor models that span a wide range in compactnesses for clarity and stress that all other models behave similarly. 
Soon after bounce, $t-t_\rbounce\lesssim0.1\unit{s}$, all approaches predict quite similar evolutions. 
At later times,  $0.1\unit{s}\lesssim t-t_\rbounce\lesssim t_\rBH$, the ``Case A'' effective potential predicts slightly higher central density than the full GR and the  ``Case A$_\phi$'' effective potential, the latter two agreeing quite well. 
Near BH formation the density evolution becomes quite similar again, especially for more compact pre-SN progenitors. }
\end{figure}

In Figure~\ref{fig:GREP}, we compare the central density of core-collapse simulations performed using \textsc{GR1D} to \textsc{FLASH} simulations using the ``Case A'' effective potential for 8 progenitor models that span a wide range of compactness.
This evolution agrees quite well in the early times after core bounce,  $t-t_\rbounce \lesssim 0.1\unit{s}$.
The ``Case A'' effective potential slightly overestimates the central density at intermediate times, $0.1\unit{s}\lesssim t-t_\rbounce\lesssim t_\rBH$, but we stress that it agrees in many other quantities quite well.
These include the neutrino luminosity, neutrino average energies, and shock radius evolution \citep{granqvist:19}.  
Near the time of BH formation the central densities generally converge again in the two cases, especially for the most compact progenitors.

We also explore an additional case, which we call ``Case A$_\phi$''.
Here, we add the additional factor of $\Gamma$ to the gravitational potential instead of the TOV mass.
The central density evolution agrees remarkably well with the full GR case, especially early on and for the most compact progenitors.
However, most other quantities, like the emergent neutrino luminosities, tend to show greater deviations from the full GR results when compared to the ``Case A'' method \citep{granqvist:19}.
Therefore we choose to continue to use the ``Case A'' potential and do not advocate for our new ``Case A$_\phi$''.

\begin{figure}[!htb]
\includegraphics[width=0.45\textwidth]{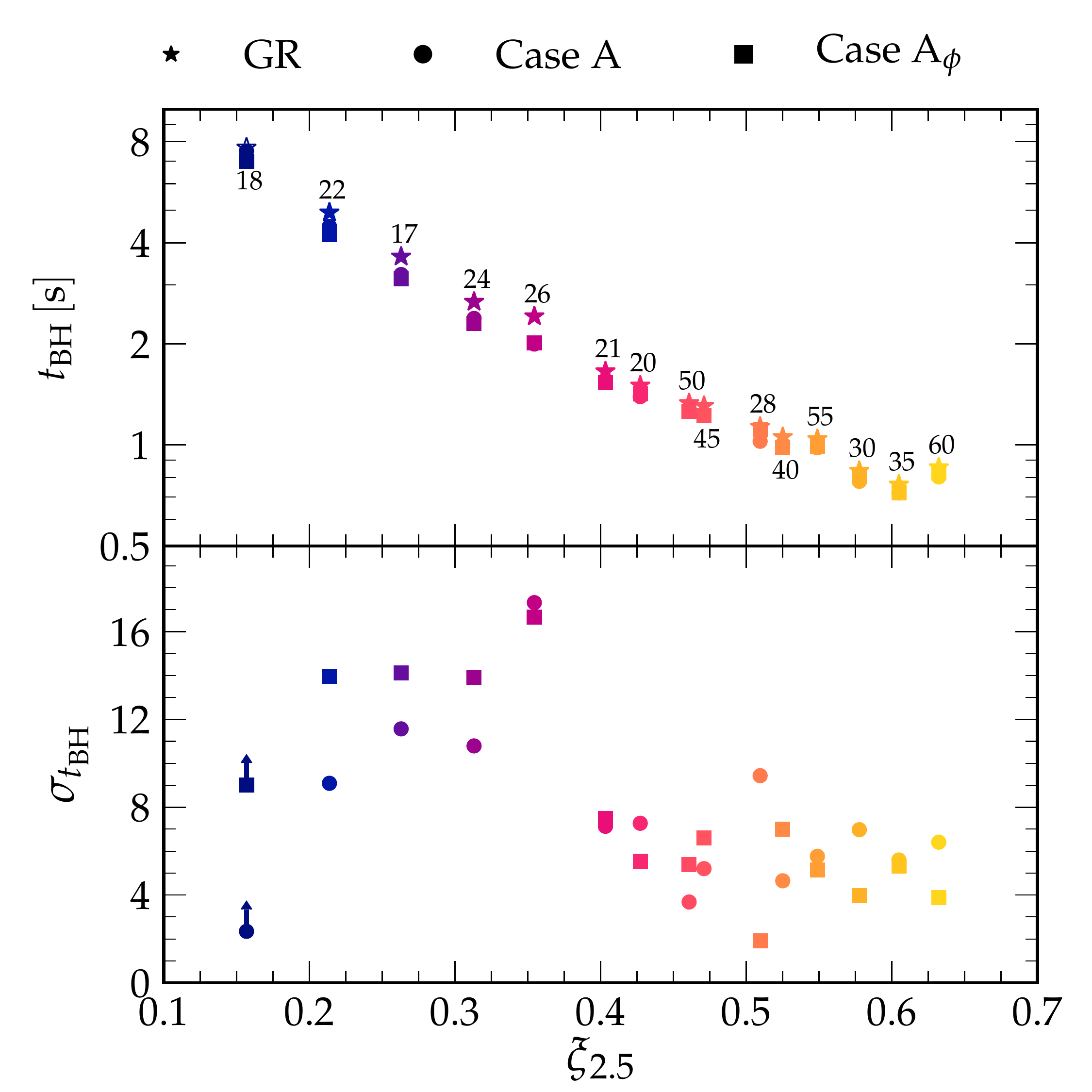}
\vspace{-0.25cm}
\caption{\label{fig:GR_GREP_tBH} Comparison of the BH formation  time for full GR simulations using \textsc{GR1D} and the ``Case A'' and ``Case A$_\phi$'' GREP approximations using \textsc{FLASH}. 
We limit the plot to 15 pre-SN progenitor models that span a wide range in compactnesses for clarity. 
Symbols are colored according to their compactness, as in Fig. \ref{fig:GREP}, and we also write explicitly the ZAMS mass of the progenitors in $M_\odot$ on the top plot.
For the $18\,M_\odot$ run, where the GR simulation did not reach BH formation due to its increasing computational cost, we replaced the $\filledstar$ symbol by a $\hollowstar$ symbol to denote the maximum simulation run time (see also Fig. \ref{fig:GREP}). 
GR simulations that formed a BH took, on average, 4 to 8\% longer to do so than their GREP counterparts for the more compact progenitors and 8 to 16\% longer for the less compact ones.}
\end{figure}

In Figure~\ref{fig:GREP}, we show that the BH formation time, $t_\rBH$, for each pre-SN progenitor is very similar for simulations that use full GR and the two different GREP approaches. 
This is in agreement with the results of \cite{couch:19} for comparisons between full GR simulations and the ``Case A'' GREP approximation using \textsc{FLASH}. 
However, as we show in Figure~\ref{fig:GR_GREP_tBH} for 15 chosen pre-SN progenitors from \citet{sukhbold:18}, full GR simulations consistently take longer to collapse than their GREP counterparts. 
Also, ``Case A$_\phi$'', despite predicting a more accurate PNS central density evolution than the ``Case A'' approach, often underestimates the BH formation time by a slightly larger margin than ``Case A''.
For the progenitors where \textsc{GR1D} runs reached BH formation, in most cases they took 3\% to 8\% (8\% to 16\%) more time after core bounce than did the \textsc{FLASH} GREP runs for progenitors with $\xi_{2.5}\gtrsim0.4$ ($\xi_{2.5}\lesssim0.4$).
For a more indepth analysis, although with a different EOS, we refer the reader to \citet{granqvist:19}.

%% file: sections/results.tex
\section{Results}
\label{sec:results}

We start our analysis detailing how pre-SN progenitor structure affects its core-collapse evolution until BH formation. 
First, we examine two extreme pre-SN progenitor cases, the high compactness progenitor \texttt{u75WHW02} ($\xi_{2.5}=0.88$) and the low compactness progenitor \texttt{s50WH07} ($\xi_{2.5}=0.22$). 
Using our baseline EOS we find, as expected, that the compact model collapses to a BH rather quickly, 447\unit{ms} after core bounce for \texttt{u75WHW02}, and takes a relatively long time to collapse for the low compactness progenitor, 4478\unit{ms} after core bounce for \texttt{s50WH07}.

\subsection{High compactness pre-supernova progenitor}
\label{ssec:high}

\begin{figure*}[htb]
\plotone{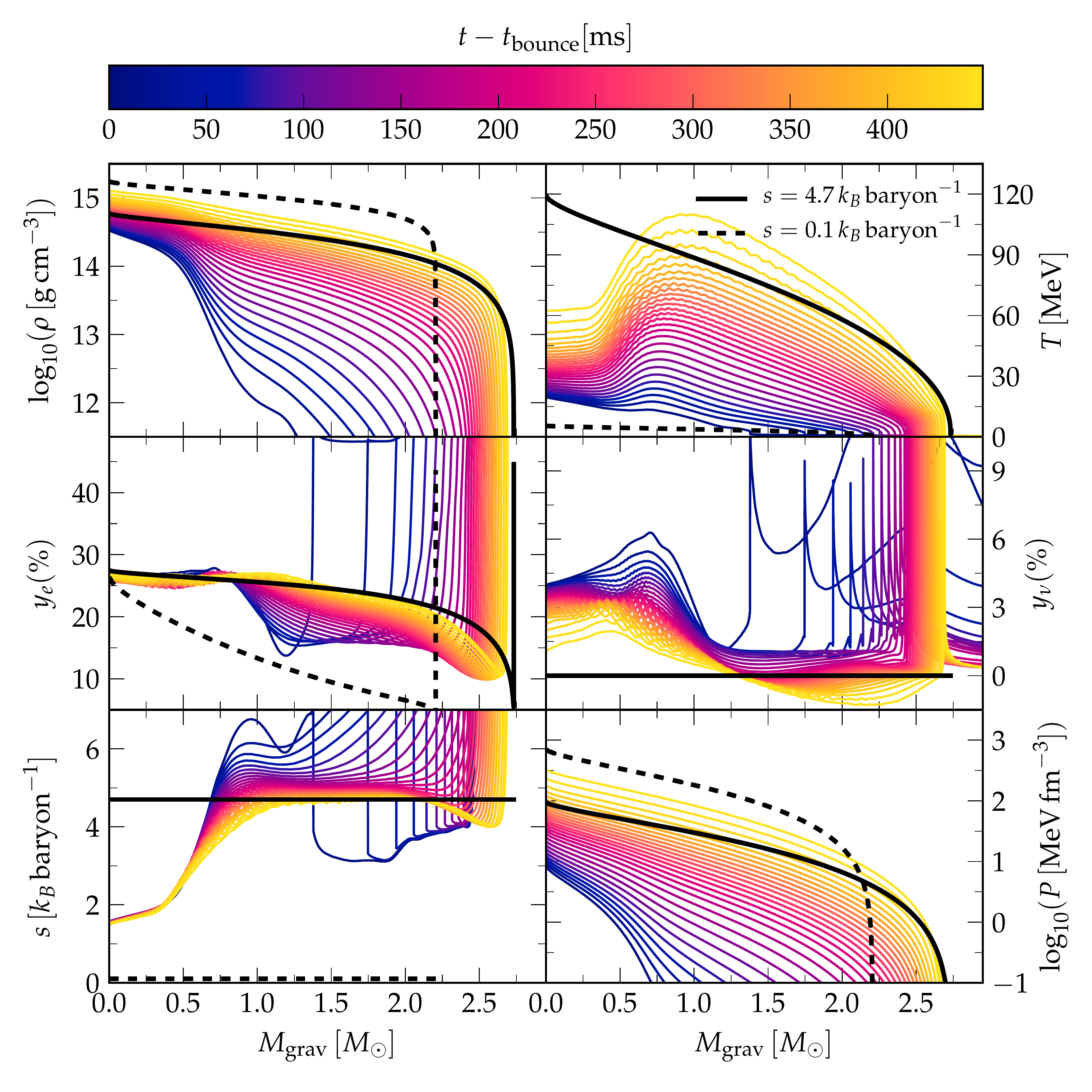}
\vspace{-0.5cm}
\caption{\label{fig:u75_evolution}  Time evolution from core bounce to BH formation ($t-t_\rbounce=447\unit{ms}$) of the pre-SN \texttt{u75WHW02} model of \citet{woosley:02}. 
We plot the baryonic mass density $\rho$ (top left), temperature $T$ (top right), lepton fraction $y_e=({n_{e^-}-n_{e^+}})/{n}$ (center left, equal to the proton fraction $y_e=y_p=n_p/n$), neutrino number fraction (center right, $y_\nu=(n_\nu-n_{\bar\nu})/n$), entropy $s$ (bottom left), and pressure $P$ as a function of gravitational mass coordinate $M_\rgrav$, i.e. Equation \eqref{eq:mTOV}. 
Additionally, we plot the properties of a cold ($s=0.1\entropy$) and a hot ($s=4.7\entropy$) non-rotating beta-equilibrated ($y_\nu=0$) NS.}
\end{figure*}

In Figure~\ref{fig:u75_evolution} we show the post bounce evolution of the \texttt{u75WHW02} progenitor star simulated using our baseline EOS. 
Evolution is shown until $1\unit{ms}$ before BH formation at $t-t_\rbounce=447\unit{ms}$, defined as the time where the core density surpasses $2.0\times10^{15}\unit{g\,cm}^{-3}$. 
At core bounce, the central density and temperature are approximately $1.5\,n_\rsat$ and $20\,\unit{MeV}$, respectively. 
After bounce the core compresses quickly, baryon number density increases almost one order of magnitude in less than $500\unit{ms}$, and its temperature rises above $50\unit{MeV}$.
Despite increasing density and temperature throughout the PNS, entropy and proton fraction below mass coordinate $M_{\rgrav}\simeq0.7\,M_\odot$ barely change. 
Meanwhile, due to the high compactness of this progenitor ($\xi_{2.5}=0.88$), over $1\,M_\odot$ is accreted onto the PNS within a few hundred milliseconds of core bounce. 
Fast accretion quickly heats up the outer layers of the PNS to temperatures of $\mathcal{O}(10\unit{MeV})$ and intermediate layers to $\mathcal{O}(100\unit{MeV})$. 
Some of the heat diffuses inwards but this is relatively limited (as compared to the lower compactness model) due to the very rapid post-bounce evolution.
At the time of BH formation, the temperature peaks at mass coordinate $M_{\rgrav}\simeq0.9\,M_\odot$. 
Neutrino and anti-neutrino emission processes and their diffusion move matter towards beta-equilibrium and decrease the average entropy of the PNS, rendering it almost uniform at $s\simeq4.7\entropy$ by BH formation time between mass coordinates $1.0 \lesssim M_{\rgrav}/M_\odot \lesssim 2.0$.

We compare the evolution of the collapsing star to the most massive beta-equilibrated NS possible. The comparison is made for two cases: (1) a PNS with constant entropy $s=0.1\entropy$, which serves as a proxy for a cold NS, and (2) for a hot NS with constant entropy $s=\tilde{s}(t_{\rBH})$. 
Here, $\tilde{s}(t)$ is the most common entropy value at a given time $t$ defined as 
\begin{equation}\label{eq:stilde}
 \tilde{s}(t) = \max(\{M_s(t)\})\,,
\end{equation}
where 

\begin{align}
M_s(t)=\int_0^{M^\rPNS}[&H(s(M)-0.01)\nonumber\\
&-H(s(M)+0.01)]dM\,,
\end{align}
with $H$ the Heaviside step function, $s(M)$ the entropy at mass coordinate $M$ in units of $k_B\unit{baryon}^{-1}$ and $M^\rPNS$ the mass enclosed by the PNS, \ie the mass within the region with density larger than $10^{11}\unit{g\,cm}^{-3}$. We find that this definition displays a smoother and more stable evolution than the maximum entropy inside the PNS within a mass or density coordinate. 
Moreover, we see only small difference between choosing the gravitational or the baryonic mass for the computation in Equation \eqref{eq:stilde}. 
For the progenitor and EOS discussed in this section we find that $\tilde{s}(t_{\rBH}) = 4.7\entropy$.

We observe that collapse to a BH occurs when the total \textit{gravitational} mass of the PNS, $M_\rgrav^\rPNS$, is within 1\% of $M_\rgrav^{\rmax,\tilde{s}}$, the maximum mass supported by a hot NS with constant entropy $\tilde{s}(t_\rBH)$.  
Here, and in what follows, a $M$ with a numeric superscript indicates the mass of a constant entropy star with the superscript denoting the entropy in units of $k_B\unit{baryon}^{-1}$. 
We note that approximately $300\unit{ms}$ after bounce, both the baryonic and gravitational masses of the PNS are already a few tenths of a solar mass larger than the maximum masses supported by a cold non-rotating beta-equilibrated NS.
As there is no mechanism for the PNS to lose baryonic mass the collapse into a BH is inevitable even if accretion were abruptly cutoff and neutrinos had enough time to carry away all of the mass associated with the excess thermal energy.

\subsection{Low compactness pre-supernova progenitor}
\label{ssec:low}

\begin{figure*}[!htb]
\plotone{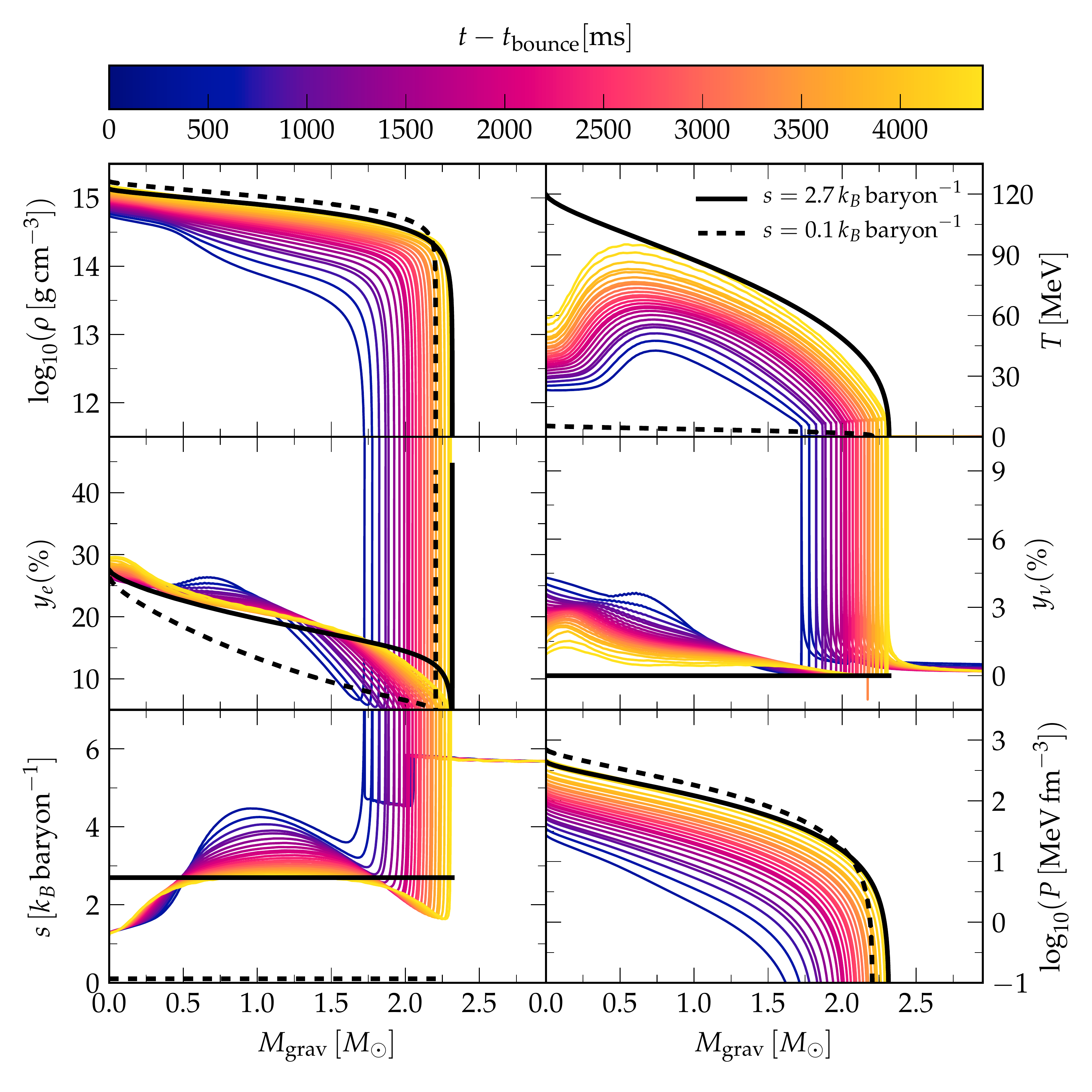}
\vspace{-0.5cm}
\caption{\label{fig:s50_evolution}   Time evolution from core bounce to BH formation ($t-t_\rbounce=4478\unit{ms}$) of the pre-SN \texttt{s50WH07} model of \citet{woosley:07}. 
Plot order as in Figure~\ref{fig:u75_evolution}. 
Additionally, we plot the properties of a cold ($s=0.1\entropy$) and a hot ($s=2.7\entropy$) non-rotating beta-equilibrated ($y_\nu=0$) NS.}
\end{figure*}

In Figure~\ref{fig:s50_evolution}, we plot the evolution of the \texttt{s50WH07} progenitor using our baseline EOS. 
Due to its relatively low compactness, this progenitor star's structure leads to a much lower accretion rate than the \texttt{u75WHW02} progenitor and, therefore, has a much longer time to equilibrate between core bounce and the collapse to a BH at $t - t_\rbounce = 4478\unit{ms}$. 
Except for the inner and outermost layers of the star, enough neutrinos have been emitted that most of the PNS is close to beta-equilibrium ($y_\nu=0$) by the end of the simulation. 
Thus, heat from the accreted layers has had time to diffuse inwards towards the PNS center and its entropy to decrease to $\tilde{s}\simeq2.7\entropy$ between mass coordinates $0.5\lesssim M_\rgrav/M_\odot \lesssim 2.0$ at the point of BH formation. 
Compared to the maximum mass of a hot non-rotating beta-equilibrated NS with constant entropy $\tilde{s}=2.7\entropy$, we observe that the PNS gravitational mass, $M^\rPNS_\rgrav = 2.30\,M_\odot$ is only slightly above the maximum gravitational mass supported by the hot star with constant entropy $2.7\entropy$, $M^{\rmax,2.7}_\rgrav = 2.29\,M_\odot$. 
Also, collapse to a BH takes place even though the PNS baryonic mass $M^\rPNS_\rbaryon = 2.55\,M_\odot$ is below the limit $M^{\rmax,2.7}_\rbaryon = 2.60\,M_\odot$. 
Furthermore, even this maximum baryonic mass supported for a hot NS with $\tilde{s}=2.7\entropy$ is below that of a cold non-rotating beta-equilibrated NS, \ie $M^{\rmax,2.7}_\rbaryon < M^{\rmax,0}_\rbaryon = 2.65\,M_\odot$. 
This implies that there is an intricate interplay between accretion, which increases the total baryonic and gravitational mass of the PNS and pushes it towards collapse into a BH, and neutrino emission, which decreases the gravitational mass of the PNS and can help revive the shock to prevent the gravitational instability that leads to BH formation. 
In the simulation of the \texttt{u75WHW02} progenitor, the PNS collapses into a BH happens once its \textit{gravitational} mass exceeds 99\% of $M_\rgrav^{\rmax,\tilde{s}}$, while in the case of the \texttt{s50WH07} progenitor the PNS \textit{gravitational} mass slightly exceeds $M_\rgrav^{\rmax,\tilde{s}}$. 
The difference is due to the contribution of the relatively colder core of the PNS formed from the core collapse of the \texttt{u75WHW02} progenitor, which cannot support as much pressure before BH formation.

\begin{figure}[!htb]
\includegraphics[trim={0.1cm 0.1cm 0.1cm 0.1cm},clip,width=0.45\textwidth]{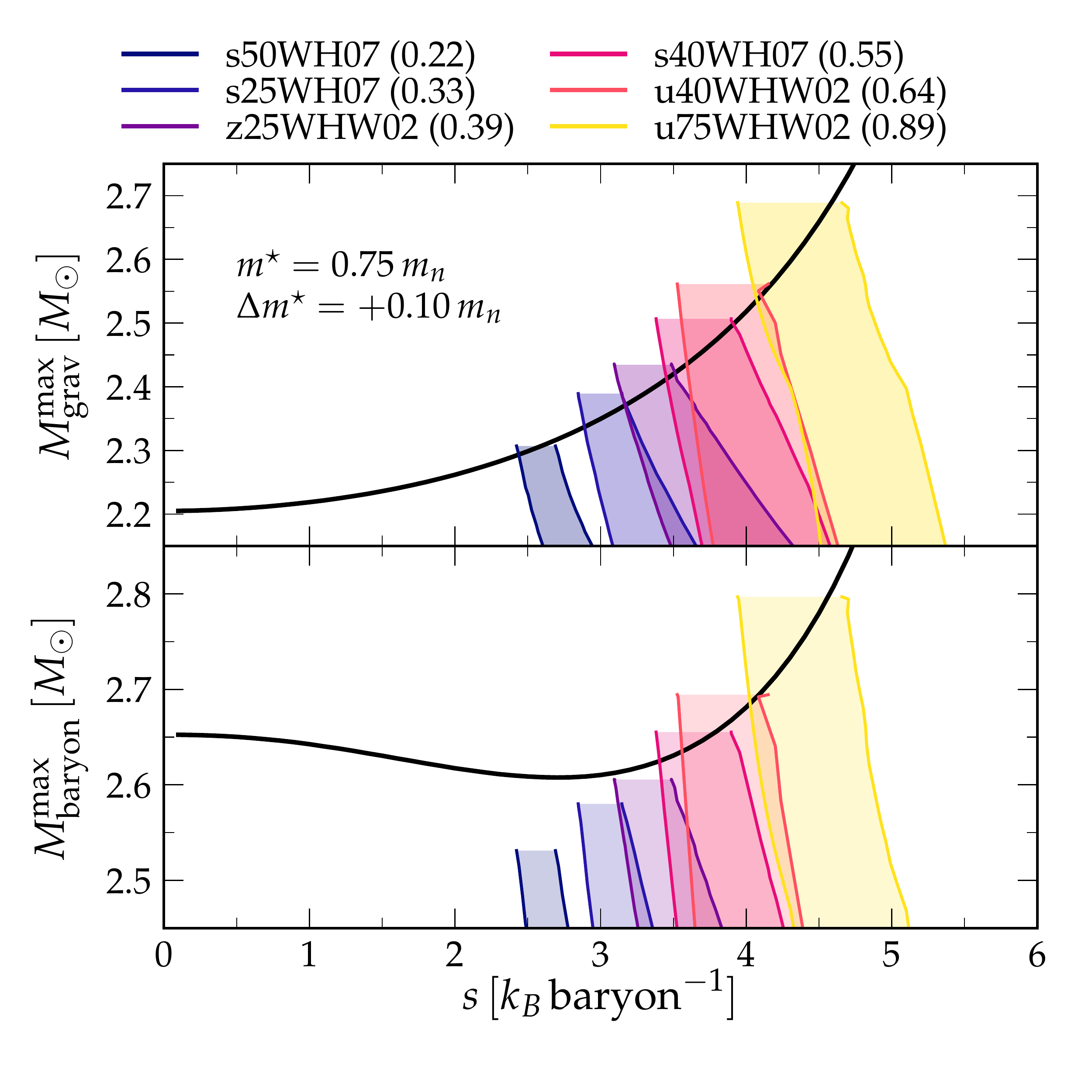}
\vspace{-0.5cm}
\caption{\label{fig:mass_entropy}  
Evolution of the gravitational (top) and baryonic (bottom) masses during the core-collapse of the six pre-SN progenitor models chosen from \citet{woosley:02} and \citet{woosley:07}. 
Shaded areas are bound left by the mass weighted average entropy $\bar{s}$, Equation \eqref{eq:sbar}, right by the most common entropy $\tilde{s}$ within the PNS, Equation \eqref{eq:stilde}, and top by the PNS mass at BH formation time. 
Black lines are the maximum possible mass $M^{\rmax,s}$ obtained solving the TOV equations for a NS with constant entropy $s$ for our baseline EOS (Section \ref{sec:model}). 
Collapse to a BH takes place, to a very good approximation, once the \textit{gravitational} mass of the PNS exceedes the maximum gravitational mass supported by a hot NS with entropy equal to the most common entropy (i.e. $s=\tilde{s}(t_{\rBH}$) inside the PNS.)}
\end{figure}

In Figure~\ref{fig:mass_entropy}, we show the evolution of the PNS baryonic and gravitational masses, $M^\rPNS_\rbaryon$ and $M^\rPNS_\rgrav$, respectively, with respect to the most common PNS entropy, $\tilde{s}$, see Equation \eqref{eq:stilde}, and the PNS mass averaged entropy 
\begin{equation}\label{eq:sbar}
 \bar{s}=\frac{1}{M^\rPNS}\int_0^{M^\rPNS} s(M) dM\, 
\end{equation}
We plot curves for the six pre-SN progenitor models chosen from \citet{woosley:02} and \citet{woosley:07}, see Table~\ref{tab:progenitors}, and simulated using our baseline EOS. 
We also include the maximum supported baryonic and gravitational masses for hot stars with a constant entropy $s$, $M^{\rmax,s}$, for the baseline EOS used in the simulations. 
The plot with respect to the gravitational mass reinforces the idea that, for a given progenitor and, as we will show later, a given EOS, BH formation occurs only once $M_\rgrav^{\rPNS} \gtrsim M^{\rmax,\tilde{s}}_\rgrav$. 
Meanwhile, the plot showing the evolution of the PNS baryonic mass emphasizes that (1) BH formation correlates with excess gravitational mass and not baryonic mass and (2) that collapse into a BH can happen for PNSs with baryonic masses lower than the maximum supported by a cold NS.

For both progenitors described here in detail, \texttt{u75WHW02} and \texttt{s50WH07}, our qualitative results for the PNS profile evolution in the first seconds after core bounce are in broad agreement with other PNS evolution simulations in spherical symmetry \citep{sumiyoshi:05, sumiyoshi:07,  sumiyoshi:08, sumiyoshi:09, martinez-pinedo:12, roberts:12a, roberts:16, fischer:09, fischer:10, fischer:12, nakazato:12, nakazato:13, nakazato:18, nakazato:19, banik:14a, char:15}. 
While some of these studies focused on the PNS path towards BH formation and EOS effects, many were only interested in long term evolutions of PNSs that would eventually form a cold NS. 
\citet{fischer:10, fischer:12, martinez-pinedo:12} simulated core collapse and PNS evolution for several seconds for pre-SN progenitor stars of 8.8, 10.8, 15, and $18\,M_\odot$. 
In these works, explosions were achieved by artificially enhancing the electronic charged current reaction rates except for the lightest progenitor, which is able to explode even in spherically symmetric simulations. 
\cite{nakazato:13} simulated PNS cooling for eight massive progenitor models using the H.~Shen EOS \citep{shen:98, shen:98a} for up to 20 seconds after core bounce by removing the stellar mantle above the shock region after 100, 200, and $300\unit{ms}$ after core bounce. 
In the one case where \cite{nakazato:13} did not remove the outer layers of the star after core bounce, for a $30\,M_\odot$ progenitor with $Z=0.004$ metalicity, the PNS collapsed into a BH within $842\unit{ms}$ and a gravitational mass $M_\rgrav^\rPNS\simeq2.53\,M_\odot$ assuming all the binding energy was radiated away in neutrinos. 
\citet{nakazato:18, nakazato:19} evolved PNSs for up to 80 seconds after removing the regions above the shock $300\unit{ms}$ after core bounce to study the effect of the EOS in the cooling timescales. 
Meanwhile, \cite{roberts:16} excised the inner $1.42\,M_\odot$ of the PNS and evolved it for 100 seconds using the \LS{220} EOS of \citet{lattimer:91}. 
Unlike these previous PNS evolution studies, however, we do not force explosions to occur. 
Instead, we allow for the overlying material above the PNS to continue to accrete, as expected in spherically symmetric simulations, and drive BH formation. 
We also do not model convective instabilities that arise within the PNS and drive convection and turbulence, even though those are sensitive to the EOS \citep{roberts:12} and render the entropy profile within the PNS monotonically increasing \citep{dessart:06, roberts:12, pan:18, gossan:19}. 
Note that this would remove the negative entropy gradients seen between  $2.1\,M_\odot\lesssim{M_\rgrav^\rPNS}\lesssim2.6\,M_\odot$ ($1.7\,M_\odot\lesssim{M_\rgrav^\rPNS}\lesssim2.3\,M_\odot$) for the \texttt{u75WHW02} (\texttt{s50WH07}) progenitor and may affect each combination of progenitor and EOS in a unique manner. 
Thus, a natural follow up to our work is a similar multidimensional study. 
An analysis of the results of \citet{sumiyoshi:06, sumiyoshi:07,  sumiyoshi:08, sumiyoshi:09, fischer:09, oconnor:11, nakazato:12, hempel:12, steiner:13, banik:14a, char:15, pan:18} in light of our conclusions are discussed in detail at the end of this section.  

\subsection{Multiple progenitors}
\label{ssec:swh18}

\begin{figure}[!htb]
\includegraphics[trim={0.1cm 0.1cm 0.1cm 0.1cm},clip,width=0.45\textwidth]{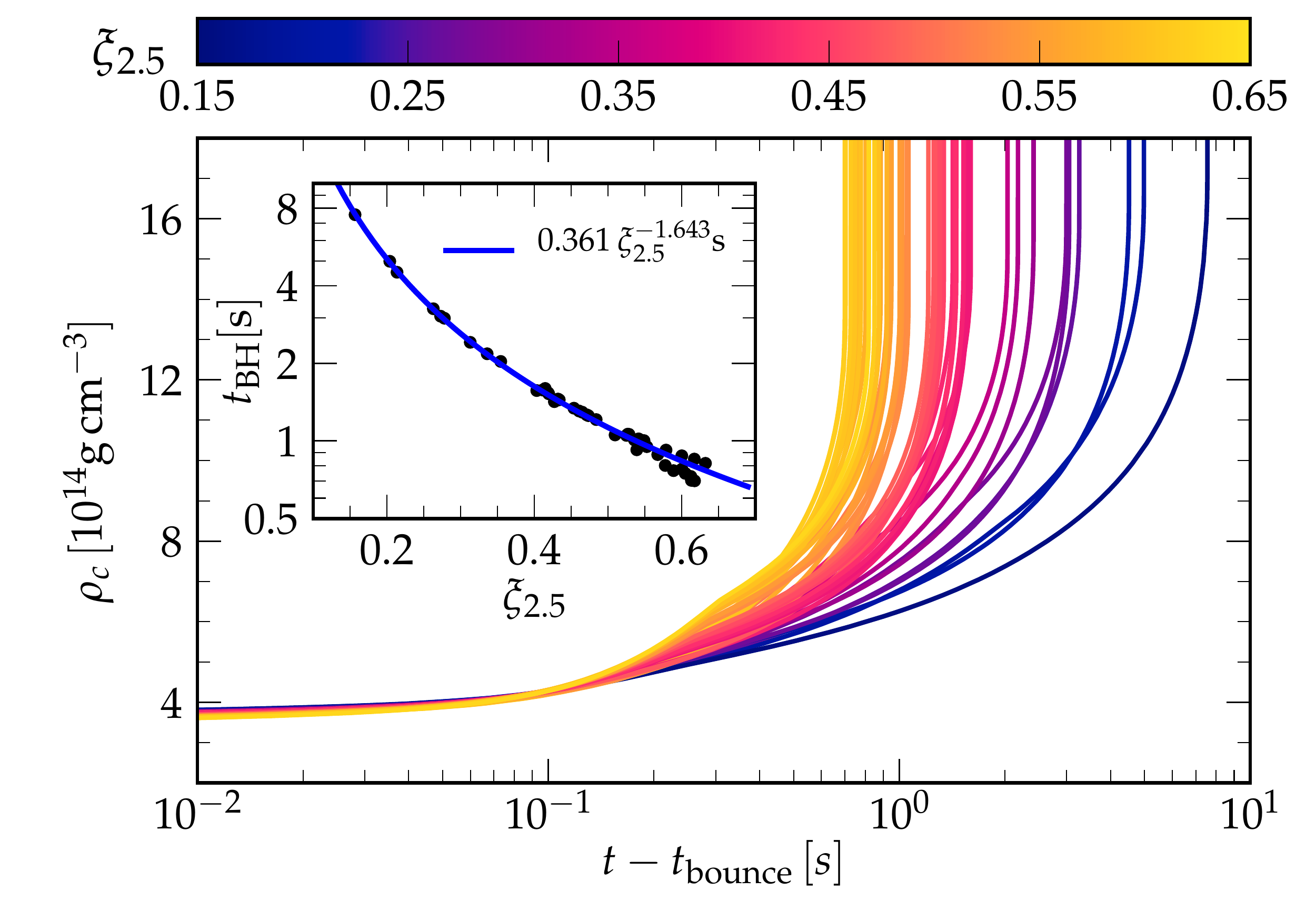}
\vspace{-0.25cm}
\caption{\label{fig:ratio} Evolution of the central density $\rho_c$ as a function of post bounce time $t-t_\rbounce$ for the 45 pre-SN progenitors of \citet{sukhbold:18}. 
In the inset we plot the BH formation time $t_{\rm BH}$ as a function of the compactness $\xi_{2.5}$, Equation \eqref{eq:compactness} and its best fitting curve, Equation \eqref{eq:tBH}.}
\end{figure}

We now explore the relationship between collapse to BH and gravitational mass of the PNS for 45 pre-SN progenitors of \citet{sukhbold:18}, Section~\ref{ssec:progenitors}.  
Again, we simulate all core collapses until BH formation occurs. 
In Figure~\ref{fig:ratio}, we plot the central density for each progenitor and observe, as expected, that the time for a BH to form decreases with increasing compactness \citep{oconnor:11}. 
The BH formation time is the time it takes to accrete the critical mass element that has a baryonic mass coordinate equal to the PNS baryonic mass at the point of BH formation.
This occurs approximately at \citep{burrows:86, oconnor:11} 
\begin{equation}\label{eq:tBH}
 t_{\rm BH} \simeq t_{\rm ff}^{M_\rBH} = A \xi_{M_\rBH}^{-b}\unit{s}\,.
\end{equation}
where $M_\rBH = M_\rPNS(t_\rBH)$ and $t_{\rm ff}^{M_\rBH}$ is the free-fall time at mass coordinate $M_\rBH$.
For realistic EOSs and progenitors $M_\rPNS(t_\rBH) \simeq 2.0 - 3.0\,M_\odot$ and, thus, setting $M_\rBH = 2.5\,M_\odot$ is a reasonable approximation. 
A straightforward calculation of $t_{\rm ff}^{2.5\,M_\odot}$ leads to $A=0.241$ and $b=3/2$ \citep{oconnor:11}. 
For our baseline EOS and the progenitors of \citet{sukhbold:18} discussed here which collapse into a BH we have $A = 0.361 \pm 0.004$ and $b = 1.643 \pm 0.008$. 
Meanwhile, for the six progenitors of \citet{woosley:02} and \citet{woosley:07}, see Table~\ref{tab:progenitors}, we obtain $A = 0.332 \pm 0.011$ and $b = 1.728 \pm 0.024$.
We observe that the exponents $b$ and amplitudes $A $ are both larger for the simulated collapses than the values obtained from the Newtonian free-fall time \citep{burrows:86, oconnor:11, couch:19}. 
These are due to the free-fall time not accounting for  existent pressure support for the infalling mass element, which depends on details of the progenitor structure and the arbitrary initial reference time, here set to be the PNS bounce time. 
Differences between the fitting for the two progenitor sets are not significant, $5\%$ for the exponent and $9\%$ for the amplitude, given that they sample the range of compactnesses at different frequencies and that one set includes only six progenitor stars.

\begin{figure*}[!htb]
\plotone{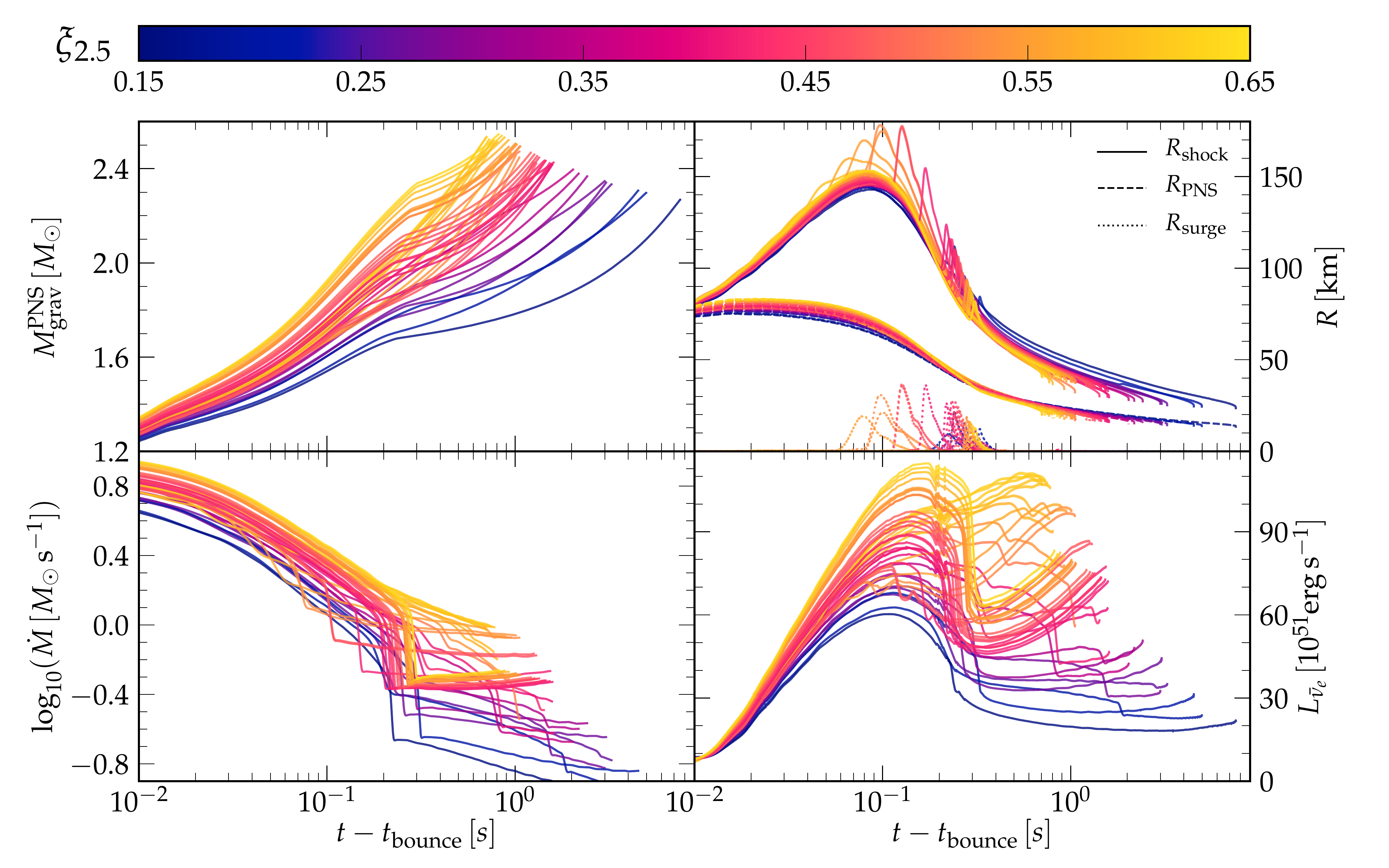}
\vspace{-0.25cm}
\caption{\label{fig:swh18} Evolution of the PNS gravitational mass $M_\rgrav^\rPNS$ (top left), accretion rate $\dot{M}$ (bottom left), shock $R_{\rm shock}$, shock surge $R_{\rm surge}$, and PNS radii $R_{\rPNS}$ (top right), and electron anti-neutrino luminosity $L_{\bar\nu_e}$ (bottom right) for core-collapse simulations of 45 pre-SN progenitor models of \citet{sukhbold:18}, see Section~\ref{ssec:progenitors}. 
We estimated the gravitational mass by computing the baryonic mass of the PNS (defined by $\rho > 10^{11}$\,g\,cm$^{-3}$) and subtracting the total integrated luminosity of the emitted neutrinos.
Large declines in accretion rates that occur due to sharp density drops in the progenitor models (see Figure~\ref{fig:preSN}) cause shock radii to expand further than expected from a smooth accretion rate, shown as $R_{\rm surge}$ in the top right plot. 
See discussion in text. 
Note that not all progenitors show sharp declines in accretion rate. 
Neutrino luminosities, represented here by the electron anti-neutrino, also show behavior that depends on the progenitor profile. }
\end{figure*}

In Figure~\ref{fig:swh18}, we plot the PNS gravitational mass, $M_\rgrav^\rPNS$;
accretion rates, $\dot{M}$; and the evolution of PNS and shock radii, $R_\rPNS$ and $R_\rshock$, respectively. 
During the first $\simeq200\unit{ms}$ post core-bounce, progenitors with higher compactness form, in general, PNSs with larger radii. 
The behavior changes at later times since accretion rates increase with compactness which, in turn, result in higher pressures on top of the PNS and force the PNS to contract faster. 
This would have implications for GW emission as the more compact and more massive a PNS is the stronger the gravitational waves and the higher the frequencies it emits \citep{pan:18, morozova:18, torres-forne:19}.

Similar to the PNS radii, the shock radii also follow a somewhat well defined trend for all progenitors. 
The exceptions occur near the times when the shock crosses steep density gradients, such as the interface between the Si and Si-O shell, and accretion rates drop dramatically. 
Lower accretion allows shock radii, if not already increasing fast due to early post-bounce dynamics, to expand significantly. 
At such times when the shock radius crosses a shell boundary, which takes place between 40\unit{ms} and 400\unit{ms} after bounce for most progenitors where this occurs, the shock radii quickly expands an extra $10\unit{km}\lesssim R_{\rm surge}\lesssim30\unit{km}$, but recedes soon after. 
This outcome is expected for spherically symmetric simulations of core collapse. 
However, in multidimensional simulations the accretion of the interface at the boundary between the Si and Si-O shell is often the driver of a successful SN \citep{summa:16, summa:18, ott:18, vartanyan:19}.

For most pre-SN progenitors, including the ones considered here, the Si/Si-O interface crosses the region where the entropy is $s=4\entropy$. 
This interface also coincides with the steepest density and entropy gradients throughout the star and occurs near mass coordinate $M\simeq2\,M_\odot$, which is discernible as a large density drop in the stellar profile, see Figure~\ref{fig:preSN}. 
Considering this, \citet{ertl:16} were able to correlate successful supernovae explosions with the mass coordinate $M(s=4\entropy)$ and how steep the entropy gradient $\mu_4=dm/dr\vert_{s=4\entropy}$ is at such location. 
This is because $M_4$ is related to when the sudden drop in accretion rate occurs while $\mu_4$ determines how significant the drop is. 
If the drop in accretion rate is close to when neutrino luminosity is at a maximum an explosion sets in \citep{ertl:16}.

We estimate the surge in shock radii caused by the shock radius crossing a steep density/entropy gradient in our simulations, $R_{\rm surge}$ in Figure~\ref{fig:swh18}. 
$R_{\rm surge}$ is computed by subtracting a smoothed shock radius from $R_\rshock$. 
The smoothed shock radius is a fit to $R_\rshock$ that excludes data points near where accretion rate drops significantly\footnote{The fit is performed with the \texttt{InterpolatedUnivariateSpline} Python module \citep{jones:01} excluding $R_\rshock$ in the region where $d\log\dot{M}/d\log(t)<-2$ and up to $\simeq100\unit{ms}$ afterwards}. 
We observe that $R_{\rm surge}$ is larger if the shock crosses a boundary layer near when neutrino luminosity is close to a maximum in its trend, $t-t_\rbounce\simeq80-200\unit{ms}$.
Thus, as in \citet{ertl:16}, we expect larger $R_{\rm surge}$ to correlate with successful supernovae explosions in multidimensional simulations. 
We reiterate that these radii surges do not correlate with compactness of the pre-SN progenitor, but with the location and density gradient of the advected boundary layer.

Finally, we also show in Figure~\ref{fig:swh18} the luminosity, $L_{\bar\nu_e}$, of electron anti-neutrinos as their count rate could be the main observable informing us of progenitor structure and possible PNS collapse to a BH during a galactic core-collapse \citep{scholberg:12}.
In the earlier stages of collapse, electron anti-neutrino luminosity rises quickly and peaks approximately $100$ to $150\unit{ms}$ after core bounce. 
For pre-SN progenitors where accretion rates drop abruptly due to the advection of a boundary shell, so do the neutrino luminosities as it is accretion that fuels the neutrino production. 
Thus, while the BH formation time is well correlated with the pre-SN progenitor compactness, 
the behavior of most other observables after the first $\sim200\unit{ms}$, such as neutrino luminosity, shock radius, and accretion rate, is not well determined solely by progenitor compactness $\xi_M$, but also depend on pre-SN progenitor structure \citep{buras:06, oconnor:13, ertl:16, ebinger:18}. 
Finally, the neutrino signal is abruptly cut-off at the moment of BH formation  due to the neutrinosphere being inside the BH event horizon \citep{sumiyoshi:06}, although we are not able to evolve our simulations until this happens.

\begin{figure}[!htb]
\includegraphics[width=0.45\textwidth]{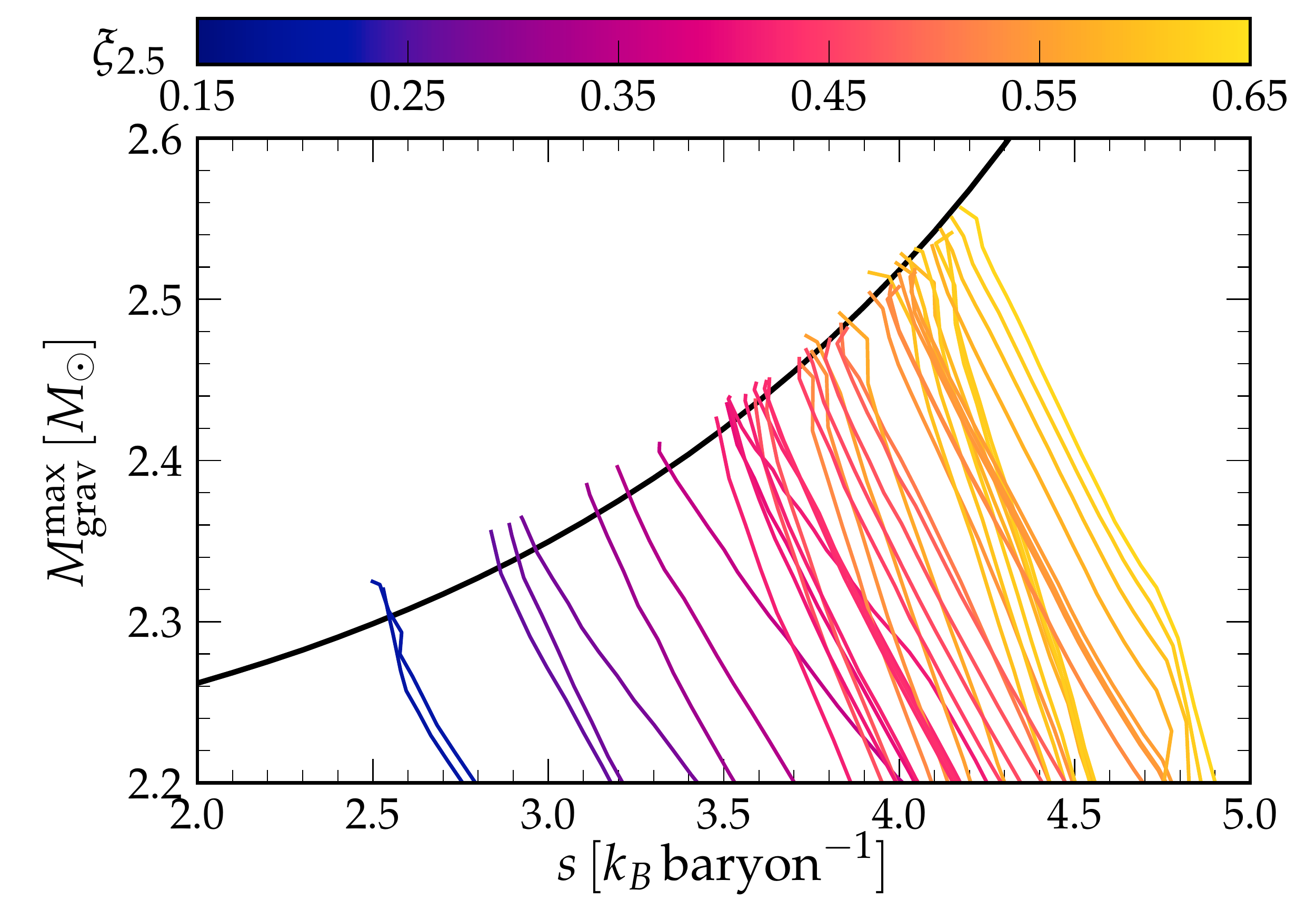}
\vspace{-0.25cm}
\caption{\label{fig:swh18_s} Evolution of the most common entropy ($\tilde{s}$) inside the PNS and its gravitational mass for the progenitors of \citet{sukhbold:18}. 
Curves are compared to the maximum gravitational mass supported by a hot NS with constant entropy. 
Collapse to a BH happens soon after the gravitational mass of the PNS is above the limit defined by the hot constant-entropy NS.}
\end{figure}

As for the \citet{woosley:02} and \citet{woosley:07} progenitors, we also consider the entropy evolution of the PNS for the \citet{sukhbold:18} progenitors. 
In Figure~\ref{fig:swh18_s}, we plot the gravitational mass of the PNS as a function of the most common entropy, $\tilde{s}$, inside the PNS. 
We compare these curves for each of the progenitors to the maximum gravitational mass supported by a hot NS with a constant entropy, which is obtained by solving the TOV equations for our baseline EOS. 
We note that in our TOV solutions we do not include the effects of thermal neutrinos, which are naturally included in our neutrino transport simulations.
We have checked that adding these to the TOV solutions for constant entropy NSs changes the maximum gravitational mass by $<1\%$ for all $s<4\,k_\mathrm{B}\,\mathrm{baryon}^{-1}$.
BH formation occurs, in almost every case, soon after the gravitational mass of the PNS is above the limit defined by the constant-entropy NS for our EOS, $M^{\rmax,s}_\rgrav$. 
Only in a few cases of high-compactness progenitor collapse BH forms with a mass below $M^{\rmax,s}_\rgrav$, but still very close to it. 

\subsection{Equation of state effects}
\label{ssec:eos}

In the discussion above we focused on the collapse of 51 pre-SN progenitor models simulated using a single EOS. 
Now we look at the effects of another 48 EOSs (Section~\ref{ssec:EOS}) on the collapse of the six progenitors of \citet{woosley:02} and \citet{woosley:07}, Section~\ref{ssec:progenitors}.

\begin{figure*}[!htb]
\includegraphics[width=0.48\textwidth]{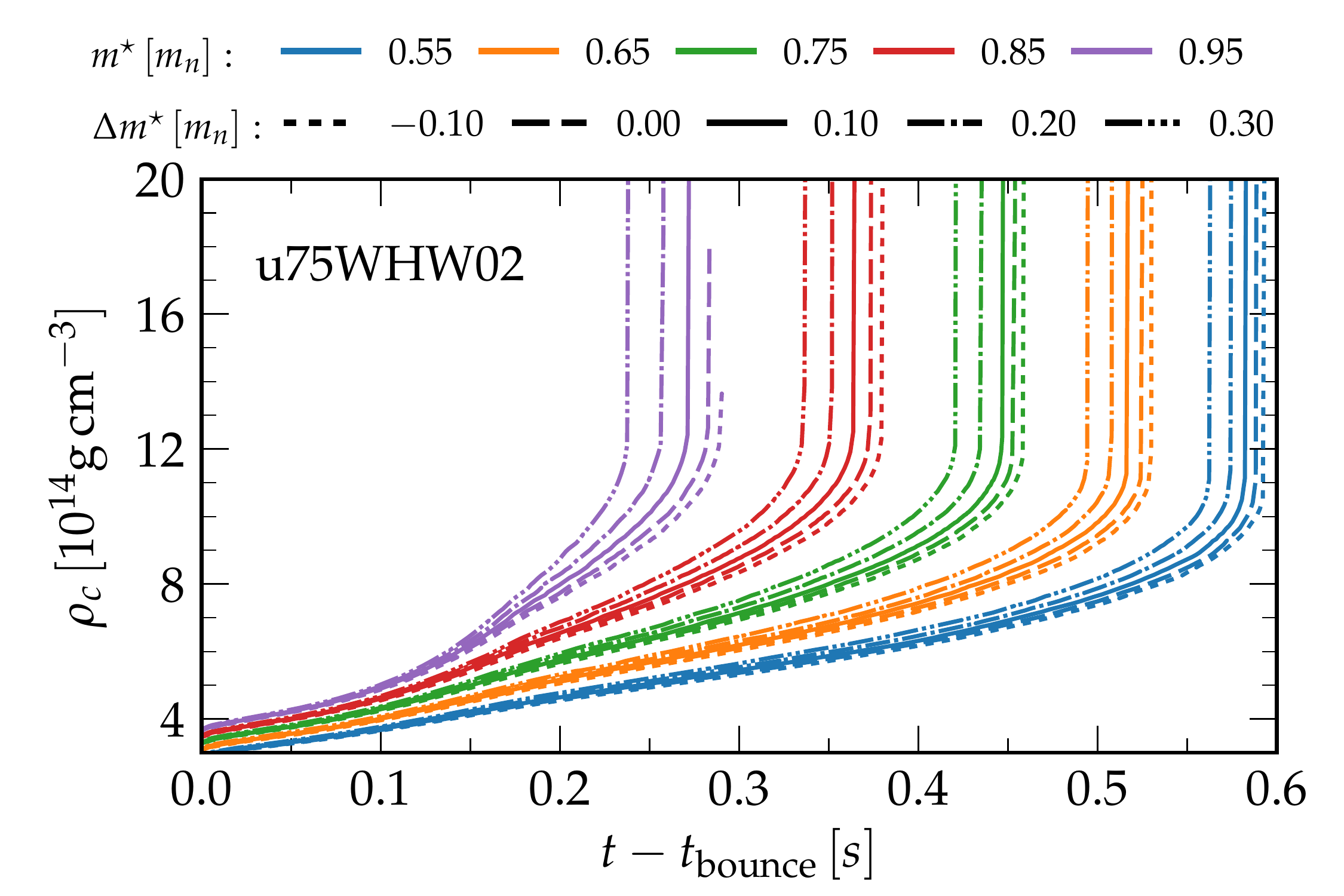}
\hspace{0.25cm}
\includegraphics[width=0.48\textwidth]{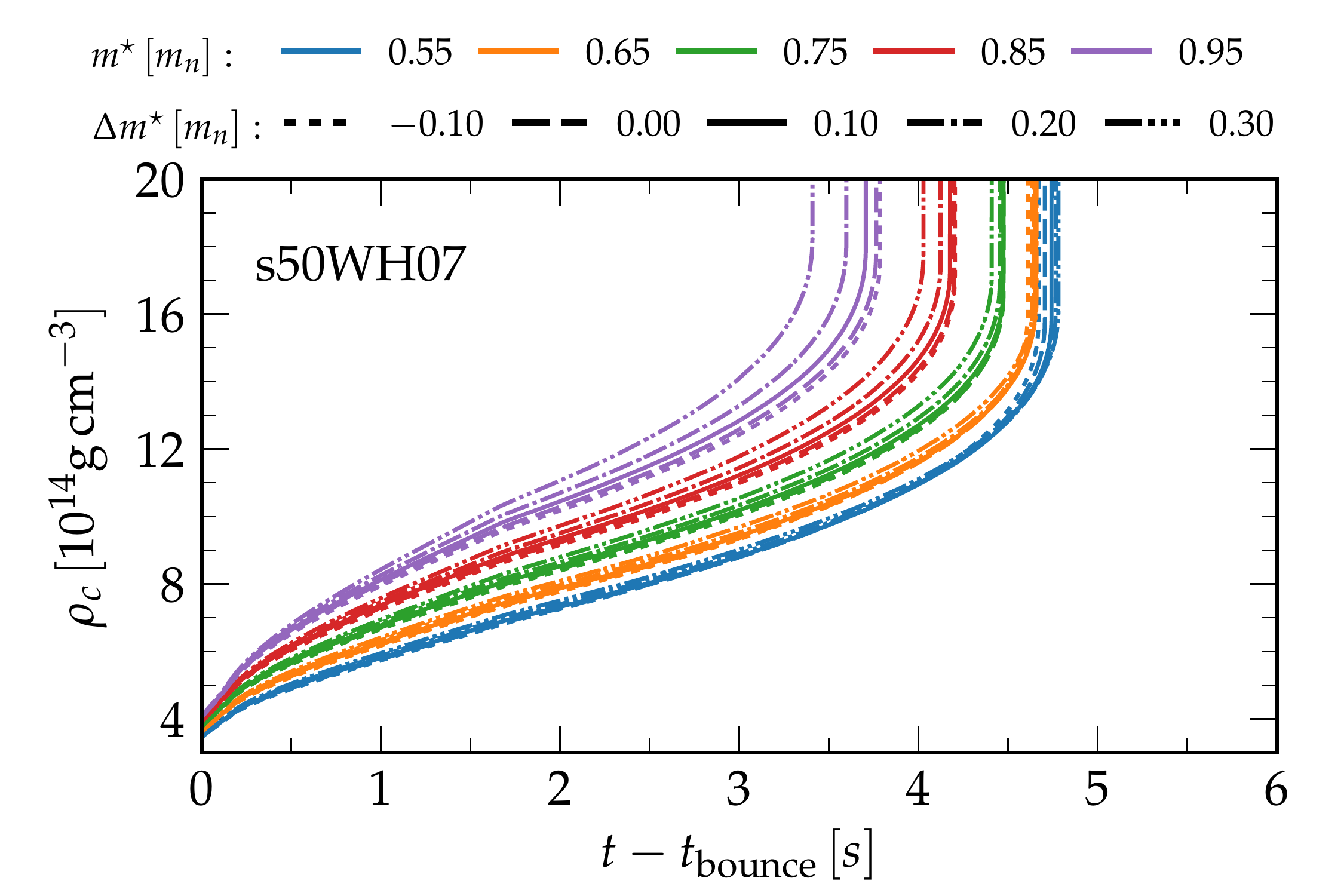}
\\
\includegraphics[width=0.48\textwidth]{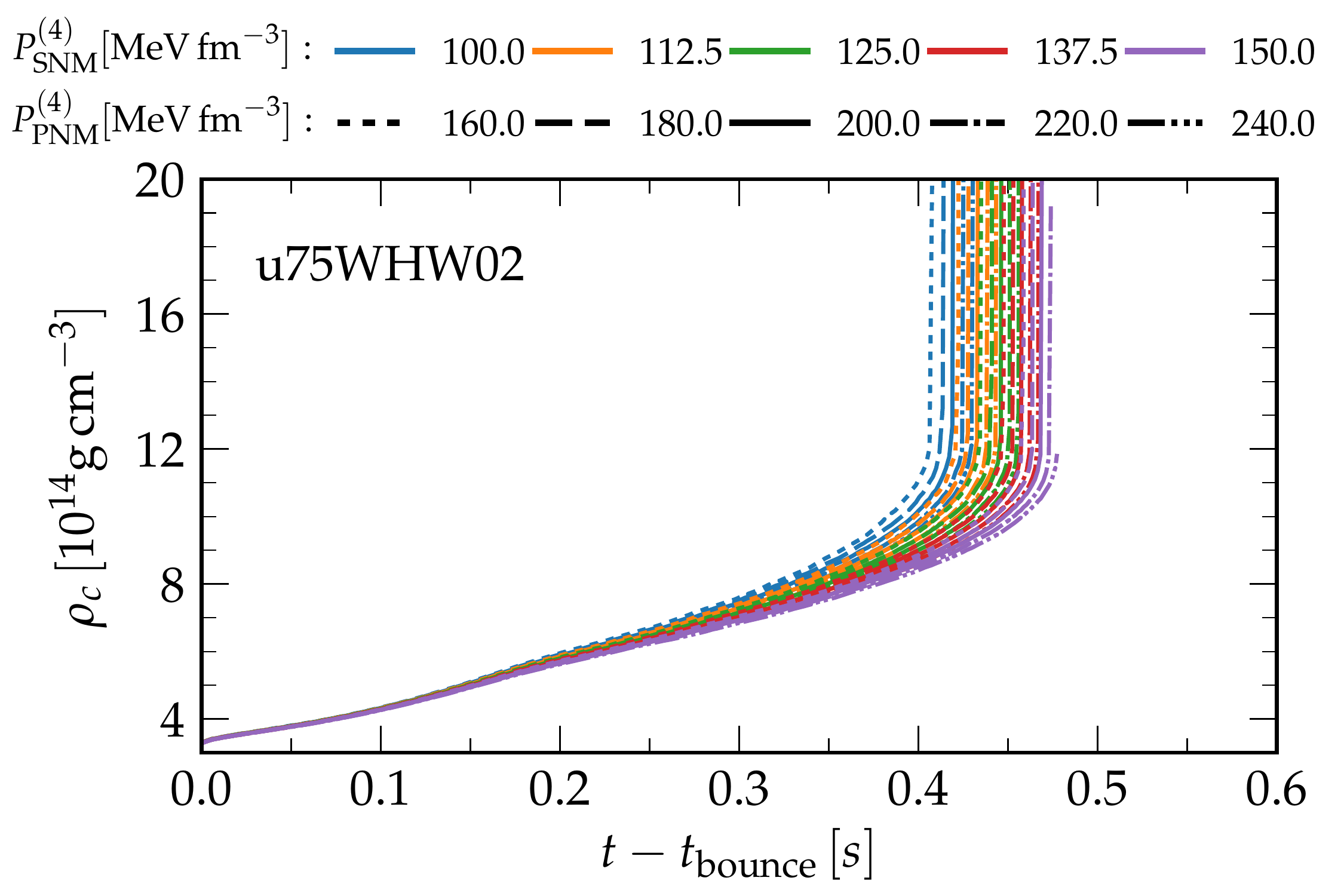}
\hspace{0.25cm}
\includegraphics[width=0.48\textwidth]{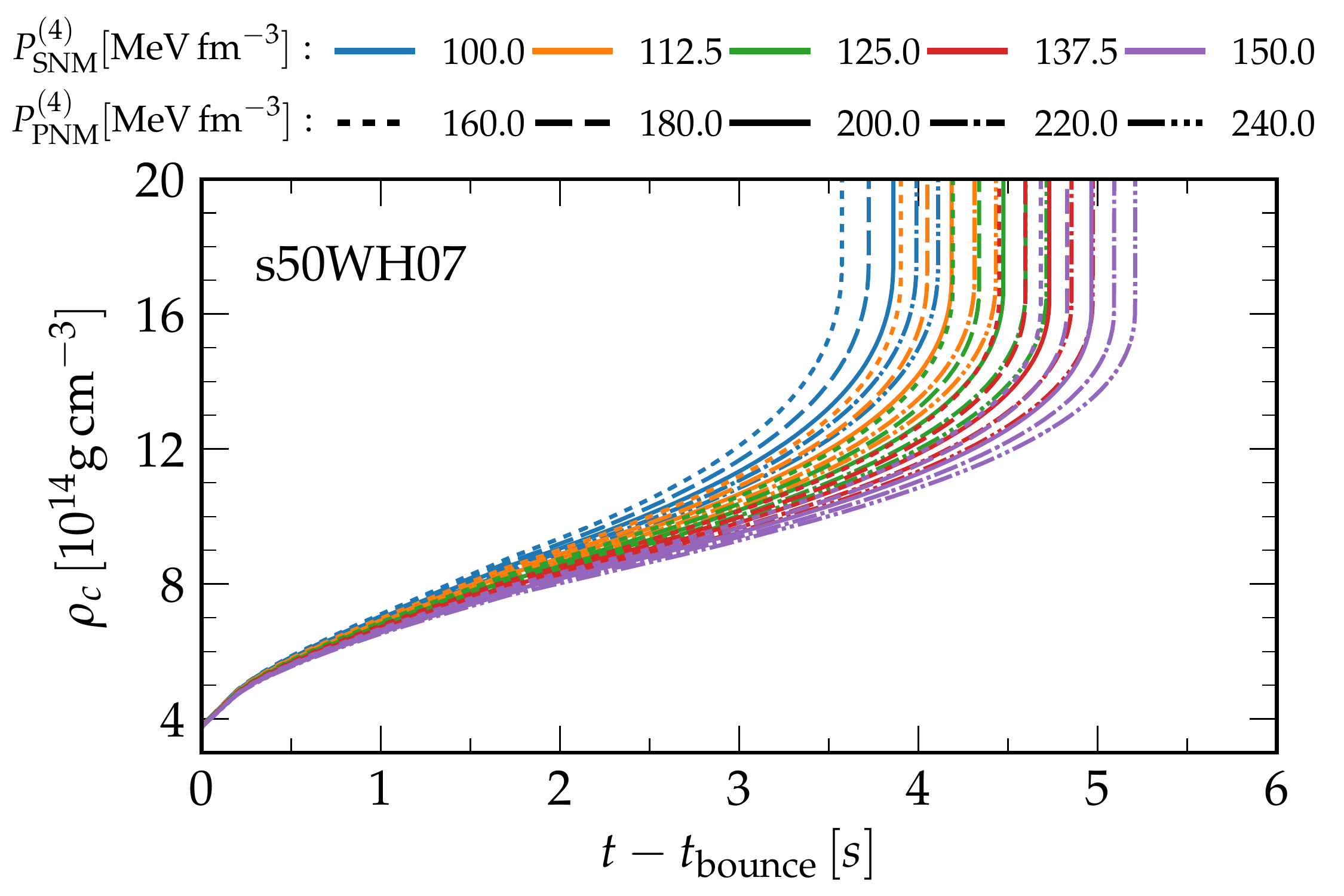}
\caption{\label{fig:rhoc}   Central density evolution for the \texttt{u75WHW02} (left) and \texttt{s50WH07} (right) progenitors, see Section~\ref{ssec:progenitors} for the EOS sets $s_M$ (top) and $s_P$ (bottom), see Section~\ref{ssec:EOS}. 
Differences in the zero temperature EOS, represented by set $s_P$, result only in small differences in central density evolution for the fast collapsing progenitor \texttt{u75WHW02}. 
Meanwhile, differences in the temperature dependence of the EOS, represented by set $s_M$, result in large differences in central density evolution for the same progenitor. 
For the less compact and slowly contracting PNS resulting from the collapse of pre-SN progenitor \texttt{s50WH07} the differences in central density between EOS sets $s_M$ and $s_P$ are not as extreme as for the \texttt{u75WHW02} progenitor.
}
\end{figure*}

In Figure~\ref{fig:rhoc} we plot the central density of the PNSs after core bounce until BH formation for the collapse of the pre-SN progenitors \texttt{u75WHW02} and \texttt{s50WH07} simulated using the EOSs from sets $s_M$ and $s_P$ of \citet{schneider:19}. 
Recall that the EOS set $s_M$ is constructed in such a way that the zero temperature component of each EOS is very similar, while their temperature dependences differ. 
Set $s_P$, on the other hand, has the same temperature dependence for all EOSs, while the zero temperature components of their EOSs differ.

For the high compactness progenitor \texttt{u75WHW02}, $\xi_{2.5}=0.88$, we observe that simulations using set $s_M$ lead to BH formation times that change by almost a factor of 2.5, from $t_\rBH=240\unit{ms}$ for for the softest EOS ($m^\star=0.95\,m_n$ and $\Delta m^\star=0.30\,m_n$) to $t_\rBH=592\unit{ms}$ for the stiffest EOS ($m^\star=0.55\,m_n$ and $\Delta m^\star=-0.10\,m_n$).  
Meanwhile, for the same progenitor, simulations performed with EOSs in set $s_P$ have BH formation times that differ by at most 15\% from each other, $t_\rBH=412\unit{ms}$ for the softest EOS ($P^{(4)}_\rSNM=100\unit{MeV\,fm}^{-3}$ and $P^{(4)}_\rPNM=160\unit{MeV\,fm}^{-3}$) and $t_\rBH=478\unit{ms}$ for the stiffest EOS ($P^{(4)}_\rSNM=150\unit{MeV\,fm}^{-3}$ and $P^{(4)}_\rPNM=240\unit{MeV\,fm}^{-3}$).

The situation changes for the low compactness progenitor \texttt{s50WH07}, $\xi_{2.5} = 0.22$. 
In this case, BH formation times for runs simulated using the $s_P$ set extends from $t_\rBH = 3.5\unit{s}$ for the softest EOS to almost $t_\rBH = 5.2\unit{s}$ for the stiffest EOS, a 48\% increase. 
For set $s_M$ and the same \texttt{s50WH07} progenitor, however, BH formation times become much less disperse than for the very compact progenitor \texttt{u75WHW02}. 
Collapse of the PNS into a BH now takes between $t_\rBH = 3.4\unit{s}$ for the softest EOS in the set to $t_\rBH = 4.7\unit{s}$ for the stiffest, only a 38\% difference.

As also shown by \citet{schneider:19}, the effective mass ($m^\star$) has a higher influence in the collapse evolution than the effective mass splitting ($\Delta m^\star$). 
Thus, for a given pre-SN progenitor, clustering of the predicted central densities soon after bounce for EOSs with similar effective masses is expected. 
Nevertheless, other parameters within their current constraints which were not studied in this work, such as the nuclear incompressibility and symmetry energy at saturation density and its slope, also play a part in setting the central density after core bounce \citep{steiner:13, yasin:18, schneider:19}. 
As expected, for EOSs that differ only in their pressure at high densities, such as the ones in set $s_P$, significant changes in the PNS evolution occur solely after central density is above $n\gtrsim3\,n_\rsat$.

\begin{figure*}[!htb]
\includegraphics[width=0.48\textwidth]{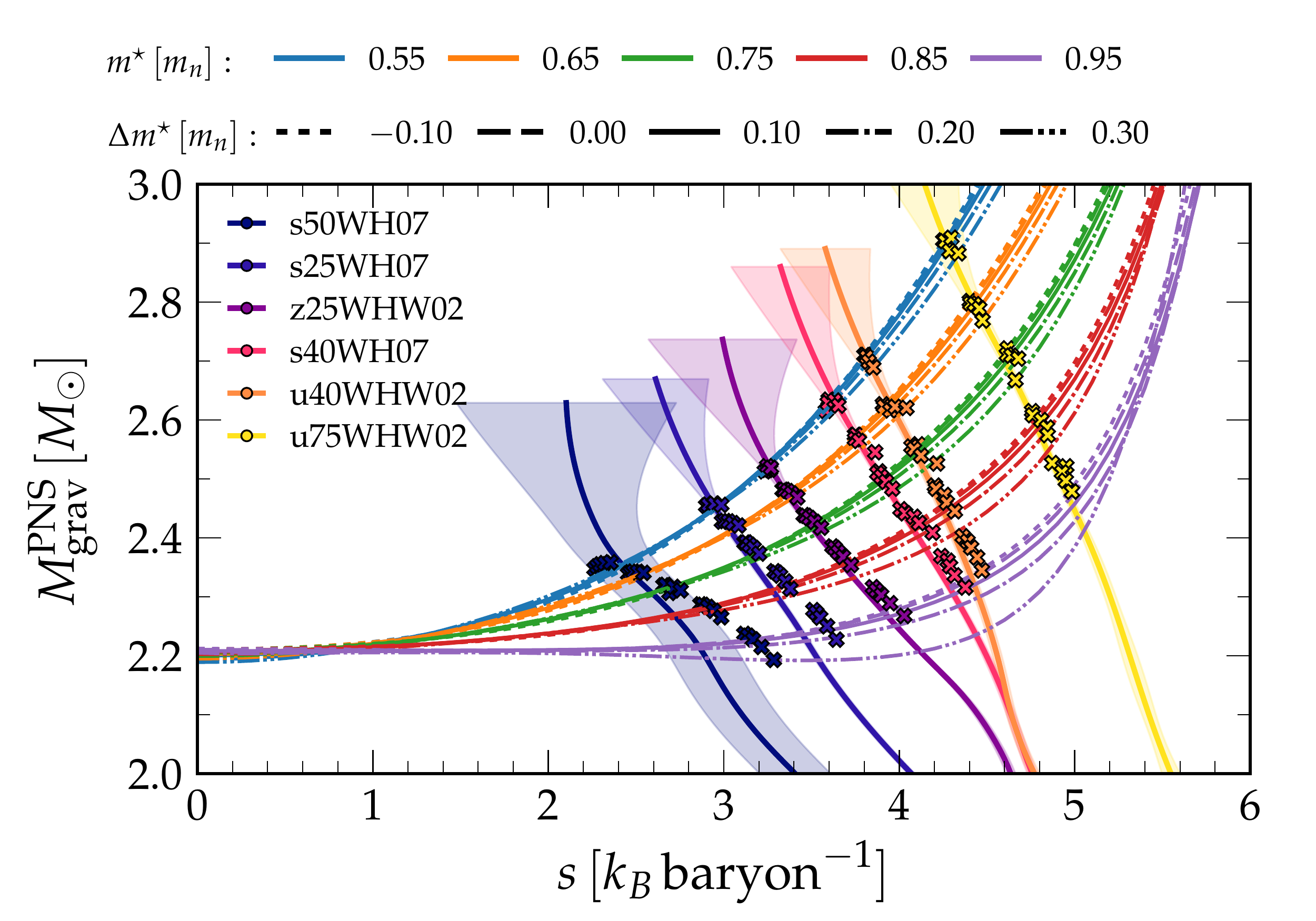}
\vspace{-0.25cm}
\includegraphics[width=0.48\textwidth]{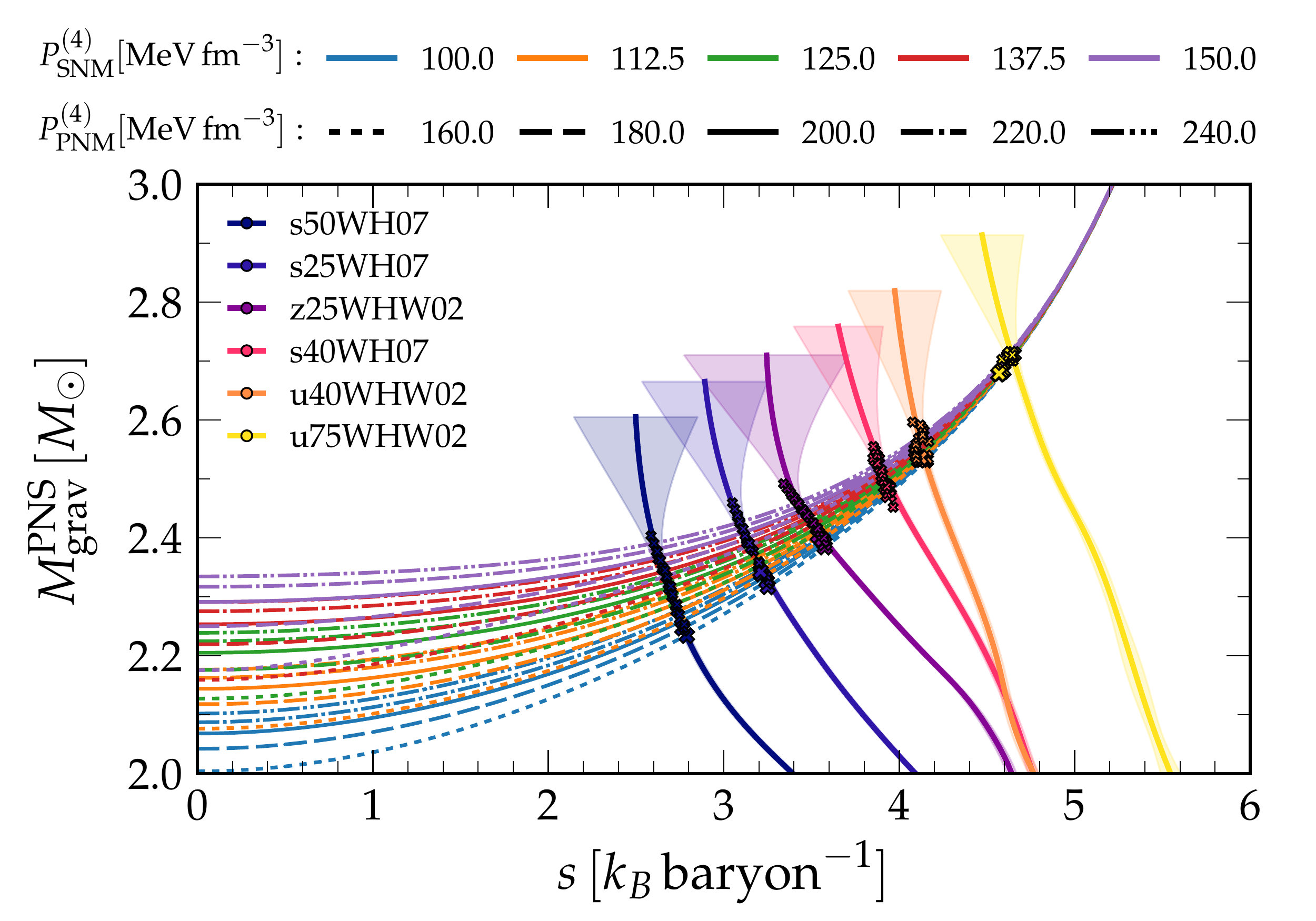}
\\
\caption{\label{fig:mmax_vs_s} Maximum gravitational mass supported by a hot NS with constant entropy for EOSs in set $s_M$ (left) and $s_P$ (right). 
Also plotted are the Gaussian averages and their 95\% confidence intervals for the core collapse 
trajectories in $\tilde{s}-M^\rPNS_\rgrav$ space averaged over all EOSs for each of the six progenitors of \citet{woosley:02, woosley:07}. 
We extend this curve up to $0.3\,M_\odot$ beyond the largest mass achieved for a given progenitor. 
We also pinpoint the moments of BH formation (crosses) of all simulations and note that they deviate little from the Gaussian average curves. 
We note that the PNS follows a similar trajectory in $\tilde{s}-M^\rPNS_\rgrav$ for each progenitor, regardless of the EOS, and collapse into a BH occurs shortly after it crosses the maximum gravitational mass threshold supported by a hot NS with constant entropy for the EOS used. 
The curve for the \texttt{s50WH07} progenitor and the $s_M$ progenitor set looks like an outlier since some of the points at BH formation time look well outside the average region. 
This occurs due to the trajectories in $\tilde{s}-M^\rPNS_\rgrav$ space for the stiffer EOSs in this set having more weight as they take significantly longer to form a BH, as well as the most common entropy, $\tilde{s}$, for runs with the softer EOSs in the set not decreasing as fast as for the stiffer ones. 
}
\end{figure*}

In Figure~\ref{fig:mass_entropy} we showed using the baseline EOS of \citet{schneider:19} and the progenitors of \citet{woosley:02} and \citet{woosley:07} that PNS was unstable against collapse into a BH only after the PNS gravitational mass overcame or was very close to $M_\rgrav^{\rmax,\tilde{s}}$, the maximum gravitational mass supported by a hot NS with constant entropy $\tilde{s}$, see Equation~\eqref{eq:stilde}. 
We verified this result to be true for all combinations of 6 progenitors and 49 EOSs discussed here.

Our approach to determine when a PNS collapses into a BH allows us to identify, for a given pre-SN progenitor, which components of the finite-temperature EOS most significantly impact the core-collapse evolution. 
In Figure~\ref{fig:mmax_vs_s}, we plot the maximum supported gravitational mass for a hot NS with a constant entropy $s$, $M_\rgrav^{\rmax,s}$, for the EOSs in set $s_M$ and $s_P$. 
As expected, EOSs in set $s_M$ produce very similar low-entropy stars, $s\lesssim2\entropy$, and very different high-entropy stars. 
This is due to their non-thermal EOS components being alike, while their thermal components are quite different. 
EOSs in set $s_P$, on the other hand, have the opposite behavior due to their identical thermal components, but distinct non-thermal high-density profiles.

Also in Figure~\ref{fig:mmax_vs_s} we show the Gaussian average and its 95\% confidence interval of the core-collapse trajectory in $\tilde{s}-M^\rPNS_\rgrav$ space\footnote{The Gaussian Process kernel includes an amplitude factor, white noise with level set to 0.1, and a Matèrn component with length scale 2 and smoothness 3/2 \citep{scikit-learn}. More details at \url{https://blog.dominodatalab.com/fitting-gaussian-process-models-python/}}.
The average is computed over all EOSs in a given set for each of the pre-SN progenitors. 
It is remarkable that the path traced in $\tilde{s}-M^\rPNS_\rgrav$ space for each progenitor is almost independent of the chosen EOS up to the point of BH formation. 
For clarity we pinpoint the moments where BH formation takes place. 

A few other remarks about the PNS evolution are important. 
First, for a given progenitor, the paths drawn in $\tilde{s}-M^\rPNS_\rgrav$ space averaged over EOSs are, to first order, close to a straight line, at least in the limited region of parameter space plotted. 
Second, these lines have similar slopes for all progenitors and, furthermore, are approximately ordered by increasing compactness. 
Nevertheless, we notice steeper slopes for progenitors that maintain large accretion rates throughout their evolution history. 
Also, we observe almost no difference between the core-collapse trajectory evolution in $\tilde{s}-M^\rPNS_\rgrav$ space simulated with EOSs in sets $s_M$ and $s_P$. 
The only exception occurs for core-collapse simulations of the \texttt{s50WH07} pre-SN progenitor, since the slow rate of accretion and distinct thermal component of the EOSs in set $s_M$ allow for different evolution of the PNS entropy near BH formation time. 
Thus, this is the combination of EOS set and progenitor with the widest confidence level region. 
All of this, allows us to conclude that for spherically-symmetric core-collapse of compact pre-SN progenitors, the path in $\tilde{s}-M^\rPNS_\rgrav$ space is almost independent of the EOS. 
Thus, this trajectory is mainly a function of the pre-SN stellar structure and, to a good approximation, only a function of progenitor compactness. 
Some of these points can also been inferred from Figure~\ref{fig:swh18_s} for multiple progenitors and a single EOS. 
Therefore, we conclude that computing $M_\rgrav^\rPNS(\tilde{s})$ for a given progenitor for a single EOS may be enough to determine if BH formation would occur faster or slower using a different EOS and what would be the BH initial mass. 
Such a plot can assist us when making comparisons between different progenitors and a single EOS or between one progenitor and different EOSs. 
This is discussed next.

The conclusion above can be used to understand the overall behavior displayed by the results shown in Figure~\ref{fig:rhoc}. 
A very compact pre-SN progenitors such as \texttt{u75WHW02}, which collapses from a PNS to a BH in a few hundred milliseconds, displays quite a wide (narrow) range of BH formation times when simulated with EOSs in set $s_M$ ($s_P$). 
This occurs mainly due to more compact progenitors accreting material fast and not leaving enough time for neutrinos to carry out the excess entropy before gravitational instability settles in. 
Thus, the path traced by this progenitor in $M_\rgrav^\rPNS-\tilde{s}$ phase space crosses the maximum supported mass by hot NSs in the high-entropy region, where EOS sets $s_M$ and $s_P$ predict very different scenarios, and lead to very different BH formation times and initial masses. 
Had we known beforehand the conclusions drawn above, we could have simulated the collapse of \texttt{u75WHW02} using a single EOS and determined with a reasonable accuracy how long it would take for the PNS to collapse into a BH had we simulated it with any other EOS. 
Moreover, the initial BH gravitational mass could also have been predicted without performing additional simulations. 
It is then clear why using EOS set $s_M$ to simulate the collapse of the pre-SN progenitor \texttt{u75WHW02} leads to quite different BH formation times and initial masses, while for set $s_P$ those times and masses are fairly similar.

Low compactness pre-SN progenitors such as \texttt{s50WH07} take a few seconds to collapse into a BH due to their much lower accretion rates. 
Lower accretion rates lead to less compressional heating of the PNS, allowing it more time to emit neutrinos and decrease the total PNS entropy. 
Due to less thermal support low compactness pre-SN progenitor stars will, for most EOSs, form BHs slower and with lower masses than those resulting from the core collapse of more compact progenitors. 
From a single simulation of such a progenitor, using one EOS, we could also determine its path along the $M_\rgrav^\rPNS-\tilde{s}$ plane, although not as accurately as for more compact progenitors. 
Yet, it should be reasonably clear now why the initial BH mass and the time it takes for it to form is as diverse for EOS set $s_M$ as it is for set $s_P$ for the  \texttt{s50WH07} progenitor.

\subsection{Other EOSs}
\label{ssec:others}

\begin{figure}[!htb]
\includegraphics[trim={0.cm 0.cm 0.cm 0.cm},clip,width=0.48\textwidth]{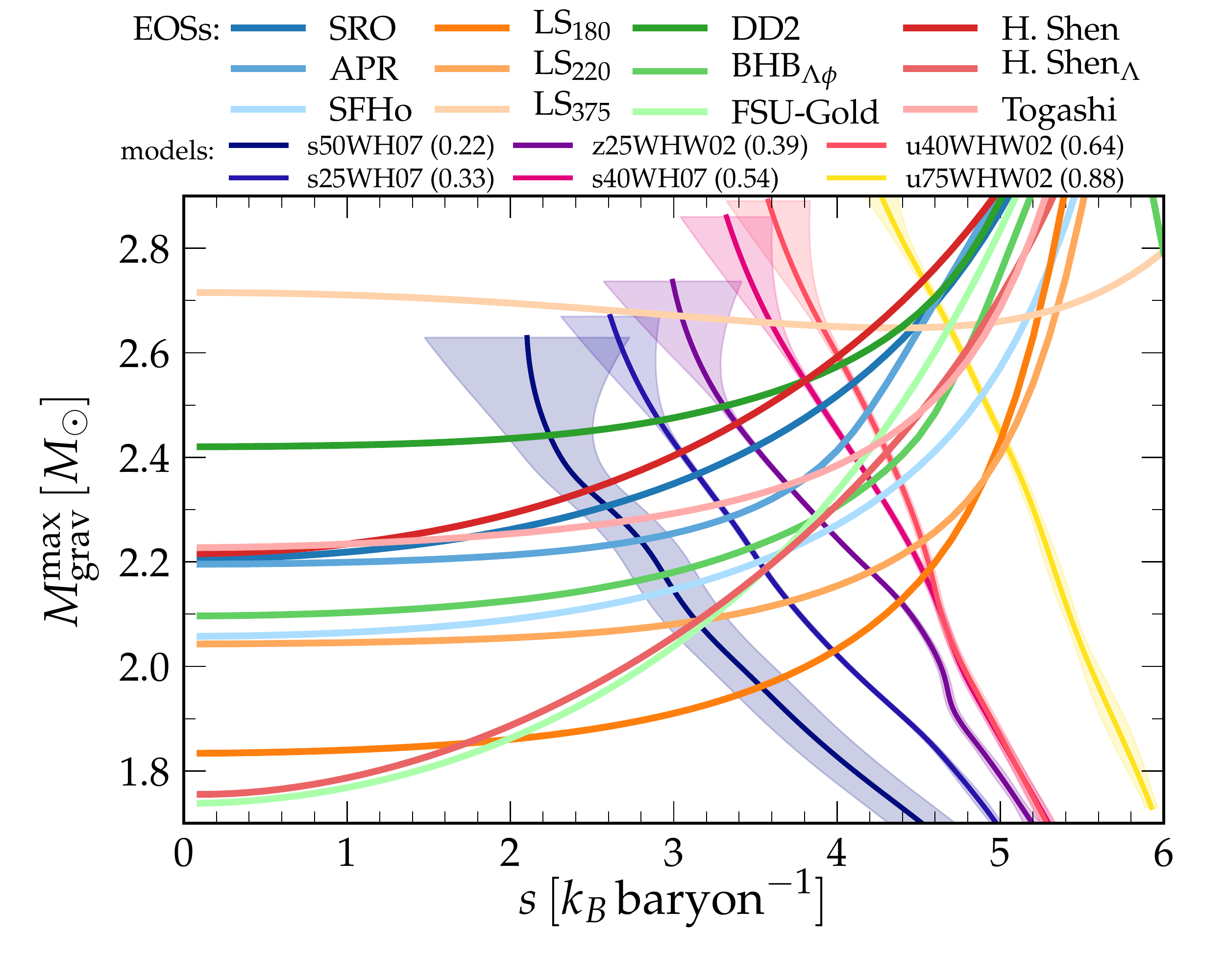}
\includegraphics[trim={0.cm 0.cm 0.cm 0.cm},clip,width=0.48\textwidth]{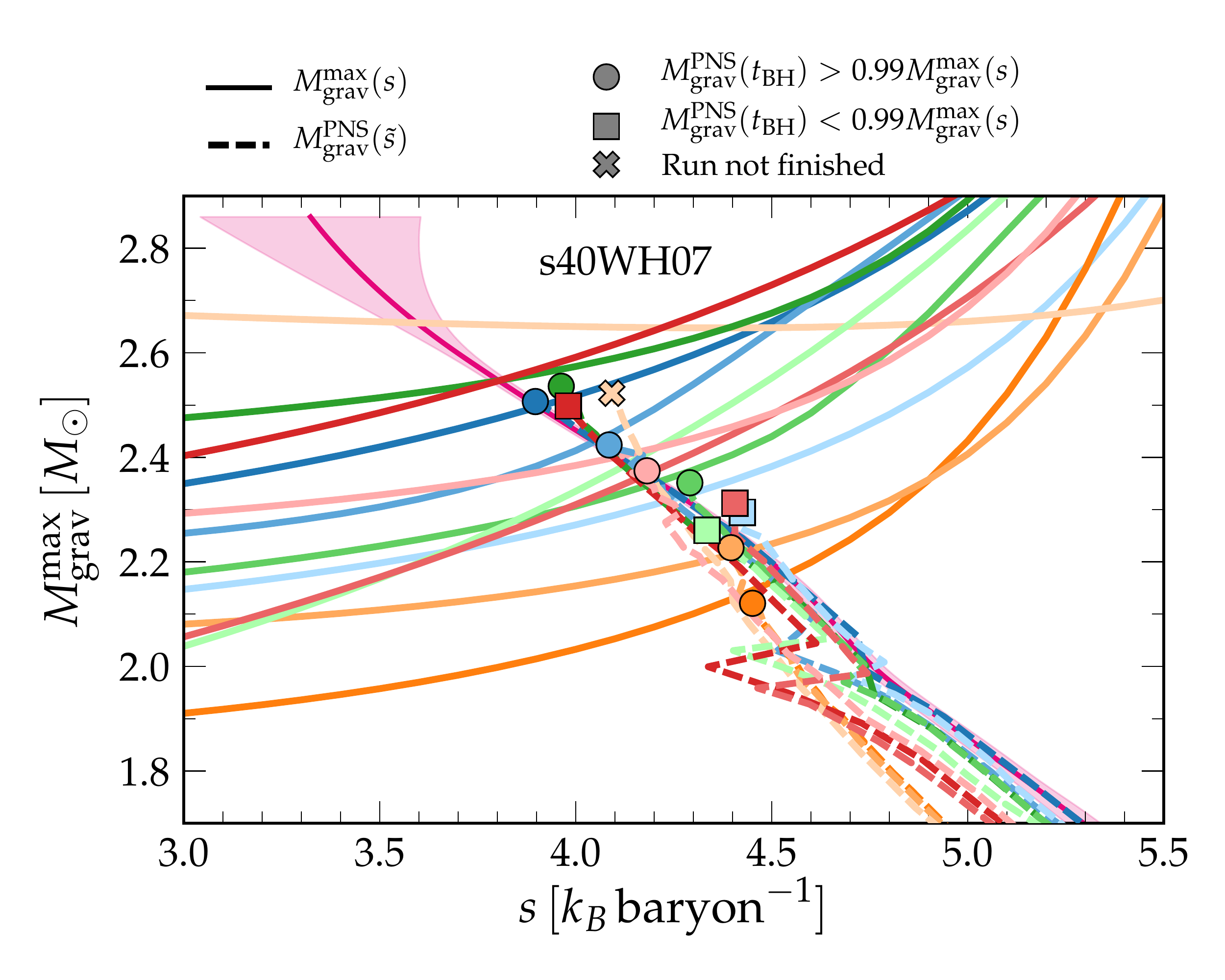}
\vspace{-1cm}
\\
\caption{\label{fig:mmax_vs_s_s40} Top: comparison of the maximum gravitational mass supported by a hot NS with constant entropy, $M_\rgrav^\rPNS$, for selected EOSs in the literature and the average trajectory of the PNS evolution in $M_\rgrav^\rPNS-\tilde{s}$ space during the core collapse of selected pre-SN progenitors models. 
Core collapse trajectory in $M_\rgrav^\rPNS-\tilde{s}$ space depends almost exclusively on the progenitor and, therefore, for a given combination of progenitor and EOS, BH forms near the point where trajectory crosses $M_\rgrav^\rmax(s)$. 
Bottom: Focus on the \texttt{s40WH07} pre-SN progenitor evolution after core bounce and simulations performed for the selected EOSs. 
We mark the BH formation point according to whether BH formed with gravitational mass $M_\rgrav^\rPNS(t_\rBH)$ above or below the one predicted by $M_\rgrav^\rmax(s)$ multiplied by a factor of 0.99. 
Only EOSs that have steep increase in $M^\rmax_\rgrav$ with increasing entropy $s$ form BHs with initial mass lower than $0.99M_\rgrav^\rmax(s)$, but still within 5\% of the mass predicted by the average trajectory $M_\rgrav^\rPNS-\tilde{s}$. 
BH formation times for the LS180, SFHo, SRO, and DD2 EOSs are, respectively, $t_\rBH=0.40$, $0.58$, $0.81$, and $1.03\unit{s}$. 
The simulation employing the \LS{375} EOS could not be evolved until BH formation due to its proton fraction being limited to $y>0.035$. 
}
\end{figure}

In Figure~\ref{fig:mmax_vs_s_s40} we plot the $M_\rgrav^\rPNS(\tilde{s})$ for many EOSs found in the literature. 
We plot the baseline SRO EOS used in this study \citep{schneider:19}, the APR EOS \citep{schneider:19a}, the three variants of the Lattimer and Swesty (LS) EOS with incompressibilities $K_\rsat = 180$, $220$ and $375\unit{MeV\,baryon}^{-1}$ \citep{lattimer:91},
the DD2 and FSU-Gold EOSs \citep{hempel:12}, the DD2 variant including hyperons BHB$_{\Lambda\phi}$ \citep{banik:14}, the H.~Shen EOS \citep{shen:98a} and its variant including $\Lambda$ hyperons \citep{shen:11}, the SFHo EOS \citep{steiner:13}, and the Togashi EOS \citep{togashi:17}. 
We also plot the Gaussian averaged trajectory in $M_\rgrav^\rPNS-\tilde{s}$ space for the six different progenitors discussed in this section for the EOS set $s_M$. 
Because the difference in maximum masses $M_\rgrav^\rPNS(s)$ changes from one EOS to the next for a given entropy $s$, we expect that BH formation times and their initial masses to also do so.
We now use the results of our simulations, which are summarized by the averaged trajectories for each progenitor in $M_\rgrav^\rPNS-\tilde{s}$ space, see Figure~\ref{fig:mmax_vs_s_s40}, to examine previous works. 

First, we note that EOSs that are stiffer at zero entropy/temperature are not necessarily stiffer at finite entropy. 
In fact, the stiffest EOS at zero temperature, \LS{375}, is the softest EOS at $s \simeq 6 \entropy$.
Also, whether an EOS is computed within a relativistic or non-relativistic model matters. 
Both the Togashi and APR EOSs are computed to reproduce the nuclear potential of \citet{akmal:98}, While \citet{togashi:17} use a relativistic formalism, \citet{schneider:19a} use non-relativistic  Skyrme-like extension to compute their APR EOS. 
Although both $M_\rgrav^\rmax(s)$ curves are quite similar, they do not completely agree and the discrepancy between them becomes larger as the entropy increases. 
In fact, $M_\rgrav^\rmax(s)$ for the APR EOS increases faster than for the Togashi EOS, a result of the non-relativistic formalism allowing very stiff EOSs which lead to superluminal sound speeds at very high densities.

Works by \citet{fischer:09, sumiyoshi:06, sumiyoshi:07} compared BH formation for a few progenitors in simulations that employed the \LS{180} EOS, or the \LS{220} EOS in the case of \citet{sumiyoshi:08}, and the H.~Shen EOS \citep{shen:98, shen:98a}. 
In all cases these studies found that for a given progenitor the H.~Shen EOS predicted PNSs to last longer than the \LS{180} EOS, and, thus, collapsed into BHs with larger initial masses. 
This is clear from Fig.~\ref{fig:mmax_vs_s_s40} where the H.~Shen EOS always displays a larger maximum PNS mass than both the \LS{180} or \LS{220} EOS.

\citet{oconnor:11} studied the EOS dependence of BH formation for over 100 pre-SN progenitors of different compactnesses using the open-source general-relativistic \textsc{GR1D} code with three-species neutrino leakage and approximate heating scheme \citep{oconnor:10}. 
They compared the core-collapse evolution for four EOSs, three Skyrme-type EOSs from \citep{lattimer:91} with incompressibilities $K_\rsat=180\unit{MeV}$ (\LS{180}), $220\unit{MeV}$ (\LS{220}) and $375\unit{MeV}$ (\LS{375}) as well as the relativistic mean field (RMF) EOS of \citet{shen:98, shen:98a} (H.~Shen). 
Although two of these EOSs are excluded by experiments, \LS{180} for being too soft and \LS{375} for being too stiff at zero temperature, they are still useful to understand the EOS dependence of BH formation. 
With these four EOSs, \citet{oconnor:11} observed, amongst other things, that BH formation times and initial masses were similar for the pairs \LS{180}-\LS{220} and \LS{375}-H.~Shen EOSs, see their Table~2 and Figure~7. 
While the \LS{180} and \LS{220} EOSs forecast that BHs will form once the PNS gravitational mass exceeds $\simeq2.0\,M_\odot$ for the less compact progenitors and $\simeq2.5\,M_\odot$ for the most compact ones, both the H.~Shen and \LS{375} EOSs predict that BHs form with initially larger masses, $\simeq2.5-2.8\,M_\odot$. 
Additionally, the differences in initial mass and collapse times between the EOS pairs, \LS{180}-\LS{220} and \LS{375}-H.~Shen, decrease with increasing progenitor compactness. 
Furthermore, \LS{375} was the only EOS to allow BH formation to take place for PNS gravitational masses lower than the maximum supported by the EOS at zero temperature.

All of the observations by \citet{oconnor:11} described above can be understood from  Figure~\ref{fig:mmax_vs_s_s40}. 
First, we see that for hot NSs with constant entropy $s$, the \LS{180} and \LS{220} EOSs predict maximum hot-NS gravitational masses that differ by approximately $0.2\,M_\odot$ for $s\lesssim3\entropy$. 
This difference in the maximum gravitational mass decreases for higher entropy and inverts for $s>5.1\entropy$. 
Thus, for spherically symmetric CCSN simulations using these two EOSs, BHs will form soon after the PNS overcomes a mass of $\simeq1.8\,M_\odot$ for the \LS{180} EOS and low-compactness progenitors and with $\simeq0.2\,M_\odot$ more mass for the \LS{220} EOS. 
The difference decreases the more compact the progenitor star is, with both EOSs predicting almost the same BH formation time and initial BH mass for the most compact progenitor tested here, \texttt{u75WHW02}. 
Nevertheless, \citet{oconnor:11} observe that the simulation using the \LS{220} EOS takes longer to collapse into a BH, 285\unit{ms} compared to 226\unit{ms} for the \LS{180} EOS. 
Due to the longer accretion time, the simulation using the \LS{220} EOS also collapses with a slightly larger gravitational mass than the one with the \LS{180} EOS, $2.50\,M_\odot$ to $2.45\,M_\odot$ of \LS{180}. 
We can understand this by referring back to Figure~\ref{fig:u75_evolution} and noticing that not all of the PNS interior has entropy $\tilde{s}$, see Equation~\eqref{eq:stilde}, as in our approach. 
Some of the PNS innermost regions have entropies lower than $\tilde{s}$, which also contribute to preventing BH formation. 
Since for entropies lower than $\tilde{s}$, the \LS{180} EOS provides lower thermal pressure than the \LS{220} EOS, we expect this to play a second order role into the core-collapse evolution. 
In fact, we arrive at a similar conclusion performing simulations of core-collapse of the \texttt{s40WH07} pre-SN progenitor using other EOSs from \citet{hempel:12}, namely the ones that use the TM1, TMA, and FSU-Gold parametrizations of the nuclear interactions, the IU-FSU EOS from \citet{fattoyev:10}, and the H.~Shen EOS \citep{shen:98, shen:98a}. 
Those EOSs predict fast increase of $M^\rmax_\rgrav$ with increasing entropy even for $s\lesssim3\entropy$, see curves for H.~Shen and FSU-Gold in Figure~\ref{fig:mmax_vs_s_s40}.
Thus, it is clear our model may needs some correction in the initial BH mass for some EOSs of order $\lesssim 0.05\,M_\odot$ due to the low entropy at the core of the PNS.

A similar pattern to that observed between the \LS{180} and \LS{220} EOSs pair is observed between the H.~Shen and \LS{375} EOSs: for pre-SN progenitors with $\xi_{2.5}\lesssim0.5$ the H.~Shen EOS predicts PNSs will collapse into BHs with initial masses $\simeq0.3\,M_\odot$ lower than the \LS{375} EOS does. 
However, for the most compact progenitor stars the BH formation times and initial masses are predicted to be similar for both EOSs, with simulations using the \LS{375} EOS forming BHs with lower initial masses than those that employ the H.~Shen EOS. 
This is another feature that we could have predicted from Figure~\ref{fig:mmax_vs_s_s40} due to the crossing of $M^\rmax_\rgrav(s)$ for the \LS{375} and H.~ Shen EOSs near $s=4.5\entropy$. 
Finally, besides our EOS with $m^\star=0.95\,m_n$ and $\Delta m^\star=0.30\,m_n$, \LS{375} is the only other EOS we have found where $M^\rmax_\rgrav(s)$ does not monotonically increase with entropy $s$.

Some works studied the effects of mesons and hyperons in BH formation. 
\citet{sumiyoshi:09, nakazato:12} studied BH formation for the EOSs of \citet{ishizuka:08}, an extension of the \cite{shen:98, shen:98a} EOS that includes pions and $\Lambda$, $\Sigma$, and $\Xi$ hyperons. 
They showed that the appearance of pions and/or hyperons soften the hot EOS, which decreases the maximum mass supported by a PNS against gravitational collapse, and gives rise to a fast increase in neutrino energies and luminosities at late times when compared to purely hadronic EOSs. 
\citet{peres:13} obtained similar results by adding pions and $\Lambda$ hyperons to the \LS{220} EOS. However, the neutrino luminosity of \citet{peres:13} for their simulated core collapse using the \LS{220}+$\Lambda$ EOS was approximately 4 times larger than for the purely hadronic \LS{220} EOS and peaked at $5\times10^{54}\unit{erg\,g}^{-1}$. 
No other studies we found show similar results.

\citet{banik:14a} compared BH formation time and initial masses for progenitors from \citet{woosley:07} for the nuclear H.~Shen EOS \citep{shen:98} and its version containing $\Lambda$ hyperons \citep{shen:98a}. 
Using the \textsc{GR1D} code and a leakage scheme for neutrinos they found that, in general and as expected, $\Lambda$ hyperons lead to lower BH formation times and initial masses. 
A follow up study, \citet{char:15}, also simulated spherically symmetric core-collapse for progenitors from \citet{woosley:07} using two EOSs containing $\Lambda$ hyperons, BHB$_{\Lambda\phi}$ and H.~Shen$_\Lambda$. 
They concluded that BH formation took longer for simulations that used the BHB$_{\Lambda\phi}$ EOS than the ones using the H.~Shen$_\Lambda$ EOS because the repulsive $\Lambda-\Lambda$ interactions included in the former and not in the latter made the BHB$_{\Lambda\phi}$ EOS stiffer. 
Nevertheless, an exception was found by \citet{char:15} for the core collapse of the \texttt{s40WH07} progenitor, which collapsed into a BH slightly faster when simulated H.~Shen$_\Lambda$ EOS. 
In Figure~\ref{fig:mmax_vs_s_s40} we note that the BHB$_{\Lambda\phi}$ EOS is stiffer than the   H.~Shen$_\Lambda$ EOS for most NSs with a fixed entropy except for a small region with $s\simeq4.5\entropy$, where the H.~Shen$_\Lambda$ EOS is stiffer. 
Not surprisingly, this is the region in $M_\rgrav^\rPNS-\tilde{s}$ parameter space we predict the \texttt{s40WH07} progenitor will cross during its collapse. 
Since all other progenitors considered by \citet{char:15} are less compact than \texttt{s40WH07} they will all collapse faster in runs that use the H.~Shen$_\Lambda$ EOS. 
In fact, \citet{char:15} observe that the less compact a progenitor is, the larger the difference in both BH formation time and its initial mass, as we expect.

\citet{pan:18} simulated core collapse of the \texttt{s40WH07} progenitor until BH formation in both 1D spherically symmetric and 2D axisymmetric geometries. 
Their runs were also performed using the \textsc{Flash} software and employed the \LS{220}, BHB$_{\Lambda\phi}$, SFHo, and DD2 EOSs. 
Although their results agree qualitative with ours, the order of collapse for their spherically symmetric runs matches what we would predict with our model, we find that the BH formation time in their runs, see their Table~1, is 13\% faster using the \LS{220} EOS and 11\% slower using the DD2 EOS. 
Also, their BHs form with baryonic masses from $0.10\,M_\odot$ to $0.15\,M_\odot$ lower than ours and PNS entropies $\tilde{s}\simeq0.5\entropy$ higher than we would expect, see their Figure~5. 
We attribute these differences to disparate neutrino treatments and resolution used in the two sets of simulations. 
More importantly, \citet{pan:18} showed that 2D simulations form BHs at later times and with higher masses than in 1D cases. 
The mass increase results from convection arising from the negative entropy and lepton fraction gradients which redistribute matter inside the PNS. 
Note that the \citet{pan:18} simulation that used the DD2 EOS did not lead to BH formation and exploded $1.3\unit{s}$ after core bounce, which was only possible due to this EOS being able to support a PNS with a large mass. 
Nevertheless, the qualitative understanding of BH formation gained from the $M_\rgrav^\rPNS(s)$ relationship is still retained for 2D runs, even if the BH initial masses are shifted to higher values due to convective motion within the PNS. 
In non-rotating full 3D simulations, it is likely that BHs form with a mass between those predicted from spherically symmetric runs, where no convection is present, and those of axisymmetric runs, where convection may be overestimated due to coherent motion of matter. 
In fact, \citet{walk:19} simulated BH formation using full 3D geometry from the collapse of the $40\,M_\odot$ pre-SN progenitor of \citet{woosley:07} using the \LS{220} EOS. 
In their run, BH formation occurs 570\unit{ms} after core bounce, a result that is between the 460\unit{ms} of \citet{pan:18} in spherical symmetry, and 530\unit{ms} in ours using the same progenitor and EOS (also in spherical symmetry), and the 704\unit{ms} predicted by the axisymmetric simulation of \citet{pan:18}.

We test further our hypothesis that a PNS collapses into a BH once its mass overcomes, or comes close to in some cases, the limit $M^\rmax_\rgrav(\tilde{s})$. 
We select a single progenitor, \texttt{s40WH07}, and simulate its core collapse using some of the EOSs found in the literature and discussed in the beginning of this Section. 
We choose this progenitor as it is expected to form a BH in nature as well as have been repeatedly studied by many groups using unique approximations \citep{oconnor:11, hempel:12, steiner:13, banik:14a, char:15, ott:18, pan:18, walk:19}. 
Results of the trajectory for each run are shown in the bottom of Figure~\ref{fig:mmax_vs_s_s40}. 
The simulations performed were evolved until BH formation except for the run using the \LS{375} EOS, which has limitations in its phase space due to the proton fraction being limited to $y>0.035$. 
The trajectory in $M_\rgrav^\rPNS-\tilde{s}$ space for all simulations compare relatively well with our estimates from the SRO EOSs in set $s_M$. 
Some differences arise, particularly for the LS EOSs, likely due to different treatments of matter at low densities, which affect the heat transfer of accreted material onto the PNS.

Lastly, we reiterate that in most cases PNSs collapse into a BH within 1\% of the gravitational mass value predicted by the crossing of the $M^\rPNS_\rgrav(\tilde{s})$ and $M^\rmax_\rgrav(s)$ curves. 
Exceptions for this exist for EOSs which show a fast increase in $M^\rmax_\rgrav(s)$ due to lower thermal pressure contributions from the low entropy innermost region of the PNS. 
For the \texttt{s40WH07} progenitor and EOSs used in this Section, see Figure~\ref{fig:mmax_vs_s}, every BH formed with entropy $\tilde{s}\simeq4\entropy$ and with a mass in the range $0.99-1.01\,M^\rmax_\rgrav(s)$. 
Exceptions exist for the H.~Shen, H.~Shen$_\Lambda$, SFHo, and FSU-Gold EOSs, which collapsed earlier, but still within $2-5\%$ of $M^\rmax_\rgrav(s)$. 
This explains the results of \citet{hempel:12, steiner:13} that correlate the time for the PNS to collapse into a BH and its mass at formation time with the maximum mass supported by a hot NS with constant $s=4\entropy$ entropy. 

%% file: sections/conclusions.tex
\section{Conclusions}
\label{sec:conclusions}

To date many works have shed light on the intricate relationships between the EOS of dense matter and the core collapse evolution of a massive star and resulting GW and neutrino signals \citep{sekiguchi:05, sumiyoshi:06, ott:11, oconnor:11, hempel:12, steiner:13, cerda-duran:13, pan:18, warren:19}. 
We build upon on these studies by performing systematic spherically symmetric CCSN simulations of 51 non-rotating pre-SN progenitors with 60 EOSs until BH formation.

For the 45 progenitors of \citet{sukhbold:18} used in this study, we simulate the collapse employing the baseline Skyrme-type SRO EOS of \citet{schneider:19} using the Newtonian \textsc{Flash} code with a modified GR potential from \citet{marek:06} and an energy-dependent, two-moment neutrino transport.
We observe that in almost every case, the recently formed PNS becomes gravitationally unstable and collapses into a BH soon after the gravitational mass of the PNS exceeds $M_\rgrav^\rmax(\tilde{s})$. 
Here $M_\rgrav^\rmax(\tilde{s})$ is the maximum mass supported by a hot PNS with constant entropy $\tilde{s}$, where $\tilde{s}$ is the most common entropy value within the PNS. 
Interestingly, the trajectory traced by the PNS in $M_\rgrav^\rPNS-\tilde{s}$ space by each  progenitor is close to a straight line where the entropy decreases as the PNS mass increases. 
These lines are approximately ordered by progenitor compactness and their slopes are slightly steeper for progenitors that maintain large accretion rates throughout collapse. 
Thus, studying the collapse of a compact pre-SN progenitor star using a single EOS may be enough to infer when BH formation will take place for most other EOSs.

For several of the compact progenitors, we test and confirm the suitability of employing the general relativistic effective potential in order to model BH formation by comparing the core-collapse evolution to fully general relativistic simulations using \textsc{GR1D}. 
We observe that employing the ``Case A'' of \citet{marek:06} the effective potential slightly underestimates the BH formation time by $\sim$4\% for the most compact progenitors are upwards of $\sim$16\% for the least compact ones.  
We tested modifications to the effective potential, and while we were able to achieve better matching to the central density evolution using ``Case A$_\phi$'', a modification of the standard ``Case A'' of \citet{marek:06}, most quantities of interest, like the neutrino luminosities, are still best reproduced with the standard ``Case A''. 
Moreover, we reaffirm that although compactness, $\xi_M$, is a good indicator of BH formation time and initial BH mass \citep{oconnor:11}, the temporal evolution of the spectra of neutrinos emitted, the accretion rate, and shock radius are not described well by this single parameter. 
As others have pointed out, at least two parameters may be necessary to characterize the core collapse outcome of a progenitor \citep{ertl:16, ebinger:18}.

To better understand EOS effects in a PNS evolution until BH formation we simulate the core collapse of six compact progenitors of \citet{woosley:02, woosley:07}. 
For each progenitor 49 simulations were performed, one simulation using the baseline SRO EOS, 24 simulations where the zero-temperature component of the EOS was altered, set $s_P$, and another 24 simulations where the thermal component of the EOS was altered, set $s_M$. 
For the least compact progenitor star in our set, \texttt{s50WHW07} with compactness $\xi_M=0.22$, both EOS sets $s_M$ and $s_P$ predicted BH would form between $3.4\unit{s}\lesssim t_\rBH \lesssim 5.2\unit{s}$ after core bounce. 
However, for the most compact progenitor in the set, \texttt{u75WH02} with compactness $\xi_M=0.88$, BH formation times were in the $0.40\unit{s}\lesssim t_\rBH \lesssim 0.46\unit{s}$ range for EOSs in set $s_P$ and in the $0.24\unit{s}\lesssim t_\rBH \lesssim 0.59\unit{s}$ range for EOSs in set $s_M$. 
Again, in almost all of these simulations, BH formation occurred only after the PNS gravitational mass surpassed $M_\rgrav^\rmax(\tilde{s})$ or was within 1\% of this value and the trajectory in $M_\rgrav^\rPNS-\tilde{s}$ space was very similar for each progenitor regardless of EOS.

Combining the results above allowed us to find a simple explanation for why BH formation times differ so much for the two EOS sets $s_M$ and $s_P$ for the core collapse of more compact progenitors while being rather similar for less compact ones. 
For EOSs in set $s_M$, which differ only in their thermal component, the differences between  $M_\rgrav^\rmax(s)$ at entropies $s\gtrsim3\entropy$ ($s\lesssim3\entropy$) are large (small).  
Meanwhile, the opposite behavior is observed for EOSs in set $s_P$. 
Because, to a good approximation, BHs from CCSNe form soon after the PNS path in $M_\rgrav^\rPNS-\tilde{s}$ space crosses $M_\rgrav^\rmax(s)$ for the EOS used in the simulation, initial BH masses will be more similar amongst EOSs that predict similar masses at the crossing location. 
This single observation allows us to also explain the patterns seen in CCSN simulations of many previous studies \citep{sumiyoshi:06, nakazato:12, char:15, pan:18}, including that of \citet{oconnor:11}, who simulated core-collapse of over 100 progenitors using 4 different EOSs, as well as to explain why BH formation times and initial gravitational masses for the \texttt{s40WH07} progenitor correlate with $M_\rgrav^\rmax(s=4\entropy)$, the maximum gravitational mass possible for a hot NS with constant $s=4\entropy$ entropy  \citep{hempel:12, steiner:13}.

Recently, \citet{yasin:18, schneider:19} showed that EOSs that are softer with increasing \textit{temperatures} increase the probability of a successful CCSN explosion since these EOSs lead to more compact PNSs which, in turn, increase neutrino luminosities and heating behind the shock. 
Here we show that these same EOSs also lead to BH formation with lower initial masses than EOSs that become stiffer at the large temperatures found within PNSs. 
In this work we focused most on Skyrme-type models, where the temperature dependence of the EOS is almost exclusively dictated by the effective mass of nucleons which are simply a function of nucleon density. 
However, more realistic EOSs will have a more intricate interplay between temperature, effective mass, and thermal pressure. 
In fact, ab-initio calculations by \citet{carbone:19, carbone:19a} show that at higher densities and temperatures, nucleon effective masses may become larger than their vacuum values.

The combined results of \citet{carbone:19, carbone:19a} and \citet{yasin:18, schneider:19} imply that successful supernovae explosions followed by BH formation may be a common evolutionary end path of the core collapse of massive stars. 
Moreover, at the high temperatures and densities found within PNSs, muons \citep{bollig:17}, pions and hyperons \citep{sumiyoshi:09, nakazato:12, peres:13, banik:14a, char:15}, and phase transitions to quark matter \citep{sagert:09, hempel:16, fischer:18, aloy:19} may play a significant role. 
The appearance of massive particles could facilitate supernovae explosions by speeding up PNS contraction and, thus, increasing neutrino emission \citep{sumiyoshi:09, nakazato:12, peres:13}. 
Also, massive leptons and hadrons, in addition to the thermal neutrinos present in the PNS, soften the EOS at high densities and temperatures, which decrease the maximum hot PNS mass. 
Thus, one may expect that for such an EOS, the maximum supported hot PNS mass $M_\rgrav^\rmax(s)$ at a large entropy could be lower than the cold NS maximum mass, as seen here for the \LS{375} EOS and the EOS in set $s_M$ with $m^\star=0.95\,m_n$ and $\Delta m^\star=0.30\,m_n$. 
This is also true for the \LS{220}+$\Lambda$ EOS as shown by \citet{oertel:16}, who studied amongst other subjects the effects of adding pions and/or hyperons on the maximum mass of cold ($T=0$) and hot PNSs ($s=4\entropy$). 
However, for all other EOSs studied here and by \citet{oertel:16} $M_\rgrav^\rmax(s=4\entropy) > M_\rgrav^\rmax(T=0)$.

We speculate on the possible core-collapse outcomes considering that the true EOS of dense matter has a large enough decrease in maximum mass supported by a hot NS. 
In such a case, there could be a combination of EOS and compact pre-SN progenitor which either triggers a neutrino-driven or magneto-rotational explosion that is both strong and early (resulting in no fallback), while producing a PNS that would collapse into BH with mass \textit{lower} than the maximum mass supported by a cold NS. 
\citet{peres:13} find a situation similar to this one in their core-collapse of a $40\,M_\odot$ progenitor star with the \LS{220}+$\Lambda$ EOS, although in their simulation the BH would continue to accrete matter.
Nevertheless, this progenitor forms a BH with initial mass $M^\rBH_\rgrav\simeq1.80\,M_\odot$.  
The initial BH mass is estimated from the baryonic mass and neutrino luminosities presented in \citet{peres:13} assuming that (1) all of the gravitational binding energy is radiated by neutrinos and (2) that the neutrino luminosity remains flat beyond 150\unit{ms} until BH formation time.
This value is lower than the maximum supported cold NS mass $M^\rPNS_\rgrav\simeq1.91\,M_\odot$ and in line with the value expected for this EOS for a hot $s=4\entropy$ NS \citep{oertel:16}.

Despite the curious picture described above, most currently available realistic EOSs predict a scenario where the maximum mass supported by a hot PNS increases with temperature and, thus, it is likely that the higher the compactness of the pre-SN progenitor is, the more massive the PNS will be when it forms a BH. 
The PNS mass evolution as well as the BH formation time and its initial mass may be deduced for a failed galactic supernova if their neutrino signal is detected \citep{kachelriess:05, sumiyoshi:07, suwa:19}. 
As we have shown here (see Figures~\ref{fig:swh18_s}, \ref{fig:mmax_vs_s}, and \ref{fig:mmax_vs_s_s40}), the PNS path in $M_\rgrav^\rPNS-\tilde{s}$ phase space is mostly a function of the compactness of the pre-SN progenitor star while the BH formation time is a function of the EOS. 
Therefore, an observation of the compactness, for example, via neutrinos \citep{horiuchi:17}, from a failed galactic supernova could then be used to impose strong constraints both on the structure of the innermost regions of the progenitor star as well as the temperature dependence of the EOS, even more so if coupled to GW detections. 
The final BH mass, on the other hand, may not be achieved for years after the core collapse and will be function of more than just the compactness of the progenitor since the total ejected and accreted masses depend on the total energy transferred by neutrinos to the shock and the structure of the progenitor star.

We emphasize that all of our simulations were performed considering non-rotating spherically symmetric stars and that CCSNe are fundamentally multidimensional systems. 
Effects due to rotation and convection, both of which are expected to increase the maximum PNS mass before its collapse into a BH \citep{cook:94, morrison:04, sekiguchi:05, oconnor:11, pan:18, obergaulinger:19, walk:19, nagakura:19}, should be addressed in the future. 
In extreme cases rotation may lead to a significant increase in the supported gravitational mass for cold NSs \citep{espino:19, szkudlarek:19}. 
Even though extreme rotation rates may be rare in stellar collapse, a picture favored by binary BH merger detections \citep{abbott:19} and stellar evolution scenarios \citep{belczynski:19, fuller:19}, its effects are not uniform across EOSs \citep{oconnor:11, richers:17}. 
Thus, an extra axis to the $M_\rgrav^\rPNS-\tilde{s}$ diagram has to be added to address its impact on PNS evolution. 
Finally, although the consequences of convection within the PNS may not be as extreme as that of rotation, it still could be the decisive factor between a successful or a failed CCSN in some cases \citep{pan:18, nagakura:19} and shift the $M_\rgrav^\rmax(s)$ relations computed for each EOS and/or progenitor.

%% file: sections/acknowledgements.tex
\begin{acknowledgments}

This work benefited from interesting and helpful discussions with M.~Prakash, C.~Constantinou, MK.L.~Warren, and H.~Yasin. 
The authors acknowledge support from the Swedish Research Council (Project No. 2018-04575).
The simulations were performed on resources provided by the Swedish National Infrastructure for Computing (SNIC) at PDC and NSC.
The software used in this work was in part developed by the DOE NNSA-ASC OASCR Flash Center
at the University of Chicago.
SMC is supported by the U.S. Department of Energy, Office of Science, Office of Nuclear Physics, under Award Numbers DE-SC0015904 and DE-SC0017955.

Software: FLASH \citep{fryxell:00,dubey:09,couch:13, oconnor:18}, GR1D \citep{oconnor:10,oconnor:15},  Matplotlib \citep{hunter:07}, NuLib \citep{oconnor:15}, SROEOS \citep{schneider:17}, scikit-learn \citep{scikit-learn}.

\end{acknowledgments}

%% file: bh_sro.bbl
\begin{thebibliography}{}
\expandafter\ifx\csname natexlab\endcsname\relax\def\natexlab#1{#1}\fi
\providecommand{\url}[1]{\href{#1}{#1}}
\providecommand{\dodoi}[1]{doi:~\href{http://doi.org/#1}{\nolinkurl{#1}}}
\providecommand{\doeprint}[1]{\href{http://ascl.net/#1}{\nolinkurl{http://ascl.net/#1}}}
\providecommand{\doarXiv}[1]{\href{https://arxiv.org/abs/#1}{\nolinkurl{https://arxiv.org/abs/#1}}}

\bibitem[{Abbott {et~al.}(2019{\natexlab{a}})Abbott, Abbott, Abbott, Abraham,
  Acernese, Ackley, Adams, Adhikari, Adya, Affeldt, Agathos, Agatsuma,
  Aggarwal, Aguiar, Aiello, Ain, Ajith, Allen, Allocca, Aloy, Altin, Amato,
  Ananyeva, Anderson, Anderson, Angelova, Antier, Appert, Arai, Araya, Areeda,
  Ar{\`e}ne, Arnaud, Arun, Ascenzi, Ashton, Aston, Astone, Aubin, Aufmuth,
  AultONeal, Austin, Avendano, {Avila-Alvarez}, Babak, Bacon, Badaracco, Bader,
  Bae, Baker, Baldaccini, Ballardin, Ballmer, Banagiri, Barayoga, Barclay,
  Barish, Barker, Barkett, Barnum, Barone, Barr, Barsotti, Barsuglia, Barta,
  Bartlett, Bartos, Bassiri, Basti, Bawaj, Bayley, Bazzan, B{\'e}csy, Bejger,
  Belahcene, Bell, Beniwal, Berger, Bergmann, Bernuzzi, Bero, Berry,
  Bersanetti, Bertolini, Betzwieser, Bhandare, Bidler, Bilenko, Bilgili,
  Billingsley, Birch, Birney, Birnholtz, Biscans, Biscoveanu, Bisht, Bitossi,
  Bizouard, Blackburn, Blackman, Blair, Blair, Blair, Bloemen, Bode, Boer,
  Boetzel, Bogaert, Bondu, Bonilla, Bonnand, Booker, Boom, Booth, Bork, Boschi,
  Bose, Bossie, Bossilkov, Bosveld, Bouffanais, Bozzi, Bradaschia, Brady,
  Bramley, Branchesi, Brau, Briant, Briggs, Brighenti, Brillet, Brinkmann,
  Brisson, Brockill, Brooks, Brown, Brunett, Buikema, Bulik, Bulten, Buonanno,
  Buskulic, Bustamante~Rosell, Buy, Byer, Cabero, Cadonati, Cagnoli, Cahillane,
  Calder{\'o}n~Bustillo, Callister, Calloni, Camp, Campbell, Canepa, Cannon,
  Cao, Cao, Capocasa, Carbognani, Caride, Carney, Carullo, Casanueva~Diaz,
  Casentini, Caudill, Cavagli{\`a}, Cavalier, Cavalieri, Cella,
  {Cerd{\'a}-Dur{\'a}n}, Cerretani, Cesarini, Chaibi, Chakravarti, Chamberlin,
  Chan, Chao, Charlton, Chase, {Chassande-Mottin}, Chatterjee, Chaturvedi,
  Chatziioannou, Cheeseboro, Chen, Chen, Chen, Cheng, Cheong, Chia, Chincarini,
  Chiummo, Cho, Cho, Cho, Christensen, Chu, Chua, Chung, Chung, Ciani, Ciobanu,
  Ciolfi, Cipriano, Cirone, Clara, Clark, Clearwater, Cleva, Cocchieri, Coccia,
  Cohadon, Cohen, Colgan, Colleoni, Collette, Collins, Cominsky, Constancio,
  Conti, Cooper, Corban, Corbitt, {Cordero-Carri{\'o}n}, Corley, Cornish,
  Corsi, Cortese, Costa, Cotesta, Coughlin, Coughlin, Coulon, Countryman,
  Couvares, Covas, Cowan, Coward, Cowart, Coyne, Coyne, Creighton, Creighton,
  Cripe, Croquette, Crowder, Cullen, Cumming, Cunningham, Cuoco, Canton,
  D{\'a}lya, Danilishin, D'Antonio, Danzmann, Dasgupta, Da~Silva~Costa,
  Datrier, Dattilo, Dave, Davier, Davis, Daw, DeBra, Deenadayalan, Degallaix,
  De~Laurentis, Del{\'e}glise, Del~Pozzo, DeMarchi, Demos, Dent, De~Pietri,
  Derby, De~Rosa, De~Rossi, DeSalvo, {de Varona}, Dhurandhar, D{\'i}az,
  Dietrich, Di~Fiore, Di~Giovanni, Di~Girolamo, Di~Lieto, Ding, Di~Pace,
  Di~Palma, Di~Renzo, Dmitriev, Doctor, Donovan, Dooley, Doravari, Dorrington,
  Downes, Drago, Driggers, Du, Ducoin, Dupej, Dwyer, Easter, Edo, Edwards,
  Effler, Ehrens, Eichholz, Eikenberry, Eisenmann, Eisenstein, Essick,
  Estelles, Estevez, Etienne, Etzel, Evans, Evans, Fafone, Fair, Fairhurst,
  Fan, Farinon, Farr, Farr, {Fauchon-Jones}, Favata, Fays, Fazio, Fee, Feicht,
  Fejer, Feng, {Fernandez-Galiana}, Ferrante, Ferreira, Ferreira, Ferrini,
  Fidecaro, Fiori, Fiorucci, Fishbach, Fisher, Fishner, {Fitz-Axen}, Flaminio,
  Fletcher, Flynn, Fong, Font, Forsyth, Fournier, Frasca, Frasconi, Frei,
  Freise, Frey, Frey, Fritschel, Frolov, Fulda, Fyffe, Gabbard, Gadre, Gaebel,
  Gair, Gammaitoni, Ganija, Gaonkar, Garcia, {Garc{\'i}a-Quir{\'o}s}, Garufi,
  Gateley, Gaudio, Gaur, Gayathri, Gemme, Genin, Gennai, George, George,
  Gergely, Germain, Ghonge, Ghosh, Ghosh, Ghosh, Giacomazzo, Giaime, Giardina,
  Giazotto, Gill, Giordano, Glover, Godwin, Goetz, Goetz, Goncharov,
  Gonz{\'a}lez, Gonzalez~Castro, Gopakumar, Gorodetsky, Gossan, Gosselin,
  Gouaty, Grado, Graef, Granata, Grant, Gras, Grassia, Gray, Gray, Greco,
  Green, Green, Gretarsson, Groot, Grote, Grunewald, Gruning, Guidi, Gulati,
  Guo, Gupta, Gupta, Gustafson, Gustafson, Haegel, Halim, Hall, Hall, Hamilton,
  Hammond, Haney, Hanke, Hanks, Hanna, Hannam, Hannuksela, Hanson, Hardwick,
  Haris, Harms, Harry, Harry, Haster, Haughian, Hayes, Healy, Heidmann,
  Heintze, Heitmann, Hello, Hemming, Hendry, Heng, Hennig, Heptonstall,
  Hernandez~Vivanco, Heurs, Hild, Hinderer, Hoak, Hochheim, Hofman, Holgado,
  Holland, Holt, Holz, Hopkins, Horst, Hough, Howell, Hoy, Hreibi, Huang,
  Huerta, Huet, Hughey, Hulko, Husa, Huttner, {Huynh-Dinh}, Idzkowski, Iess,
  Ingram, Inta, Intini, Irwin, Isa, Isac, Isi, Iyer, Izumi, Jacqmin, Jadhav,
  Jani, Janthalur, Jaranowski, Jenkins, Jiang, Johnson, {Johnson-McDaniel},
  Jones, Jones, Jones, Jonker, Ju, Junker, Kalaghatgi, Kalogera, Kamai,
  Kandhasamy, Kang, Kanner, Kapadia, Karki, Karvinen, Kashyap, Kasprzack,
  Katsanevas, Katsavounidis, Katzman, Kaufer, Kawabe, Keerthana,
  K{\'e}f{\'e}lian, Keitel, Kennedy, Key, Khalili, Khan, Khan, Khan, Khan,
  Khazanov, Khursheed, Kijbunchoo, Kim, Kim, Kim, Kim, Kim, Kim, Kimball, King,
  King, {Kinley-Hanlon}, Kirchhoff, Kissel, Kleybolte, Klika, Klimenko,
  Knowles, Koch, Koehlenbeck, Koekoek, Koley, Kondrashov, Kontos, Koper,
  Korobko, Korth, Kowalska, Kozak, Kringel, Krishnendu, Kr{\'o}lak, Kuehn,
  Kumar, Kumar, Kumar, Kumar, Kuo, Kutynia, Kwang, Lackey, Lai, Lam, Landry,
  Lane, Lang, Lange, Lantz, Lanza, {Lartaux-Vollard}, Lasky, Laxen, Lazzarini,
  Lazzaro, Leaci, Leavey, Lecoeuche, Lee, Lee, Lee, Lee, Lee, Lee, Lehmann,
  Lenon, Leroy, Letendre, Levin, Li, Li, Li, Li, Lin, Linde, Linker,
  Littenberg, Liu, Liu, Lo, Lockerbie, London, Longo, Lorenzini, Loriette,
  Lormand, Losurdo, Lough, Lousto, Lovelace, Lower, L{\"u}ck, Lumaca, Lundgren,
  Lynch, Ma, Macas, Macfoy, MacInnis, Macleod, Macquet, {Maga{\~n}a-Sandoval},
  Maga{\~n}a~Zertuche, Magee, Majorana, Maksimovic, Malik, Man, Mandic,
  Mangano, Mansell, Manske, Mantovani, Marchesoni, Marion, M{\'a}rka,
  M{\'a}rka, Markakis, Markosyan, Markowitz, Maros, Marquina, Marsat, Martelli,
  Martin, Martin, Martynov, Mason, Massera, Masserot, Massinger, {Masso-Reid},
  Mastrogiovanni, Matas, Matichard, Matone, Mavalvala, Mazumder, McCann,
  McCarthy, McClelland, McCormick, McCuller, McGuire, McIver, McManus, McRae,
  McWilliams, Meacher, Meadors, Mehmet, Mehta, Meidam, Melatos, Mendell,
  Mercer, Mereni, Merilh, Merzougui, Meshkov, Messenger, Messick, Metzdorff,
  Meyers, Miao, Michel, Middleton, Mikhailov, Milano, Miller, Miller,
  Millhouse, Mills, {Milovich-Goff}, Minazzoli, Minenkov, Mishkin, Mishra,
  Mistry, Mitra, Mitrofanov, Mitselmakher, Mittleman, Mo, Moffa, Mogushi,
  Mohapatra, Montani, Moore, Moraru, Moreno, Morisaki, Mours, {Mow-Lowry},
  Mukherjee, Mukherjee, Mukherjee, Mukund, Mullavey, Munch, Mu{\~n}iz,
  Muratore, Murray, Nagar, Nardecchia, Naticchioni, Nayak, Neilson, Nelemans,
  Nelson, Nery, Neunzert, Ng, Ng, Nguyen, Nichols, Nielsen, Nissanke, Nitz,
  Nocera, North, Nuttall, Obergaulinger, Oberling, O'Brien, O'Dea, Ogin, Oh,
  Oh, Ohme, Ohta, Okada, Oliver, Oppermann, Oram, O'Reilly, Ormiston, Ortega,
  O'Shaughnessy, Ossokine, Ottaway, Overmier, Owen, Pace, Pagano, Page, Pai,
  Pai, Palamos, Palashov, Palomba, {Pal-Singh}, Pan, Pang, Pang, Pankow,
  Pannarale, Pant, Paoletti, Paoli, Papa, Parida, Parker, Pascucci,
  Pasqualetti, Passaquieti, Passuello, Patil, Patricelli, Pearlstone, Pedersen,
  Pedraza, Pedurand, Pele, Penn, Perego, Perez, Perreca, Pfeiffer, Phelps,
  Phukon, Piccinni, Pichot, Piergiovanni, Pillant, Pinard, Pirello, Pitkin,
  Poggiani, Pong, Ponrathnam, Popolizio, Porter, Powell, Prajapati, Prasad,
  Prasai, Prasanna, Pratten, Prestegard, Privitera, Prodi, Prokhorov, Puncken,
  Punturo, Puppo, P{\"u}rrer, Qi, Quetschke, Quinonez, Quintero,
  {Quitzow-James}, Raab, Radkins, Radulescu, Raffai, Raja, Rajan, Rajbhandari,
  Rakhmanov, Ramirez, {Ramos-Buades}, Rana, Rao, Rapagnani, Raymond, Razzano,
  Read, Regimbau, Rei, Reid, Reitze, Ren, Ricci, Richardson, Richardson,
  Ricker, Riemenschneider, Riles, Rizzo, Robertson, Robie, Robinet, Rocchi,
  Rolland, Rollins, Roma, Romanelli, Romano, Romel, Romie, Rose, Rosi{\'n}ska,
  Rosofsky, Ross, Rowan, R{\"u}diger, Ruggi, Rutins, Ryan, Sachdev, Sadecki,
  Sakellariadou, Salafia, Salconi, Saleem, Salemi, Samajdar, Sammut, Sanchez,
  Sanchez, {Sanchis-Gual}, Sandberg, Sanders, Santiago, Sarin, Sassolas,
  Sathyaprakash, Saulson, Sauter, Savage, Schale, Scheel, Scheuer, Schmidt,
  Schnabel, Schofield, Sch{\"o}nbeck, Schreiber, Schulte, Schutz, Schwalbe,
  Scott, Scott, Seidel, Sellers, Sengupta, Sennett, Sentenac, Sequino, Sergeev,
  Setyawati, Shaddock, Shaffer, Shahriar, Shaner, Shao, Sharma, Shawhan, Shen,
  Shink, Shoemaker, Shoemaker, ShyamSundar, Siellez, Sieniawska, Sigg, Silva,
  Singer, Singh, Singhal, Sintes, Sitmukhambetov, Skliris, Slagmolen,
  {Slaven-Blair}, Smith, Smith, Somala, Son, Sorazu, Sorrentino, Souradeep,
  Sowell, Spencer, Srivastava, Srivastava, Staats, Stachie, Standke, Steer,
  Steinke, Steinlechner, Steinlechner, Steinmeyer, Stevenson, Stocks, Stone,
  Stops, Strain, Stratta, Strigin, Strunk, Sturani, Stuver, Sudhir,
  Summerscales, Sun, Sunil, Suresh, Sutton, Swinkels, Szczepa{\'n}czyk, Tacca,
  Tait, Talbot, Talukder, Tanner, T{\'a}pai, Taracchini, Tasson, Taylor, Thies,
  Thomas, Thomas, Thondapu, Thorne, Thrane, Tiwari, Tiwari, Tiwari, Toland,
  Tonelli, Tornasi, {Torres-Forn{\'e}}, Torrie, T{\"o}yr{\"a}, Travasso,
  Traylor, Tringali, Trovato, Trozzo, Trudeau, Tsang, Tse, Tso, Tsukada, Tsuna,
  Tuyenbayev, Ueno, Ugolini, Unnikrishnan, Urban, Usman, Vahlbruch, Vajente,
  Valdes, {van Bakel}, {van Beuzekom}, {van den Brand}, Van Den~Broeck,
  {Vander-Hyde}, {van Heijningen}, {van der Schaaf}, {van Veggel}, Vardaro,
  Varma, Vass, Vas{\'u}th, Vecchio, Vedovato, Veitch, Veitch, Venkateswara,
  Venugopalan, Verkindt, Vetrano, Vicer{\'e}, Viets, Vine, Vinet, Vitale, Vo,
  Vocca, Vorvick, Vyatchanin, Wade, Wade, Wade, Walet, Walker, Wallace, Walsh,
  Wang, Wang, Wang, Wang, Wang, Ward, Warden, Warner, Was, Watchi, Weaver, Wei,
  Weinert, Weinstein, Weiss, Wellmann, Wen, Wessel, We{\ss}els, Westhouse,
  Wette, Whelan, White, Whiting, Whittle, Wilken, Williams, Williamson, Willis,
  Willke, Wimmer, Winkler, Wipf, Wittel, Woan, Woehler, Wofford, Worden,
  Wright, Wu, Wysocki, Xiao, Yamamoto, Yancey, Yang, Yap, Yazback, Yeeles, Yu,
  Yu, Yuen, Yvert, Zadro{\.z}ny, Zanolin, Zappa, Zelenova, Zendri, Zevin,
  Zhang, Zhang, Zhang, Zhao, Zhou, Zhou, Zhu, Zimmerman, Zlochower, Zucker,
  Zweizig, \& {LIGO Scientific Collaboration and Virgo
  Collaboration}}]{abbott:19}
Abbott, B.~P., Abbott, R., Abbott, T.~D., {et~al.} 2019{\natexlab{a}}, PhRvX,
  9, 031040, \dodoi{10.1103/PhysRevX.9.031040}

\bibitem[{Abbott {et~al.}(2019{\natexlab{b}})Abbott, Abbott, Abbott, Acernese,
  Ackley, Adams, Adams, Addesso, Adhikari, Adya, Affeldt, Agarwal, Agathos,
  Agatsuma, Aggarwal, Aguiar, Aiello, Ain, Ajith, Allen, Allen, Allocca, Aloy,
  Altin, Amato, Ananyeva, Anderson, Anderson, Angelova, Antier, Appert, Arai,
  Araya, Areeda, Ar{\`e}ne, Arnaud, Arun, Ascenzi, Ashton, Ast, Aston, Astone,
  Atallah, Aubin, Aufmuth, Aulbert, AultONeal, Austin, {Avila-Alvarez}, Babak,
  Bacon, Badaracco, Bader, Bae, Baker, Baldaccini, Ballardin, Ballmer,
  Banagiri, Barayoga, Barclay, Barish, Barker, Barkett, Barnum, Barone, Barr,
  Barsotti, Barsuglia, Barta, Bartlett, Bartos, Bassiri, Basti, Batch, Bawaj,
  Bayley, Bazzan, B{\'e}csy, Beer, Bejger, Belahcene, Bell, Beniwal, Bensch,
  Berger, Bergmann, Bernuzzi, Bero, Berry, Bersanetti, Bertolini, Betzwieser,
  Bhandare, Bilenko, Bilgili, Billingsley, Billman, Birch, Birney, Birnholtz,
  Biscans, Biscoveanu, Bisht, Bitossi, Bizouard, Blackburn, Blackman, Blair,
  Blair, Blair, Bloemen, Bock, Bode, Boer, Boetzel, Bogaert, Bohe, Bondu,
  Bonilla, Bonnand, Booker, Boom, Booth, Bork, Boschi, Bose, Bossie, Bossilkov,
  Bosveld, Bouffanais, Bozzi, Bradaschia, Brady, Bramley, Branchesi, Brau,
  Briant, Brighenti, Brillet, Brinkmann, Brisson, Brockill, Brooks, Brown,
  Brunett, Buchanan, Buikema, Bulik, Bulten, Buonanno, Buskulic, Buy, Byer,
  Cabero, Cadonati, Cagnoli, Cahillane, Bustillo, Callister, Calloni, Camp,
  Canepa, Canizares, Cannon, Cao, Cao, Capano, Capocasa, Carbognani, Caride,
  Carney, Carullo, Diaz, Casentini, Caudill, Cavagli{\`a}, Cavalier, Cavalieri,
  Cella, Cepeda, {Cerd{\'a}-Dur{\'a}n}, Cerretani, Cesarini, Chaibi,
  Chamberlin, Chan, Chao, Charlton, Chase, {Chassande-Mottin}, Chatterjee,
  Chatziioannou, Cheeseboro, Chen, Chen, Chen, Cheng, Chia, Chincarini,
  Chiummo, Chmiel, Cho, Cho, Chow, Christensen, Chu, Chua, Chua, Chung, Chung,
  Ciani, Ciobanu, Ciolfi, Cipriano, Cirelli, Cirone, Clara, Clark, Clearwater,
  Cleva, Cocchieri, Coccia, Cohadon, Cohen, Colla, Collette, Collins, Cominsky,
  Constancio, Conti, Cooper, Corban, Corbitt, {Cordero-Carri{\'o}n}, Corley,
  Cornish, Corsi, Cortese, Costa, Cotesta, Coughlin, Coughlin, Coulon,
  Countryman, Couvares, Covas, Cowan, Coward, Cowart, Coyne, Coyne, Creighton,
  Creighton, Cripe, Crowder, Cullen, Cumming, Cunningham, Cuoco, Canton,
  D{\'a}lya, Danilishin, D'Antonio, Danzmann, Dasgupta, Costa, Dattilo, Dave,
  Davier, Davis, Daw, Day, DeBra, Deenadayalan, Degallaix, De~Laurentis,
  Del{\'e}glise, Del~Pozzo, Demos, Denker, Dent, De~Pietri, Derby, Dergachev,
  De~Rosa, De~Rossi, DeSalvo, {de Varona}, Dhurandhar, D{\'i}az, Dietrich,
  Di~Fiore, Di~Giovanni, Di~Girolamo, Di~Lieto, Ding, Di~Pace, Di~Palma,
  Di~Renzo, Dmitriev, Doctor, Dolique, Donovan, Dooley, Doravari, Dorrington,
  {\'A}lvarez, Downes, Drago, Dreissigacker, Driggers, Du, Dudi, Dupej, Dwyer,
  Easter, Edo, Edwards, Effler, Eggenstein, Ehrens, Eichholz, Eikenberry,
  Eisenmann, Eisenstein, Essick, Estelles, Estevez, Etienne, Etzel, Evans,
  Evans, Fafone, Fair, Fairhurst, Fan, Farinon, Farr, Farr, {Fauchon-Jones},
  Favata, Fays, Fee, Fehrmann, Feicht, Fejer, Feng, {Fernandez-Galiana},
  Ferrante, Ferreira, Ferrini, Fidecaro, Fiori, Fiorucci, Fishbach, Fisher,
  Fishner, {Fitz-Axen}, Flaminio, Fletcher, Fong, Font, Forsyth, Forsyth,
  Fournier, Frasca, Frasconi, Frei, Freise, Frey, Frey, Fritschel, Frolov,
  Fulda, Fyffe, Gabbard, Gadre, Gaebel, Gair, Gammaitoni, Ganija, Gaonkar,
  Garcia, {Garc{\'i}a-Quir{\'o}s}, Garufi, Gateley, Gaudio, Gaur, Gayathri,
  Gemme, Genin, Gennai, George, George, Gergely, Germain, Ghonge, Ghosh, Ghosh,
  Ghosh, Giacomazzo, Giaime, Giardina, Giazotto, Gill, Giordano, Glover, Goetz,
  Goetz, Goncharov, Gonz{\'a}lez, Castro, Gopakumar, Gorodetsky, Gossan,
  Gosselin, Gouaty, Grado, Graef, Granata, Grant, Gras, Gray, Greco, Green,
  Green, Gretarsson, Groot, Grote, Grunewald, Gruning, Guidi, Gulati, Guo,
  Gupta, Gupta, Gushwa, Gustafson, Gustafson, Halim, Hall, Hall, Hamilton,
  Hamilton, Hammond, Haney, Hanke, Hanks, Hanna, Hannam, Hannuksela, Hanson,
  Hardwick, Harms, Harry, Harry, Hart, Haster, Haughian, Healy, Heidmann,
  Heintze, Heitmann, Hello, Hemming, Hendry, Heng, Hennig, Heptonstall,
  Hernandez, Heurs, Hild, Hinderer, Hoak, Hochheim, Hofman, Holland, Holt,
  Holz, Hopkins, Horst, Hough, Houston, Howell, Hreibi, Huerta, Huet, Hughey,
  Hulko, Husa, Huttner, {Huynh-Dinh}, Iess, Indik, Ingram, Inta, Intini, Isa,
  Isac, Isi, Iyer, Izumi, Jacqmin, Jani, Jaranowski, Johnson, Johnson, Jones,
  Jones, Jonker, Ju, Junker, Kalaghatgi, Kalogera, Kamai, Kandhasamy, Kang,
  Kanner, Kapadia, Karki, Karvinen, Kasprzack, Kastaun, Katolik, Katsanevas,
  Katsavounidis, Katzman, Kaufer, Kawabe, Keerthana, K{\'e}f{\'e}lian, Keitel,
  Kemball, Kennedy, Key, Khalili, Khamesra, Khan, Khan, Khan, Khan, Khazanov,
  Kijbunchoo, Kim, Kim, Kim, Kim, Kim, Kim, King, King, {Kinley-Hanlon},
  Kirchhoff, Kissel, Kleybolte, Klimenko, Knowles, Koch, Koehlenbeck, Koley,
  Kondrashov, Kontos, Korobko, Korth, Kowalska, Kozak, Kr{\"a}mer, Kringel,
  Krishnan, Kr{\'o}lak, Kuehn, Kumar, Kumar, Kumar, Kuo, Kutynia, Kwang,
  Lackey, Lai, Landry, Landry, Lang, Lange, Lantz, Lanza, {Lartaux-Vollard},
  Lasky, Laxen, Lazzarini, Lazzaro, Leaci, Leavey, Lee, Lee, Lee, Lee, Lee,
  Lehmann, Lenon, Leonardi, Leroy, Letendre, Levin, Li, Li, Li, Linker,
  Littenberg, Liu, Liu, Lo, Lockerbie, London, Longo, Lorenzini, Loriette,
  Lormand, Losurdo, Lough, Lousto, Lovelace, L{\"u}ck, Lumaca, Lundgren, Lynch,
  Ma, Macas, Macfoy, Machenschalk, MacInnis, Macleod, Hernandez,
  {Maga{\~n}a-Sandoval}, Zertuche, Magee, Majorana, Maksimovic, Man, Mandic,
  Mangano, Mansell, Manske, Mantovani, Marchesoni, Marion, M{\'a}rka,
  M{\'a}rka, Markakis, Markosyan, Markowitz, Maros, Marquina, Martelli,
  Martellini, Martin, Martin, Martynov, Mason, Massera, Masserot, Massinger,
  {Masso-Reid}, Mastrogiovanni, Matas, Matichard, Matone, Mavalvala, Mazumder,
  McCann, McCarthy, McClelland, McCormick, McCuller, McGuire, McIver, McManus,
  McRae, McWilliams, Meacher, Meadors, Mehmet, Meidam, {Mejuto-Villa}, Melatos,
  Mendell, {Mendoza-Gandara}, Mercer, Mereni, Merilh, Merzougui, Meshkov,
  Messenger, Messick, Metzdorff, Meyers, Miao, Michel, Middleton, Mikhailov,
  Milano, Miller, Miller, Miller, Miller, Millhouse, Mills, {Milovich-Goff},
  Minazzoli, Minenkov, Ming, Mishra, Mitra, Mitrofanov, Mitselmakher,
  Mittleman, Moffa, Mogushi, Mohan, Mohapatra, Montani, Moore, Moraru, Moreno,
  Morisaki, Mours, {Mow-Lowry}, Mueller, Muir, Mukherjee, Mukherjee, Mukherjee,
  Mukund, Mullavey, Munch, Mu{\~n}iz, Muratore, Murray, Nagar, Napier,
  Nardecchia, Naticchioni, Nayak, Neilson, Nelemans, Nelson, Nery, Neunzert,
  Nevin, Newport, Ng, Ng, Nguyen, Nguyen, Nichols, Nielsen, Nissanke, Nitz,
  Nocera, Nolting, North, Nuttall, Obergaulinger, Oberling, O'Brien, O'Dea,
  Ogin, Oh, Oh, Ohme, Ohta, Okada, Oliver, Oppermann, Oram, O'Reilly, Ormiston,
  Ortega, O'Shaughnessy, Ossokine, Ottaway, Overmier, Owen, Pace, Pagano, Page,
  Page, Pai, Pai, Palamos, Palashov, Palomba, {Pal-Singh}, Pan, Pan, Pang,
  Pang, Pankow, Pannarale, Pant, Paoletti, Paoli, Papa, Parida, Parker,
  Pascucci, Pasqualetti, Passaquieti, Passuello, Patil, Patricelli, Pearlstone,
  Pedersen, Pedraza, Pedurand, Pekowsky, Pele, Penn, Perez, Perreca, Perri,
  Pfeiffer, Phelps, Phukon, Piccinni, Pichot, Piergiovanni, Pierro, Pillant,
  Pinard, Pinto, Pirello, Pitkin, Poggiani, Popolizio, Porter, Possenti, Post,
  Powell, Prasad, Pratt, Pratten, Predoi, Prestegard, Principe, Privitera,
  Prodi, Prokhorov, Puncken, Punturo, Puppo, P{\"u}rrer, Qi, Quetschke,
  Quintero, {Quitzow-James}, Raab, Rabeling, Radkins, Raffai, Raja, Rajan,
  Rajbhandari, Rakhmanov, Ramirez, {Ramos-Buades}, Rana, Rapagnani, Raymond,
  Razzano, Read, Regimbau, Rei, Reid, Reitze, Ren, Ricci, Ricker,
  Riemenschneider, Riles, Rizzo, Robertson, Robie, Robinet, Robson, Rocchi,
  Rolland, Rollins, Roma, Romano, Romel, Romie, Rosi{\'n}ska, Ross, Rowan,
  R{\"u}diger, Ruggi, Rutins, Ryan, Sachdev, Sadecki, Sakellariadou, Salconi,
  Saleem, Salemi, Samajdar, Sammut, Sampson, Sanchez, Sanchez, {Sanchis-Gual},
  Sandberg, Sanders, Sarin, Sassolas, Sathyaprakash, Saulson, Sauter, Savage,
  Sawadsky, Schale, Scheel, Scheuer, Schmidt, Schnabel, Schofield,
  Sch{\"o}nbeck, Schreiber, Schuette, Schulte, Schutz, Schwalbe, Scott, Scott,
  Seidel, Sellers, Sengupta, Sentenac, Sequino, Sergeev, Setyawati, Shaddock,
  Shaffer, Shah, Shahriar, Shaner, Shao, Shapiro, Shawhan, Shen, Shoemaker,
  Shoemaker, Siellez, Siemens, Sieniawska, Sigg, Silva, Singer, Singh, Singhal,
  Sintes, Slagmolen, {Slaven-Blair}, Smith, Smith, Smith, Somala, Son, Sorazu,
  Sorrentino, Souradeep, Spencer, Srivastava, Staats, Steinke, Steinlechner,
  Steinlechner, Steinmeyer, Steltner, Stevenson, Stocks, Stone, Stops, Strain,
  Stratta, Strigin, Strunk, Sturani, Stuver, Summerscales, Sun, Sunil, Suresh,
  Sutton, Swinkels, Szczepa{\'n}czyk, Tacca, Tait, Talbot, Talukder, Tanner,
  T{\'a}pai, Taracchini, Tasson, Taylor, Taylor, Tewari, Theeg, Thies, Thomas,
  Thomas, Thomas, Thorne, Thrane, Tiwari, Tiwari, Tokmakov, Toland, Tonelli,
  Tornasi, {Torres-Forn{\'e}}, Torrie, T{\"o}yr{\"a}, Travasso, Traylor,
  Trinastic, Tringali, Trozzo, Tsang, Tse, Tso, Tsuna, Tsukada, Tuyenbayev,
  Ueno, Ugolini, Urban, Usman, Vahlbruch, Vajente, Valdes, {van Bakel}, {van
  Beuzekom}, {van den Brand}, Van Den~Broeck, {Vander-Hyde}, {van der Schaaf},
  {van Heijningen}, {van Veggel}, Vardaro, Varma, Vass, Vas{\'u}th, Vecchio,
  Vedovato, Veitch, Veitch, Venkateswara, Venugopalan, Verkindt, Vetrano,
  Vicer{\'e}, Viets, Vinciguerra, Vine, Vinet, Vitale, Vo, Vocca, Vorvick,
  Vyatchanin, Wade, Wade, Wade, Walet, Walker, Wallace, Walsh, Wang, Wang,
  Wang, Wang, Wang, Ward, Warner, Was, Watchi, Weaver, Wei, Weinert, Weinstein,
  Weiss, Wellmann, Wen, Wessel, We{\ss}els, Westerweck, Wette, Whelan, Whiting,
  Whittle, Wilken, Williams, Williams, Williamson, Willis, Willke, Wimmer,
  Winkler, Wipf, Wittel, Woan, Woehler, Wofford, Wong, Worden, Wright, Wu,
  Wysocki, Xiao, Yam, Yamamoto, Yancey, Yang, Yap, Yazback, Yu, Yu, Yvert,
  Zadro{\.z}ny, Zanolin, Zelenova, Zendri, Zevin, Zhang, Zhang, Zhang, Zhang,
  Zhang, Zhao, Zhou, Zhou, Zhu, Zhu, Zimmerman, Zlochower, Zucker, Zweizig, \&
  {LIGO Scientific Collaboration and Virgo Collaboration}}]{abbott:19a}
---. 2019{\natexlab{b}}, PhRvX, 9, 011001, \dodoi{10.1103/PhysRevX.9.011001}

\bibitem[{Akmal {et~al.}(1998)Akmal, Pandharipande, \& Ravenhall}]{akmal:98}
Akmal, A., Pandharipande, V.~R., \& Ravenhall, D.~G. 1998, PhRvC, 58, 1804,
  \dodoi{10.1103/PhysRevC.58.1804}

\bibitem[{Aloy {et~al.}(2019)Aloy, Ib{\'a}{\~n}ez, {Sanchis-Gual},
  Obergaulinger, Font, Serna, \& Marquina}]{aloy:19}
Aloy, M.~A., Ib{\'a}{\~n}ez, J.~M., {Sanchis-Gual}, N., {et~al.} 2019, MNRAS,
  484, 4980, \dodoi{10.1093/mnras/stz293}

\bibitem[{Antoniadis {et~al.}(2013)Antoniadis, Freire, Wex, Tauris, Lynch, {van
  Kerkwijk}, Kramer, Bassa, Dhillon, Driebe, Hessels, Kaspi, Kondratiev,
  Langer, Marsh, McLaughlin, Pennucci, Ransom, Stairs, {van Leeuwen}, Verbiest,
  \& Whelan}]{antoniadis:13}
Antoniadis, J., Freire, P. C.~C., Wex, N., {et~al.} 2013, Sci, 340, 1233232,
  \dodoi{10.1126/science.1233232}

\bibitem[{Banik(2014)}]{banik:14}
Banik, S. 2014, PhRvC, 89, 035807, \dodoi{10.1103/PhysRevC.89.035807}

\bibitem[{Banik {et~al.}(2014)Banik, Hempel, \& Bandyopadhyay}]{banik:14a}
Banik, S., Hempel, M., \& Bandyopadhyay, D. 2014, ApJS, 214, 22,
  \dodoi{10.1088/0067-0049/214/2/22}

\bibitem[{Belczynski {et~al.}(2016)Belczynski, Holz, Bulik, \&
  O'Shaughnessy}]{belczynski:16}
Belczynski, K., Holz, D.~E., Bulik, T., \& O'Shaughnessy, R. 2016, Natur, 534,
  512, \dodoi{10.1038/nature18322}

\bibitem[{Belczynski {et~al.}(2012)Belczynski, Wiktorowicz, Fryer, Holz, \&
  Kalogera}]{belczynski:12}
Belczynski, K., Wiktorowicz, G., Fryer, C.~L., Holz, D.~E., \& Kalogera, V.
  2012, ApJ, 757, 91, \dodoi{10.1088/0004-637X/757/1/91}

\bibitem[{Belczynski {et~al.}(2019)Belczynski, Klencki, Fields, Olejak, Berti,
  Meynet, Fryer, Holz, O'Shaughnessy, Brown, Bulik, Leung, Nomoto, Madau,
  Hirschi, Jones, Mondal, Chruslinska, Drozda, Gerosa, Doctor, Giersz, Ekstrom,
  Georgy, Askar, Wysocki, Natan, Farr, Wiktorowicz, Miller, Farr, \&
  Lasota}]{belczynski:19}
Belczynski, K., Klencki, J., Fields, C.~E., {et~al.} 2019, arXiv:1706.07053
  [astro-ph, physics:gr-qc].
\newblock \doarXiv{1706.07053}

\bibitem[{Bethe(1990)}]{bethe:90}
Bethe, H.~A. 1990, RvMP, 62, 801, \dodoi{10.1103/RevModPhys.62.801}

\bibitem[{Bethe \& Wilson(1985)}]{bethe:85}
Bethe, H.~A., \& Wilson, J.~R. 1985, ApJ, 295, 14, \dodoi{10.1086/163343}

\bibitem[{Bollig {et~al.}(2017)Bollig, Janka, Lohs, {Mart{\'i}nez-Pinedo},
  Horowitz, \& Melson}]{bollig:17}
Bollig, R., Janka, H.-T., Lohs, A., {et~al.} 2017, PhRvL, 119, 242702,
  \dodoi{10.1103/PhysRevLett.119.242702}

\bibitem[{Bruenn(1985)}]{bruenn:85}
Bruenn, S.~W. 1985, ApJS, 58, 771, \dodoi{10.1086/191056}

\bibitem[{Buras {et~al.}(2006)Buras, Janka, Rampp, \& Kifonidis}]{buras:06}
Buras, R., Janka, H.-T., Rampp, M., \& Kifonidis, K. 2006, A\&A, 457, 281,
  \dodoi{10.1051/0004-6361:20054654}

\bibitem[{Burrows(1986)}]{burrows:86}
Burrows, A. 1986, ApJ, 300, 488, \dodoi{10.1086/163826}

\bibitem[{Burrows(1988)}]{burrows:88}
---. 1988, ApJ, 334, 891, \dodoi{10.1086/166885}

\bibitem[{Burrows {et~al.}(2020)Burrows, Radice, Vartanyan, Nagakura, Skinner,
  \& Dolence}]{burrows:20}
Burrows, A., Radice, D., Vartanyan, D., {et~al.} 2020, MNRAS, 491, 2715,
  \dodoi{10.1093/mnras/stz3223}

\bibitem[{Burrows {et~al.}(2006)Burrows, Reddy, \& Thompson}]{burrows:06}
Burrows, A., Reddy, S., \& Thompson, T.~A. 2006, NuPhA, 777, 356,
  \dodoi{10.1016/j.nuclphysa.2004.06.012}

\bibitem[{Burrows {et~al.}(2018)Burrows, Vartanyan, Dolence, Skinner, \&
  Radice}]{burrows:18}
Burrows, A., Vartanyan, D., Dolence, J.~C., Skinner, M.~A., \& Radice, D. 2018,
  SSRv, 214, 33, \dodoi{10.1007/s11214-017-0450-9}

\bibitem[{Carbone(2019)}]{carbone:19}
Carbone, A. 2019, arXiv:1908.04736 [astro-ph, physics:gr-qc, physics:nucl-ex,
  physics:nucl-th].
\newblock \doarXiv{1908.04736}

\bibitem[{Carbone \& Schwenk(2019)}]{carbone:19a}
Carbone, A., \& Schwenk, A. 2019, PhRvC, 100, 025805,
  \dodoi{10.1103/PhysRevC.100.025805}

\bibitem[{Cardall {et~al.}(2013)Cardall, Endeve, \& Mezzacappa}]{cardall:13}
Cardall, C.~Y., Endeve, E., \& Mezzacappa, A. 2013, PhRvD, 88, 023011,
  \dodoi{10.1103/PhysRevD.88.023011}

\bibitem[{{Cerd{\'a}-Dur{\'a}n} {et~al.}(2013){Cerd{\'a}-Dur{\'a}n}, DeBrye,
  Aloy, Font, \& Obergaulinger}]{cerda-duran:13}
{Cerd{\'a}-Dur{\'a}n}, P., DeBrye, N., Aloy, M.~A., Font, J.~A., \&
  Obergaulinger, M. 2013, ApJ, 779, L18, \dodoi{10.1088/2041-8205/779/2/L18}

\bibitem[{Char {et~al.}(2015)Char, Banik, \& Bandyopadhyay}]{char:15}
Char, P., Banik, S., \& Bandyopadhyay, D. 2015, ApJ, 809, 116,
  \dodoi{10.1088/0004-637X/809/2/116}

\bibitem[{Colgate \& White(1966)}]{colgate:66}
Colgate, S.~A., \& White, R.~H. 1966, ApJ, 143, 626, \dodoi{10.1086/148549}

\bibitem[{Constantinou {et~al.}(2014)Constantinou, Muccioli, Prakash, \&
  Lattimer}]{constantinou:14}
Constantinou, C., Muccioli, B., Prakash, M., \& Lattimer, J.~M. 2014, PhRvC,
  89, 065802, \dodoi{10.1103/PhysRevC.89.065802}

\bibitem[{Cook {et~al.}(1994)Cook, Shapiro, \& Teukolsky}]{cook:94}
Cook, G.~B., Shapiro, S.~L., \& Teukolsky, S.~A. 1994, ApJ, 424, 823,
  \dodoi{10.1086/173934}

\bibitem[{Couch(2019)}]{couch:19a}
Couch, S. 2019, in prep.

\bibitem[{Couch(2013)}]{couch:13}
Couch, S.~M. 2013, ApJ, 765, 29, \dodoi{10.1088/0004-637X/765/1/29}

\bibitem[{Couch {et~al.}(2019)Couch, Warren, \& O'Connor}]{couch:19}
Couch, S.~M., Warren, M.~L., \& O'Connor, E.~P. 2019, arXiv:1902.01340
  [astro-ph].
\newblock \doarXiv{1902.01340}

\bibitem[{Danielewicz(2002)}]{danielewicz:02}
Danielewicz, P. 2002, Sci, 298, 1592, \dodoi{10.1126/science.1078070}

\bibitem[{Dessart {et~al.}(2006)Dessart, Burrows, Livne, \& Ott}]{dessart:06}
Dessart, L., Burrows, A., Livne, E., \& Ott, C.~D. 2006, ApJ, 645, 534,
  \dodoi{10.1086/504068}

\bibitem[{Dimmelmeier {et~al.}(2002{\natexlab{a}})Dimmelmeier, Font, \&
  M{\"u}ller}]{dimmelmeier:02}
Dimmelmeier, H., Font, J.~A., \& M{\"u}ller, E. 2002{\natexlab{a}}, A\&A, 388,
  917, \dodoi{10.1051/0004-6361:20020563}

\bibitem[{Dimmelmeier {et~al.}(2002{\natexlab{b}})Dimmelmeier, Font, \&
  M{\"u}ller}]{dimmelmeier:02a}
---. 2002{\natexlab{b}}, A\&A, 393, 523, \dodoi{10.1051/0004-6361:20021053}

\bibitem[{Dimmelmeier {et~al.}(2005)Dimmelmeier, Novak, Font, Ib{\'a}{\~n}ez,
  \& M{\"u}ller}]{dimmelmeier:05}
Dimmelmeier, H., Novak, J., Font, J.~A., Ib{\'a}{\~n}ez, J.~M., \& M{\"u}ller,
  E. 2005, PhRvD, 71, 064023, \dodoi{10.1103/PhysRevD.71.064023}

\bibitem[{Dubey {et~al.}(2009)Dubey, Antypas, Ganapathy, Reid, Riley, Sheeler,
  Siegel, \& Weide}]{dubey:09}
Dubey, A., Antypas, K., Ganapathy, M.~K., {et~al.} 2009, ParC, 35, 512,
  \dodoi{10.1016/j.parco.2009.08.001}

\bibitem[{Ebinger {et~al.}(2018)Ebinger, Curtis, Fr{\"o}hlich, Hempel, Perego,
  Liebend{\"o}rfer, \& Thielemann}]{ebinger:18}
Ebinger, K., Curtis, S., Fr{\"o}hlich, C., {et~al.} 2018, ApJ, 870, 1,
  \dodoi{10.3847/1538-4357/aae7c9}

\bibitem[{Ertl {et~al.}(2016)Ertl, Janka, Woosley, Sukhbold, \&
  Ugliano}]{ertl:16}
Ertl, T., Janka, H.-T., Woosley, S.~E., Sukhbold, T., \& Ugliano, M. 2016, ApJ,
  818, 124, \dodoi{10.3847/0004-637X/818/2/124}

\bibitem[{Esin {et~al.}(1998)Esin, Narayan, Cui, Grove, \& Zhang}]{esin:98}
Esin, A.~A., Narayan, R., Cui, W., Grove, J.~E., \& Zhang, S.-N. 1998, ApJ,
  505, 854, \dodoi{10.1086/306186}

\bibitem[{Espino \& Paschalidis(2019)}]{espino:19}
Espino, P.~L., \& Paschalidis, V. 2019, PhRvD, 99, 083017,
  \dodoi{10.1103/PhysRevD.99.083017}

\bibitem[{Fattoyev {et~al.}(2010)Fattoyev, Horowitz, Piekarewicz, \&
  Shen}]{fattoyev:10}
Fattoyev, F.~J., Horowitz, C.~J., Piekarewicz, J., \& Shen, G. 2010, PhRvC, 82,
  055803, \dodoi{10.1103/PhysRevC.82.055803}

\bibitem[{Fischer {et~al.}(2012)Fischer, {Mart{\'i}nez-Pinedo}, Hempel, \&
  Liebend{\"o}rfer}]{fischer:12}
Fischer, T., {Mart{\'i}nez-Pinedo}, G., Hempel, M., \& Liebend{\"o}rfer, M.
  2012, PhRvD, 85, 083003, \dodoi{10.1103/PhysRevD.85.083003}

\bibitem[{Fischer {et~al.}(2009)Fischer, Whitehouse, Mezzacappa, Thielemann, \&
  Liebend{\"o}rfer}]{fischer:09}
Fischer, T., Whitehouse, S.~C., Mezzacappa, A., Thielemann, F.-K., \&
  Liebend{\"o}rfer, M. 2009, A\&A, 499, 1, \dodoi{10.1051/0004-6361/200811055}

\bibitem[{Fischer {et~al.}(2010)Fischer, Whitehouse, Mezzacappa, Thielemann, \&
  Liebend{\"o}rfer}]{fischer:10}
---. 2010, A\&A, 517, A80, \dodoi{10.1051/0004-6361/200913106}

\bibitem[{Fischer {et~al.}(2018)Fischer, Bastian, Wu, Baklanov, Sorokina,
  Blinnikov, Typel, Kl{\"a}hn, \& Blaschke}]{fischer:18}
Fischer, T., Bastian, N.-U.~F., Wu, M.-R., {et~al.} 2018, NatAs, 2, 980,
  \dodoi{10.1038/s41550-018-0583-0}

\bibitem[{Fryxell {et~al.}(2000)Fryxell, Olson, Ricker, Timmes, Zingale, Lamb,
  MacNeice, Rosner, Truran, \& Tufo}]{fryxell:00}
Fryxell, B., Olson, K., Ricker, P., {et~al.} 2000, ApJS, 131, 273,
  \dodoi{10.1086/317361}

\bibitem[{Fuller \& Ma(2019)}]{fuller:19}
Fuller, J., \& Ma, L. 2019, ApJ, 881, L1, \dodoi{10.3847/2041-8213/ab339b}

\bibitem[{Gossan {et~al.}(2019)Gossan, Fuller, \& Roberts}]{gossan:19}
Gossan, S.~E., Fuller, J., \& Roberts, L.~F. 2019, MNRAS, stz3243,
  \dodoi{10.1093/mnras/stz3243}

\bibitem[{Granqvist(2019)}]{granqvist:19}
Granqvist, E. 2019, Stockholm University Bachelor thesis: Approximating General
  Relativistic Effects in {{Newtonian}} Hydrodynamic Supernova Simulations,
  http://urn.kb.se/resolve?urn=urn:nbn:se:su:diva-169517

\bibitem[{Hempel {et~al.}(2012)Hempel, Fischer, {Schaffner-Bielich}, \&
  Liebend{\"o}rfer}]{hempel:12}
Hempel, M., Fischer, T., {Schaffner-Bielich}, J., \& Liebend{\"o}rfer, M. 2012,
  ApJ, 748, 70, \dodoi{10.1088/0004-637X/748/1/70}

\bibitem[{Hempel {et~al.}(2016)Hempel, Heinimann, Yudin, Iosilevskiy,
  Liebend{\"o}rfer, \& Thielemann}]{hempel:16}
Hempel, M., Heinimann, O., Yudin, A., {et~al.} 2016, PhRvD, 94, 103001,
  \dodoi{10.1103/PhysRevD.94.103001}

\bibitem[{Horiuchi {et~al.}(2017)Horiuchi, Nakamura, Takiwaki, \&
  Kotake}]{horiuchi:17}
Horiuchi, S., Nakamura, K., Takiwaki, T., \& Kotake, K. 2017, JPhG, 44, 114001,
  \dodoi{10.1088/1361-6471/aa8f1f}

\bibitem[{Horowitz(2002)}]{horowitz:02}
Horowitz, C.~J. 2002, PhRvD, 65, 043001, \dodoi{10.1103/PhysRevD.65.043001}

\bibitem[{H{\"u}depohl(2014)}]{hudepohlthesis:14}
H{\"u}depohl, L. 2014, Dissertation, Technische Universit{\"a}t M{\"u}nchen,
  M{\"u}nchen

\bibitem[{Hunter(2007)}]{hunter:07}
Hunter, J.~D. 2007, CSE, 9, 90, \dodoi{10.1109/MCSE.2007.55}

\bibitem[{Ishizuka {et~al.}(2008)Ishizuka, Ohnishi, Tsubakihara, Sumiyoshi, \&
  Yamada}]{ishizuka:08}
Ishizuka, C., Ohnishi, A., Tsubakihara, K., Sumiyoshi, K., \& Yamada, S. 2008,
  JPhG, 35, 085201, \dodoi{10.1088/0954-3899/35/8/085201}

\bibitem[{Janka(2001)}]{janka:01}
Janka, H.-T. 2001, A\&A, 368, 527, \dodoi{10.1051/0004-6361:20010012}

\bibitem[{Janka(2012)}]{janka:12}
---. 2012, ARNPS, 62, 407, \dodoi{10.1146/annurev-nucl-102711-094901}

\bibitem[{Jones {et~al.}(2001)Jones, Oliphant, Peterson, {et~al.}}]{jones:01}
Jones, E., Oliphant, T., Peterson, P., {et~al.} 2001, {{SciPy}}: {{Open}}
  Source Scientific Tools for {{Python}}.
\newblock \url{http://www.scipy.org/}

\bibitem[{Kachelrie{\ss} {et~al.}(2005)Kachelrie{\ss}, Tom{\`a}s, Buras, Janka,
  Marek, \& Rampp}]{kachelriess:05}
Kachelrie{\ss}, M., Tom{\`a}s, R., Buras, R., {et~al.} 2005, PhRvD, 71, 063003,
  \dodoi{10.1103/PhysRevD.71.063003}

\bibitem[{Kitaura {et~al.}(2006)Kitaura, Janka, \& Hillebrandt}]{kitaura:06}
Kitaura, F.~S., Janka, H.-T., \& Hillebrandt, W. 2006, A\&A, 450, 345,
  \dodoi{10.1051/0004-6361:20054703}

\bibitem[{Kubota {et~al.}(1998)Kubota, Tanaka, Makishima, Ueda, Dotani, Inoue,
  \& Yamaoka}]{kubota:98}
Kubota, A., Tanaka, Y., Makishima, K., {et~al.} 1998, PASJ, 50, 667,
  \dodoi{10.1093/pasj/50.6.667}

\bibitem[{Lattimer(1981)}]{lattimer:81}
Lattimer, J.~M. 1981, ARNPS, 31, 337,
  \dodoi{10.1146/annurev.ns.31.120181.002005}

\bibitem[{Lattimer \& Prakash(2000)}]{lattimer:00}
Lattimer, J.~M., \& Prakash, M. 2000, PhR, 333-334, 121,
  \dodoi{10.1016/S0370-1573(00)00019-3}

\bibitem[{Lattimer \& Swesty(1991)}]{lattimer:91}
Lattimer, J.~M., \& Swesty, F.~D. 1991, NuPhA, 535, 331,
  \dodoi{10.1016/0375-9474(91)90452-C}

\bibitem[{Liebend{\"o}rfer {et~al.}(2004)Liebend{\"o}rfer, Messer, Mezzacappa,
  Bruenn, Cardall, \& Thielemann}]{liebendorfer:04}
Liebend{\"o}rfer, M., Messer, O. E.~B., Mezzacappa, A., {et~al.} 2004, ApJS,
  150, 263, \dodoi{10.1086/380191}

\bibitem[{Liebend{\"o}rfer {et~al.}(2001)Liebend{\"o}rfer, Mezzacappa, \&
  Thielemann}]{liebendorfer:01}
Liebend{\"o}rfer, M., Mezzacappa, A., \& Thielemann, F.-K. 2001, PhRvD, 63,
  104003, \dodoi{10.1103/PhysRevD.63.104003}

\bibitem[{Liebend{\"o}rfer {et~al.}(2005)Liebend{\"o}rfer, Rampp, Janka, \&
  Mezzacappa}]{liebendorfer:05}
Liebend{\"o}rfer, M., Rampp, M., Janka, H.-T., \& Mezzacappa, A. 2005, ApJ,
  620, 840, \dodoi{10.1086/427203}

\bibitem[{Liebend{\"o}rfer {et~al.}(2002)Liebend{\"o}rfer, Rosswog, \&
  Thielemann}]{liebendorfer:02}
Liebend{\"o}rfer, M., Rosswog, S., \& Thielemann, F.-K. 2002, ApJS, 141, 229,
  \dodoi{10.1086/339872}

\bibitem[{Mandel \& {de Mink}(2016)}]{mandel:16}
Mandel, I., \& {de Mink}, S.~E. 2016, MNRAS, 458, 2634,
  \dodoi{10.1093/mnras/stw379}

\bibitem[{Marek {et~al.}(2006)Marek, Dimmelmeier, Janka, M{\"u}ller, \&
  Buras}]{marek:06}
Marek, A., Dimmelmeier, H., Janka, H.-T., M{\"u}ller, E., \& Buras, R. 2006,
  A\&A, 445, 273, \dodoi{10.1051/0004-6361:20052840}

\bibitem[{Margueron {et~al.}(2018)Margueron, Hoffmann~Casali, \&
  Gulminelli}]{margueron:18}
Margueron, J., Hoffmann~Casali, R., \& Gulminelli, F. 2018, PhRvC, 97, 025805,
  \dodoi{10.1103/PhysRevC.97.025805}

\bibitem[{{Mart{\'i}nez-Pinedo} {et~al.}(2012){Mart{\'i}nez-Pinedo}, Fischer,
  Lohs, \& Huther}]{martinez-pinedo:12}
{Mart{\'i}nez-Pinedo}, G., Fischer, T., Lohs, A., \& Huther, L. 2012, PhRvL,
  109, 251104, \dodoi{10.1103/PhysRevLett.109.251104}

\bibitem[{Melson {et~al.}(2015)Melson, Janka, Bollig, Hanke, Marek, \&
  M{\"u}ller}]{melson:15}
Melson, T., Janka, H.-T., Bollig, R., {et~al.} 2015, ApJ, 808, L42,
  \dodoi{10.1088/2041-8205/808/2/L42}

\bibitem[{Miller {et~al.}(2019)Miller, Lamb, Dittmann, Bogdanov, Arzoumanian,
  Gendreau, Guillot, Harding, Ho, Lattimer, Ludlam, Mahmoodifar, Morsink, Ray,
  Strohmayer, Wood, Enoto, Foster, Okajima, Prigozhin, \& Soong}]{miller:19}
Miller, M.~C., Lamb, F.~K., Dittmann, A.~J., {et~al.} 2019, ApJ, 887, L24,
  \dodoi{10.3847/2041-8213/ab50c5}

\bibitem[{Morozova {et~al.}(2018)Morozova, Radice, Burrows, \&
  Vartanyan}]{morozova:18}
Morozova, V., Radice, D., Burrows, A., \& Vartanyan, D. 2018, ApJ, 861, 10,
  \dodoi{10.3847/1538-4357/aac5f1}

\bibitem[{Morrison {et~al.}(2004)Morrison, Baumgarte, \& Shapiro}]{morrison:04}
Morrison, I.~A., Baumgarte, T.~W., \& Shapiro, S.~L. 2004, ApJ, 610, 941,
  \dodoi{10.1086/421897}

\bibitem[{Nagakura {et~al.}(2019)Nagakura, Burrows, Radice, \&
  Vartanyan}]{nagakura:19}
Nagakura, H., Burrows, A., Radice, D., \& Vartanyan, D. 2019, arXiv:1912.07615
  [astro-ph].
\newblock \doarXiv{1912.07615}

\bibitem[{Nakazato {et~al.}(2012)Nakazato, Furusawa, Sumiyoshi, Ohnishi,
  Yamada, \& Suzuki}]{nakazato:12}
Nakazato, K., Furusawa, S., Sumiyoshi, K., {et~al.} 2012, ApJ, 745, 197,
  \dodoi{10.1088/0004-637X/745/2/197}

\bibitem[{Nakazato {et~al.}(2013)Nakazato, Sumiyoshi, Suzuki, Totani, Umeda, \&
  Yamada}]{nakazato:13}
Nakazato, K., Sumiyoshi, K., Suzuki, H., {et~al.} 2013, ApJS, 205, 2,
  \dodoi{10.1088/0067-0049/205/1/2}

\bibitem[{Nakazato \& Suzuki(2019)}]{nakazato:19}
Nakazato, K., \& Suzuki, H. 2019, ApJ, 878, 25,
  \dodoi{10.3847/1538-4357/ab1d4b}

\bibitem[{Nakazato {et~al.}(2018)Nakazato, Suzuki, \& Togashi}]{nakazato:18}
Nakazato, K., Suzuki, H., \& Togashi, H. 2018, PhRvC, 97, 035804,
  \dodoi{10.1103/PhysRevC.97.035804}

\bibitem[{N{\"a}ttil{\"a} {et~al.}(2016)N{\"a}ttil{\"a}, Steiner, Kajava,
  Suleimanov, \& Poutanen}]{nattila:16}
N{\"a}ttil{\"a}, J., Steiner, A.~W., Kajava, J. J.~E., Suleimanov, V.~F., \&
  Poutanen, J. 2016, A\&A, 591, A25, \dodoi{10.1051/0004-6361/201527416}

\bibitem[{Obergaulinger \& Aloy(2019)}]{obergaulinger:19}
Obergaulinger, M., \& Aloy, M.~{\'A}. 2019, arXiv:1909.01105 [astro-ph].
\newblock \doarXiv{1909.01105}

\bibitem[{O'Connor(2015)}]{oconnor:15}
O'Connor, E. 2015, ApJS, 219, 24, \dodoi{10.1088/0067-0049/219/2/24}

\bibitem[{O'Connor \& Ott(2010)}]{oconnor:10}
O'Connor, E., \& Ott, C.~D. 2010, CQGra, 27, 114103,
  \dodoi{10.1088/0264-9381/27/11/114103}

\bibitem[{O'Connor \& Ott(2011)}]{oconnor:11}
---. 2011, ApJ, 730, 70, \dodoi{10.1088/0004-637X/730/2/70}

\bibitem[{O'Connor \& Ott(2013)}]{oconnor:13}
---. 2013, ApJ, 762, 126, \dodoi{10.1088/0004-637X/762/2/126}

\bibitem[{O'Connor {et~al.}(2018)O'Connor, Bollig, Burrows, Couch, Fischer,
  Janka, Kotake, Lentz, Liebend{\"o}rfer, Messer, Mezzacappa, Takiwaki, \&
  Vartanyan}]{oconnor:18}
O'Connor, E., Bollig, R., Burrows, A., {et~al.} 2018, JPhG, 45, 104001,
  \dodoi{10.1088/1361-6471/aadeae}

\bibitem[{O'Connor \& Couch(2018{\natexlab{a}})}]{oconnor:18a}
O'Connor, E.~P., \& Couch, S.~M. 2018{\natexlab{a}}, ApJ, 854, 63,
  \dodoi{10.3847/1538-4357/aaa893}

\bibitem[{O'Connor \& Couch(2018{\natexlab{b}})}]{oconnor:18b}
---. 2018{\natexlab{b}}, ApJ, 865, 81, \dodoi{10.3847/1538-4357/aadcf7}

\bibitem[{Oertel {et~al.}(2016)Oertel, Gulminelli, Provid{\^e}ncia, \&
  Raduta}]{oertel:16}
Oertel, M., Gulminelli, F., Provid{\^e}ncia, C., \& Raduta, A.~R. 2016, EPJA,
  52, 50, \dodoi{10.1140/epja/i2016-16050-1}

\bibitem[{Ott {et~al.}(2018)Ott, Roberts, {da Silva Schneider}, Fedrow, Haas,
  \& Schnetter}]{ott:18}
Ott, C.~D., Roberts, L.~F., {da Silva Schneider}, A., {et~al.} 2018, ApJ, 855,
  L3, \dodoi{10.3847/2041-8213/aaa967}

\bibitem[{Ott {et~al.}(2011)Ott, Reisswig, Schnetter, O'Connor, Sperhake,
  L{\"o}ffler, Diener, Abdikamalov, Hawke, \& Burrows}]{ott:11}
Ott, C.~D., Reisswig, C., Schnetter, E., {et~al.} 2011, PhRvL, 106, 161103,
  \dodoi{10.1103/PhysRevLett.106.161103}

\bibitem[{Pan {et~al.}(2018)Pan, Liebend{\"o}rfer, Couch, \&
  Thielemann}]{pan:18}
Pan, K.-C., Liebend{\"o}rfer, M., Couch, S.~M., \& Thielemann, F.-K. 2018, ApJ,
  857, 13, \dodoi{10.3847/1538-4357/aab71d}

\bibitem[{Pedregosa {et~al.}(2011)Pedregosa, Varoquaux, Gramfort, Michel,
  Thirion, Grisel, Blondel, Prettenhofer, Weiss, Dubourg, Vanderplas, Passos,
  Cournapeau, Brucher, Perrot, \& Duchesnay}]{scikit-learn}
Pedregosa, F., Varoquaux, G., Gramfort, A., {et~al.} 2011, JMLR, 12, 2825

\bibitem[{Peres {et~al.}(2013)Peres, Oertel, \& Novak}]{peres:13}
Peres, B., Oertel, M., \& Novak, J. 2013, PhRvD, 87, 043006,
  \dodoi{10.1103/PhysRevD.87.043006}

\bibitem[{Portegies~Zwart \& McMillan(2000)}]{portegieszwart:00}
Portegies~Zwart, S.~F., \& McMillan, S. L.~W. 2000, ApJ, 528, L17,
  \dodoi{10.1086/312422}

\bibitem[{Prakash {et~al.}(1997)Prakash, Bombaci, Prakash, Ellis, Lattimer, \&
  Knorren}]{prakash:97}
Prakash, M., Bombaci, I., Prakash, M., {et~al.} 1997, PhR, 280, 1,
  \dodoi{10.1016/S0370-1573(96)00023-3}

\bibitem[{Rampp \& Janka(2002)}]{rampp:02}
Rampp, M., \& Janka, H.-T. 2002, A\&A, 396, 361,
  \dodoi{10.1051/0004-6361:20021398}

\bibitem[{Richers {et~al.}(2017)Richers, Ott, Abdikamalov, O'Connor, \&
  Sullivan}]{richers:17}
Richers, S., Ott, C.~D., Abdikamalov, E., O'Connor, E., \& Sullivan, C. 2017,
  PhRvD, 95, 063019, \dodoi{10.1103/PhysRevD.95.063019}

\bibitem[{Roberts(2012)}]{roberts:12}
Roberts, L.~F. 2012, ApJ, 755, 126, \dodoi{10.1088/0004-637X/755/2/126}

\bibitem[{Roberts \& Reddy(2016)}]{roberts:16}
Roberts, L.~F., \& Reddy, S. 2016, in Handbook of {{Supernovae}}, ed. A.~W.
  Alsabti \& P.~Murdin ({Cham}: {Springer International Publishing}), 1--31.
\newblock \url{http://link.springer.com/10.1007/978-3-319-20794-0_5-1}

\bibitem[{Roberts {et~al.}(2012)Roberts, Shen, Cirigliano, Pons, Reddy, \&
  Woosley}]{roberts:12a}
Roberts, L.~F., Shen, G., Cirigliano, V., {et~al.} 2012, PhRvL, 108, 061103,
  \dodoi{10.1103/PhysRevLett.108.061103}

\bibitem[{Rodriguez {et~al.}(2015)Rodriguez, Morscher, Pattabiraman,
  Chatterjee, Haster, \& Rasio}]{rodriguez:15}
Rodriguez, C.~L., Morscher, M., Pattabiraman, B., {et~al.} 2015, PhRvL, 115,
  051101, \dodoi{10.1103/PhysRevLett.115.051101}

\bibitem[{Sagert {et~al.}(2009)Sagert, Fischer, Hempel, Pagliara,
  {Schaffner-Bielich}, Mezzacappa, Thielemann, \& Liebend{\"o}rfer}]{sagert:09}
Sagert, I., Fischer, T., Hempel, M., {et~al.} 2009, PhRvL, 102, 081101,
  \dodoi{10.1103/PhysRevLett.102.081101}

\bibitem[{Schneider {et~al.}(2019{\natexlab{a}})Schneider, Constantinou,
  Muccioli, \& Prakash}]{schneider:19}
Schneider, A.~S., Constantinou, C., Muccioli, B., \& Prakash, M.
  2019{\natexlab{a}}, PhRvC, 100, 025803, \dodoi{10.1103/PhysRevC.100.025803}

\bibitem[{Schneider {et~al.}(2017)Schneider, Roberts, \& Ott}]{schneider:17}
Schneider, A.~S., Roberts, L.~F., \& Ott, C.~D. 2017, PhRvC, 96, 065802,
  \dodoi{10.1103/PhysRevC.96.065802}

\bibitem[{Schneider {et~al.}(2019{\natexlab{b}})Schneider, Roberts, Ott, \&
  O'Connor}]{schneider:19a}
Schneider, A.~S., Roberts, L.~F., Ott, C.~D., \& O'Connor, E.
  2019{\natexlab{b}}, PhRvC, 100, 055802, \dodoi{10.1103/PhysRevC.100.055802}

\bibitem[{Scholberg(2012)}]{scholberg:12}
Scholberg, K. 2012, ARNPS, 62, 81, \dodoi{10.1146/annurev-nucl-102711-095006}

\bibitem[{Sekiguchi \& Shibata(2005)}]{sekiguchi:05}
Sekiguchi, Y.-i., \& Shibata, M. 2005, PhRvD, 71, 084013,
  \dodoi{10.1103/PhysRevD.71.084013}

\bibitem[{Shen {et~al.}(1998{\natexlab{a}})Shen, Toki, Oyamatsu, \&
  Sumiyoshi}]{shen:98}
Shen, H., Toki, H., Oyamatsu, K., \& Sumiyoshi, K. 1998{\natexlab{a}}, NuPhA,
  637, 435, \dodoi{10.1016/S0375-9474(98)00236-X}

\bibitem[{Shen {et~al.}(1998{\natexlab{b}})Shen, Toki, Oyamatsu, \&
  Sumiyoshi}]{shen:98a}
---. 1998{\natexlab{b}}, PThPh, 100, 1013, \dodoi{10.1143/PTP.100.1013}

\bibitem[{Shen {et~al.}(2011)Shen, Toki, Oyamatsu, \& Sumiyoshi}]{shen:11}
---. 2011, ApJS, 197, 20, \dodoi{10.1088/0067-0049/197/2/20}

\bibitem[{Shibata {et~al.}(2011)Shibata, Kiuchi, Sekiguchi, \&
  Suwa}]{shibata:11}
Shibata, M., Kiuchi, K., Sekiguchi, Y.-i., \& Suwa, Y. 2011, PThPh, 125, 1255,
  \dodoi{10.1143/PTP.125.1255}

\bibitem[{Steiner {et~al.}(2013)Steiner, Hempel, \& Fischer}]{steiner:13}
Steiner, A.~W., Hempel, M., \& Fischer, T. 2013, ApJ, 774, 17,
  \dodoi{10.1088/0004-637X/774/1/17}

\bibitem[{Stevenson {et~al.}(2017)Stevenson, {Vigna-G{\'o}mez}, Mandel,
  Barrett, Neijssel, Perkins, \& {de Mink}}]{stevenson:17}
Stevenson, S., {Vigna-G{\'o}mez}, A., Mandel, I., {et~al.} 2017, NatCo, 8,
  14906, \dodoi{10.1038/ncomms14906}

\bibitem[{Sukhbold {et~al.}(2018)Sukhbold, Woosley, \& Heger}]{sukhbold:18}
Sukhbold, T., Woosley, S.~E., \& Heger, A. 2018, ApJ, 860, 93,
  \dodoi{10.3847/1538-4357/aac2da}

\bibitem[{Sumiyoshi {et~al.}(2009)Sumiyoshi, Ishizuka, Ohnishi, Yamada, \&
  Suzuki}]{sumiyoshi:09}
Sumiyoshi, K., Ishizuka, C., Ohnishi, A., Yamada, S., \& Suzuki, H. 2009, ApJ,
  690, L43, \dodoi{10.1088/0004-637X/690/1/L43}

\bibitem[{Sumiyoshi {et~al.}(2007)Sumiyoshi, Yamada, \& Suzuki}]{sumiyoshi:07}
Sumiyoshi, K., Yamada, S., \& Suzuki, H. 2007, ApJ, 667, 382,
  \dodoi{10.1086/520876}

\bibitem[{Sumiyoshi {et~al.}(2008)Sumiyoshi, Yamada, \& Suzuki}]{sumiyoshi:08}
---. 2008, ApJ, 688, 1176, \dodoi{10.1086/592183}

\bibitem[{Sumiyoshi {et~al.}(2006)Sumiyoshi, Yamada, Suzuki, \&
  Chiba}]{sumiyoshi:06}
Sumiyoshi, K., Yamada, S., Suzuki, H., \& Chiba, S. 2006, PhRvL, 97, 091101,
  \dodoi{10.1103/PhysRevLett.97.091101}

\bibitem[{Sumiyoshi {et~al.}(2005)Sumiyoshi, Yamada, Suzuki, Shen, Chiba, \&
  Toki}]{sumiyoshi:05}
Sumiyoshi, K., Yamada, S., Suzuki, H., {et~al.} 2005, ApJ, 629, 922,
  \dodoi{10.1086/431788}

\bibitem[{Summa {et~al.}(2016)Summa, Hanke, Janka, Melson, Marek, \&
  M{\"u}ller}]{summa:16}
Summa, A., Hanke, F., Janka, H.-T., {et~al.} 2016, ApJ, 825, 6,
  \dodoi{10.3847/0004-637X/825/1/6}

\bibitem[{Summa {et~al.}(2018)Summa, Janka, Melson, \& Marek}]{summa:18}
Summa, A., Janka, H.-T., Melson, T., \& Marek, A. 2018, ApJ, 852, 28,
  \dodoi{10.3847/1538-4357/aa9ce8}

\bibitem[{Suwa {et~al.}(2019)Suwa, Sumiyoshi, Nakazato, Takahira, Koshio, Mori,
  \& Wendell}]{suwa:19}
Suwa, Y., Sumiyoshi, K., Nakazato, K., {et~al.} 2019, ApJ, 881, 139,
  \dodoi{10.3847/1538-4357/ab2e05}

\bibitem[{Swesty {et~al.}(1994)Swesty, Lattimer, \& Myra}]{swesty:94}
Swesty, F.~D., Lattimer, J.~M., \& Myra, E.~S. 1994, ApJ, 425, 195,
  \dodoi{10.1086/173974}

\bibitem[{Szkudlarek {et~al.}(2019)Szkudlarek, {Gondek-Rosi{\'n}ska}, Villain,
  \& Ansorg}]{szkudlarek:19}
Szkudlarek, M., {Gondek-Rosi{\'n}ska}, D., Villain, L., \& Ansorg, M. 2019,
  ApJ, 879, 44, \dodoi{10.3847/1538-4357/ab1752}

\bibitem[{Togashi {et~al.}(2017)Togashi, Nakazato, Takehara, Yamamuro, Suzuki,
  \& Takano}]{togashi:17}
Togashi, H., Nakazato, K., Takehara, Y., {et~al.} 2017, NuPhA, 961, 78,
  \dodoi{10.1016/j.nuclphysa.2017.02.010}

\bibitem[{{Torres-Forn{\'e}} {et~al.}(2019){Torres-Forn{\'e}},
  {Cerd{\'a}-Dur{\'a}n}, Obergaulinger, M{\"u}ller, \& Font}]{torres-forne:19}
{Torres-Forn{\'e}}, A., {Cerd{\'a}-Dur{\'a}n}, P., Obergaulinger, M.,
  M{\"u}ller, B., \& Font, J.~A. 2019, PhRvL, 123, 051102,
  \dodoi{10.1103/PhysRevLett.123.051102}

\bibitem[{Ugliano {et~al.}(2012)Ugliano, Janka, Marek, \& Arcones}]{ugliano:12}
Ugliano, M., Janka, H.-T., Marek, A., \& Arcones, A. 2012, ApJ, 757, 69,
  \dodoi{10.1088/0004-637X/757/1/69}

\bibitem[{Vartanyan {et~al.}(2019)Vartanyan, Burrows, \& Radice}]{vartanyan:19}
Vartanyan, D., Burrows, A., \& Radice, D. 2019, MNRAS, 489, 2227,
  \dodoi{10.1093/mnras/stz2307}

\bibitem[{Vitale {et~al.}(2017)Vitale, Lynch, Sturani, \& Graff}]{vitale:17}
Vitale, S., Lynch, R., Sturani, R., \& Graff, P. 2017, CQGra, 34, 03LT01,
  \dodoi{10.1088/1361-6382/aa552e}

\bibitem[{Walk {et~al.}(2019)Walk, Tamborra, Janka, \& Summa}]{walk:19}
Walk, L., Tamborra, I., Janka, H.-T., \& Summa, A. 2019, arXiv:1910.12971
  [astro-ph, physics:hep-ph].
\newblock \doarXiv{1910.12971}

\bibitem[{Warren {et~al.}(2019)Warren, Couch, O'Connor, \&
  Morozova}]{warren:19}
Warren, M.~L., Couch, S.~M., O'Connor, E.~P., \& Morozova, V. 2019,
  arXiv:1912.03328 [astro-ph].
\newblock \doarXiv{1912.03328}

\bibitem[{Wiktorowicz {et~al.}(2013)Wiktorowicz, Belczynski, \&
  Maccarone}]{wiktorowicz:13}
Wiktorowicz, G., Belczynski, K., \& Maccarone, T.~J. 2013, arXiv:1312.5924
  [astro-ph].
\newblock \doarXiv{1312.5924}

\bibitem[{Woosley \& Heger(2007)}]{woosley:07}
Woosley, S., \& Heger, A. 2007, PhR, 442, 269,
  \dodoi{10.1016/j.physrep.2007.02.009}

\bibitem[{Woosley {et~al.}(2002)Woosley, Heger, \& Weaver}]{woosley:02}
Woosley, S.~E., Heger, A., \& Weaver, T.~A. 2002, RvMP, 74, 1015,
  \dodoi{10.1103/RevModPhys.74.1015}

\bibitem[{Woosley \& Weaver(1995)}]{woosley:95}
Woosley, S.~E., \& Weaver, T.~A. 1995, ApJS, 101, 181, \dodoi{10.1086/192237}

\bibitem[{Yasin {et~al.}(2018)Yasin, Sch{\"a}fer, Arcones, \&
  Schwenk}]{yasin:18}
Yasin, H., Sch{\"a}fer, S., Arcones, A., \& Schwenk, A. 2018, arXiv:1812.02002
  [astro-ph, physics:nucl-ex, physics:nucl-th].
\newblock \doarXiv{1812.02002}

\end{thebibliography}
